\documentclass[]{article}
\usepackage{amssymb}
\usepackage{epsfig}
\voffset= - 0.3in
\hoffset= - 1.0in         

\catcode`\@=11
\def\marginnote#1{}

\newcount\hour
\newcount\minute
\newtoks\amorpm
\hour=\time\divide\hour by60
\minute=\time{\multiply\hour by60 \global\advance\minute by-\hour}
\edef\standardtime{{\ifnum\hour<12 \global\amorpm={am}%
        \else\global\amorpm={pm}\advance\hour by-12 \fi
        \ifnum\hour=0 \hour=12 \fi
        \number\hour:\ifnum\minute<10 0\fi\number\minute\the\amorpm}}
\edef\militarytime{\number\hour:\ifnum\minute<10 0\fi\number\minute}

\def\draftlabel#1{{\@bsphack\if@filesw {\let\thepage\relax
   \xdef\@gtempa{\write\@auxout{\string
      \newlabel{#1}{{\@currentlabel}{\thepage}}}}}\@gtempa
   \if@nobreak \ifvmode\nobreak\fi\fi\fi\@esphack}
        \gdef\@eqnlabel{#1}}
\def\@eqnlabel{}
\def\@vacuum{}
\def\draftmarginnote#1{\marginpar{\raggedright\scriptsize\tt#1}}

\def\draft{\oddsidemargin -.5truein
        \def\@oddfoot{\sl preliminary draft \hfil
        \rm\thepage\hfil\sl\today\quad\militarytime}
        \let\@evenfoot\@oddfoot \overfullrule 3pt
        \let\label=\draftlabel
        \let\marginnote=\draftmarginnote
   \def\@eqnnum{(\theequation)\rlap{\kern\marginparsep\tt\@eqnlabel}%
\global\let\@eqnlabel\@vacuum}  }

\makeatletter
\newdimen\normalarrayskip            
\newdimen\minarrayskip               
\normalarrayskip\baselineskip
\minarrayskip\jot
\newif\ifold             \oldtrue            \def\new{\oldfalse}
\def\arraymode{\ifold\relax\else\displaystyle\fi}
\def\eqnumphantom{\phantom{(\theequation)}} 
\def\@arrayskip{\ifold\baselineskip\z@\lineskip\z@
     \else
     \baselineskip\minarrayskip\lineskip1\baselineskip\fi}

\def\@arrayclassz{\ifcase \@lastchclass \@acolampacol \or
\@ampacol \or \or \or \@addamp \or
   \@acolampacol \or \@firstampfalse \@acol \fi
\edef\@preamble{\@preamble
  \ifcase \@chnum
     \hfil$\relax\arraymode\@sharp$\hfil
     \or $\relax\arraymode\@sharp$\hfil
     \or \hfil$\relax\arraymode\@sharp$\fi}}

\def\@array[#1]#2{\setbox\@arstrutbox=\hbox{\vrule
     height\arraystretch \ht\strutbox
     depth\arraystretch \dp\strutbox
width\z@}\@mkpream{#2}\edef\@preamble{\halign \noexpand\@halignto
\bgroup \tabskip\z@ \@arstrut \@preamble \tabskip\z@ \cr}%
\let\@startpbox\@@startpbox \let\@endpbox\@@endpbox
  \if #1t\vtop \else \if#1b\vbox \else \vcenter \fi\fi
  \bgroup \let\par\relax
  \let\@sharp##\let\protect\relax
  \@arrayskip\@preamble}
\def\eqnarray{\stepcounter{equation}%
              \let\@currentlabel=\theequation
              \global\@eqnswtrue
              \global\@eqcnt\z@
              \tabskip\@centering              
              \let\\=\@eqncr
              $$%
            \halign to \displaywidth  \bgroup
             \eqnumphantom \@eqnsel
      \hskip\@centering                               
    $\displaystyle  \tabskip\z@ {##}$%
    &\global\@eqcnt\@ne \hskip 2\arraycolsep
         $ \displaystyle  \arraymode{##}$\hfil
    &\global\@eqcnt\tw@ \hskip 2\arraycolsep
         $\displaystyle\tabskip\z@{##}$\hfil
         \tabskip\@centering
    &{##}\tabskip\z@\cr}
\makeatother

\newfont{\hr}{msbm10}
\newfont{\ams}{msam10}

\font\numbers=cmss12
\font\upright=cmu10 scaled\magstep1
\def\stroke{\vrule height8pt width0.4pt depth-0.1pt}
\def\topfleck{\vrule height8pt width0.5pt depth-5.9pt}
\def\botfleck{\vrule height2pt width0.5pt depth0.1pt}
\def\Zmath{\vcenter{\hbox{\numbers\rlap{\rlap{Z}\kern 0.8pt\topfleck}\kern 2.2pt
                   \rlap Z\kern 6pt\botfleck\kern 1pt}}}
\def\Qmath{\vcenter{\hbox{\upright\rlap{\rlap{Q}\kern
                   3.8pt\stroke}\phantom{Q}}}}
\def\Nmath{\vcenter{\hbox{\upright\rlap{I}\kern 1.7pt N}}}
\def\Cmath{\vcenter{\hbox{\upright\rlap{\rlap{C}\kern
                   3.8pt\stroke}\phantom{C}}}}
\def\Rmath{\vcenter{\hbox{\upright\rlap{I}\kern 1.7pt R}}}
\def\Z{\ifmmode\Zmath\else$\Zmath$\fi}
\def\Q{\ifmmode\Qmath\else$\Qmath$\fi}
\def\N{\ifmmode\Nmath\else$\Nmath$\fi}
\def\C{\ifmmode\Cmath\else$\Cmath$\fi}
\def\R{\ifmmode\Rmath\else$\Rmath$\fi}

\def\d{\partial}
\def\p{\partial}
\def\bea{\begin{eqnarray}}
\def\eea{\end{eqnarray}}

\def\beq{\begin{equation}}
\def\eeq{\end{equation}}
\def\ba{\beq\new\begin{array}{c}}
\def\ea{\end{array}\eeq}
\def\be{\ba}
\def\ee{\ea}
\def\F{{\cal F}}

\def\stackreb#1#2{\mathrel{\mathop{#2}\limits_{#1}}}
\def\Tr{{\rm Tr}}
\def\res{{\rm res}}
\def\Bf#1{\mbox{\boldmath $#1$}}

\def\bgamma{{\Bf\gamma}}

\def\bphi{{\Bf\phi}}

\def\Im{{\rm Im}}
\def\Re{{\rm Re}}

\def\half{{\textstyle{1\over2}}}
\def\ha{{1\over 2}}
\def\N2{${\cal N}=2$}
\def\4N{${\cal N}=4$}
\def\1N{${\cal N}=1$}
\def\1N*{${\cal N}=1^*$}

\def\CV{{\cal V}}
\def\CS{{\cal S}}

\def\vpint{{\lefteqn{\int}{\,-}}}
\def\nin{{\lefteqn{\ \in}{\,\ /\ }}}
\def\theequation{\thesection.\arabic{equation}}
\newcommand{\rf}[1]{(\ref{#1})}

\begin{document}


\begin{flushright}
FIAN/TD-17/05\\
ITEP/TH-63/05
\end{flushright}
\vspace{1.0 cm}

\renewcommand{\thefootnote}{\fnsymbol{footnote}}
\begin{center}
{\Large\bf
Matrix Models, Complex Geometry and Integrable Systems. I
\footnote{Based on lectures presented at
several schools on mathematical physics and the talks at
"Complex geometry and string theory"
and the Polivanov memorial seminar.}
}\\
\vspace{1.0 cm}
{\large A.Marshakov}\\
\vspace{0.6 cm}
{\em
Theory Department, P.N.Lebedev Physics Institute,\\
Institute of Theoretical and Experimental Physics\\ Moscow, Russia
}\\
\vspace{0.3 cm}
{e-mail:\ \ mars@lpi.ru,\ \ mars@itep.ru}
\end{center}
\begin{quotation}
\noindent
We consider the simplest gauge theories given by one- and two- matrix integrals
and concentrate on their stringy and geometric properties. We remind general
integrable structure behind the matrix integrals and turn to the
geometric properties of planar matrix models, demonstrating that
they are universally described in terms of integrable systems directly related to
the theory of complex curves.
We study the main ingredients of this geometric picture, suggesting that it can be
generalized beyond one complex dimension, and formulate them in terms of the
quasiclassical integrable systems, solved by construction of
tau-functions or prepotentials. The complex curves and tau-functions of one- and two-
matrix models are discussed in detail.
\end{quotation}

\renewcommand{\thefootnote}{\arabic{footnote}}
\setcounter{section}{0}
\setcounter{footnote}{0}
\setcounter{equation}0

\section{Gauge fields, geometry and string theory}

The relation between gauge fields and strings is still one of the most interesting
problems of modern theoretical physics, see e.g. \cite{Pol}. At present it goes
far beyond the well-known fact, that spectrum of open strings in flat
background contains the vector field, which
couples to the simplest open string excitation, and reparameterization
invariance of the quantum mechanics of string modes requires it to be massless.

The study of moduli space of closed string theories suggests that even there
it is natural to introduce open strings, ending only on certain hypersurfaces
or D-branes \cite{Dbr}. The stack of $N$ D-branes leads in this way to appearence
of the matrix or non-Abelian gauge
field, where the indices of the matrix label the particular branes from the
stack so that the corresponding string starts from and ends on. This is a
geometric way to formulate the $SU(N)$ Yang-Mills theory, and one finds
in this way that the gauge theory, or the stack of D-branes, modifies
the background geometry of the closed string theory, see detailed discussion
of these issues, say in \cite{Polchb,Maldarev,MUFN}.
This allows to conjecture that gauge theory can be
reformulated in terms of certain gravity picture, presumably in some
multi-dimensional target space.

The non-Abelian gauge theory must not have colored states (confinement). In
close string theory language it means that, when closed-string background is modified,
the open sting states should disappear
from the spectrum. In terms of partition functions (or generating
functions for the string amplitudes) this requirement acquires the form of
equality
\be
\label{ZF}
Z_{\rm gauge} = \exp\left(F_{\rm string} \right)
\ee
where $Z_{\rm gauge}$ is statistical sum in some gauge theory, while its free energy
$F_{\rm string}$ should be a partition function of a certain {\em closed}
string theory.

The most simple (but not the only!) geometric interpretation of the relation
(\ref{ZF}) is the $1\over N$-expansion \cite{1/N} in the inverse size of matrices
from the side of matrix gauge theory. The perturbation series
in quantum field theory has vanishing radius of convergency, if one sums over
all possible diagrams. However, in {\em matrix} theory
the Feynman graphs are fat, and each Feynman diagram can be
uniquely assigned a two-dimensional surface of particular topology where it
can be drawn without self intersections. Hence, an extra parameter -- the genus
of graph or
of such two dimensional Riemann surface appears -- and the series of the
graphs of {\em given} topology can be possibly summed up within some finite radius of
convergency.

This means that for any matrix theory of the $N\times N$ matrices one gets
in fact a "double" expansion
\be
\label{expan}
F_{\rm string} = \sum_{g=0}^\infty N^{2-2g}\sum_h F_{g,h}\curlywedge^h
\ee
where $g$ is the introduced genus of Riemann surface, corresponding the Feynman
graph with $V$ vertices $E$ edges and $F$ faces with the
Euler characteristics $\chi_g = V-E+F=2-2g$. An extra sum
over the holes $h=2g-2+F$ or faces of graphs, drawn on a given
Riemann surface, can be thought
of as summing (with a generating parameter $\curlywedge$ to be identified slightly later)
over all contributions of open the strings, which,
being summed up, disappear from the spectrum of the theory, modifying
the geometry og closed strings.

A sample example of gauge theory we study below is the
zero-dimensional theory, or the matrix integral (considered, first, in
this context in \cite{maint})
\be
\label{mamo}
Z = \int {\rm d}\Phi e^{-{1\over\hbar}\Tr W(\Phi)}
\ee
where $\Phi$ is $N\times N$ matrix (for example, Hermitian),
and $W(\Phi)$ is some generic polynomial
\be
\label{mmpot}
W(\Phi) = \sum_{k>0} t_k \Phi^k
\ee
For the matrices of finite size $N\times N$ this partition function is just $N^2$-tuple
integral, which nevertheless possesses almost all known nontrivial properties of
the quantum gauge theory path integrals.
Obviously the "action" $\Tr W(\Phi)$ in
\rf{mamo} is invariant under the zero-dimensional gauge transformations (where
the derivative term is absent)
$\Phi\to  U^\dagger\Phi U$. The integration measure in
(\ref{mamo}) is just a certainly normalized product of the differentials of all
matrix elements
\be
\label{measure1mm}
{\rm d}\Phi = {1\over V_N}\prod d\Phi_{ij}
\ee
where, as in any gauge theory, the normalization factor $V_N$ contains
the volume of
"orbit" of the unitary group $U(N)/ U(1)^N$ together with the "eigenvalue-permutation"
factor $N!$. This normalization becomes natural after separation of the
physical and gauge degrees of freedom - the diagonalization of the matrix $\Phi$ by
unitary gauge transformation $\Phi = U^\dagger\bphi U$, where $\bphi$ is
diagonal. The volume of the orbit $U(N)/ U(1)^N$
is then cancelled from denominator after trivial integration over the gauge
degrees of freedom.

Taking, as an example the potential
$W(\Phi) = \half \Phi^2 + {g_{YM}\over 3}\Phi^3 + \dots$ (where in analogy
with the Yang-Mills theory we denoted the coupling constant by $g_{YM}$),
introducing the 't Hooft coupling
$\curlywedge=g_{YM}^2N$
and expanding the integral \rf{mamo}
after rescaling of the field $\Phi$, one gets exactly expansion of the form
(\ref{expan}), since for a graph
with $V$ vertices, $E$ propagators and $H$ closed loops or holes the
corresponding contribution is weighted by
\be
\label{count}
\left(N\over\curlywedge\right)^V\left(\curlywedge\over N\right)^E N^H =
N^{V-E+H}\curlywedge^{E-V}
\ee
This counting is universal and does not depend on
particular model, i.e. the same is true for realistic
four-dimensional gauge theory with an $SU(N)$ gauge group
(direct generalization of phenomenological $SU(2)$ or $SU(3)$).

Now, as we already mentioned, the series in $\curlywedge$ over the contribution of
open strings (at fixed genus $g$!)
can be in principle summed up. This leads to the formula
\be
\label{CLOSTR}
F_{\rm string} = \sum_{g=0}^\infty N^{2-2g}F_g(\curlywedge)
\ee
where $F_g$ can be thought of as genus-$g$ partition functions of some {\em closed}
string theory.
More strictly, it means that parameter $\curlywedge$ can be interpreted as a
parameter of closed string background. It is very important to point out
that string interpretation requires {\em smooth} Riemann surfaces in \rf{CLOSTR},
achieved only in the regime when contribution of all gauge diagrams is essential,
i.e. when $\curlywedge \gg 1$ or in the {\em non-perturbative} regime for the gauge
theory. It is also necessary to point out that there is no direct geometric
language, describing in this way gauge theory
at weak coupling, since the corresponding limit on string theory side is
very singular\footnote{See, however, \cite{Witampl} where the weak-coupling expansion
in gauge theory is compared to the non-perturbative expansion in string theory
over the world-sheet instantons.}.

This is both good and bad. Good, since the gauge/string duality conjecture allows in
principle, if we trust it, to predict non-trivial effects in the non-perturbative
phase of gauge theory, i.e. in the regime which is mostly interesting for present
theory of elementary particles. And it is bad, since it becomes very hard to test
the conjecture itself, since it is not easy to provide good examples, for which both
sides of equality \rf{ZF} can be calculated by independent methods. Basically it
is the reason, why study of the matrix integrals \rf{mamo} is so
important -- since this is the simplest test example of the gauge/string correspondence.
As we see below, already in this case it is a nontrivial problem, with lots of
nice structures hidden inside, which are not only interesting by themselves, but,
as we try to demonstrate in the very end, can be also applicable to testing
higher-dimensional examples of the gauge/string correspondence.

The ${1\over N}$-expansion can be also understood as quasiclassical
computation of the matrix integral \rf{mamo}, where the square of gauge
coupling $g_{YM}^2=\hbar$ can be identified with the closed string coupling $\hbar=g_s$
or the Planck constant, which governs an expansion over the closed string loops.
In what follows it is useful to rename the 't~Hooft coupling $\curlywedge$, introducing
the variable
\be
\label{t0}
t_0 = \hbar N
\ee
which, in particular, will remain fixed and finite
under quasiclassical limit $N\to\infty$ together with
$\hbar\to 0$. The ${1\over N}$-expansion \rf{CLOSTR} then can be
equivalently written as a quasiclassical one
\be
\label{clostrh}
F_{\rm string} = \sum_{g=0}^\infty \hbar^{2g-2}F_g(t_0)
\ee
where the coefficients differ from \rf{CLOSTR} by different scaling in
$t_0=\curlywedge$.

\paragraph{Example: the gaussian case}\

For the potential $W(\Phi)=\ha\Phi^2-t_1\Phi$ the integral
(\ref{mamo}) can be immediately computed:
\be
\label{gauss}
Z_{\rm gauss} =
\int {\rm d}\Phi e^{-{1\over\hbar}\Tr\left(\ha\Phi^2-t_1\Phi\right)}\
\stackreb{\Phi\to t_1+\phi}{=}\
e^{-{Nt_1^2\over 2\hbar}}
\int {\rm d}\Phi e^{-{1\over 2\hbar}\Tr\Phi^2} =
C_N e^{-{Nt_1^2\over 2\hbar}}\left(2\pi\hbar\right)^{N^2\over 2}
\ee
with $C_N={1\over V_N}$. The $N$-dependence of the volume of unitary
group ensure for (\ref{gauss}) that both
${\d^2\over\d t_1^2}\log Z_{\rm gauss}(N)$ and the ratio
${Z_{\rm gauss}(N+1)Z_{\rm gauss}(N-1)/ Z_{\rm gauss}(N)^2}$ are
linear functions of the matrix size $N$. This is the first sign of
appearance in the context of matrix models of the Toda lattice equations,
to be considered below. The quasiclassical expansion \rf{clostrh} of \rf{gauss} reads
\be
\label{qgauss}
F_{\rm gauss} = {1\over \hbar^2}\left({t_0^2\over 2}\left(\log t_0 - {3\over 2}\right)
+ \half t_0t_1^2\right) + O(\hbar^0)
\ee
where the first logarithmic term in the r.h.s. comes from carefully treated large $N$
asymptotic of the volume factor $V_N$. Expression in brackets in the r.h.s. of
\rf{qgauss}, proportional to $1/\hbar^2$, is a "mostly trivialized" example of
quasiclassical tau-functions to be in the center of our interest below.

\setcounter{equation}0
\section{Matrix models and integrability}

Before discussing the quasiclassical limit of the matrix integral \rf{mamo}, let us
remind some its general properties. The $N^2$-tuple integral can be computed, first
using gauge symmetry, being reduced to the only $N$-tuple eigenvalue integral.
The key properties of the last one
are in fact encoded into a single (contour) integral, due to the fact that such
integrals are directly related with the integrable systems of Toda family.

\subsection{Matrix ensembles and Toda lattice}

Let us start here will the well-known and widely
used statement of \cite{GMMMO}, that matrix model partition (or
generating) functions satisfy equations of the Toda hierarchy, or, more
strictly, that certain
integrals over random matrices are particular tau-functions of the Toda
lattice hierarchy. Using the gauge invariance $\Phi\rightarrow U^\dagger\Phi U$
of the "single-trace" potential \rf{mmpot} of general form, one can
rewrite formula \rf{mamo} as an {\em eigenvalue} integral (in this section
we put $\hbar=1$), i.e. the partition function \rf{mamo} equals $Z = \tau_N(t)$, where
\be
\label{mamoeig}
\tau_N(t) =
{1\over N!}\int \prod_{i=1}^N \left(d\phi_i e^{-W(\phi_i)}\right)
\Delta^2(\phi)
\ee
and
\be
\label{vdm}
\Delta (\phi) = \prod_{i<j}(\phi_i-\phi_j) = \det_{ij} \|\phi_i^{j-1}\|
\ee
is the Van-der-Monde determinant.

In order to verify \rf{mamoeig} one has to use the gauge transformation,
which allows to diagonalize the matrix $\Phi$ by
$\Phi = U^\dagger{\bphi}U$, where $U\in U(N)/U(1)^N$ belongs to some
"orbit" of the unitary group $U(N)$ and $\bphi = {\rm diag}(\phi_1,\dots,
\phi_N)$ is a diagonal matrix. The Van-der-Monde determinant comes into
the integration measure of (\ref{mamoeig}) as a Jacobian of transformation
\be
\left.\delta\Phi_{ij}\right|_{\Phi=\bphi} = \left(U^\dagger\bphi U-\bphi\right)_{ij}\
\stackreb{U=\exp\left(i\delta\Omega\right)}{=}\
-i[\delta\Omega,\bphi]_{ij} = -i\delta\Omega_{ij}(\phi_i-\phi_j)
\ee
Analogous to (\ref{mamoeig}) eigenvalue representations exist also in the ``multimatrix''
case, where the integral is taken over some ensemble of random matrices with special
single-trace potentials. For
example, in the two-matrix model (the "complex-conjugated" version to be mostly considered
below) one introduces two matrices $\Phi$ and $\Phi^\dagger$ with the interaction
\be
\label{V2mm}
V(\Phi, \Phi^\dagger) = \Phi^\dagger\Phi - W(\Phi;t) -
{\tilde W}({\Phi^\dagger};{\bar t})
\ee
and the matrix integrals reduce to the integrals
over their eigenvalues $\int d\Phi d\Phi^\dagger
\exp\left(-V(\Phi,\Phi^\dagger)\right)=\tau_N(t,{\bar t})$, where
similar to \rf{mamoeig}
\be
\label{mamocompl}
\tau_N(t,{\bar t}) =
{1\over N!}\int \prod_{i=1}^N \left(d^2z_i e^{-V(z_i,\bar z_i)}\right)
\left|\Delta(z)\right|^2
\ee
due to the Harish-Chandra-Itzykson-Zuber formula. In the representations
\rf{mamoeig} and \rf{mamocompl} the $N^2$-tuple matrix integrals are reduced
to the $N$-tuple eigenvalue integrals (in the last case the integration over the
eigenvalues $(z_i,\bar z_i)\in\mathbb{C}\subset\mathbb{C}^2$ is taken generally
over some submanifold of $\mathbb{C}^2$ of real dimension two), and this is
a particular case of quite common in gauge theories localization procedure.

The most effective way to calculate the eigenvalue integral is to use the
method of orthogonal polynomials. The Van-der-Monde determinant \rf{vdm} can
be equally written as
\be
\label{vdmpoly}
\Delta(\phi) = \det_{ij} \|\phi_i^{j-1}\| =
\det_{ij} \|P_{j-1}(\phi_i)\|
\ee
with {\em any} polynomials, normalized with unit higher coefficient
\be
\label{defpol}
P_k(\phi) = \phi^k + O(\phi^{k-1})
\ee
and for the computation of matrix integral \rf{mamoeig} it is convenient to
choose particular $P_j(\phi)$ being orthogonal polynomials with the weight,
determined by the matrix model potential $W(\Phi)$
\be\label{sp}
\langle i|j\rangle \equiv
\int d\phi e^{-W(\phi)}P_i(\phi)P_j(\phi) =
\delta _{ij}e^{q_i(t)}
\ee
Substitute the Van-der-Monde determinant in the form (\ref{vdmpoly}),
where the polynomials
satisfy \rf{sp}, into \rf{mamoeig}. After expanding the determinant and using
orthogonality (\ref{sp}), the partition function (\ref{mamo}) acquires the form of
\be\label{tautc}
\tau_N(t) = \prod ^{N-1}_{i=0}e^{q_i(t)} = \det _{ij} H_{ij}(t)
\ee
a product of their norms \rf{sp} or the determinant of the so called ``moment matrix''
\cite{KMMOZ}
\be\label{moma}
H_{ij}(t) = \int  \phi^{i+j}e^{-W(\phi)}d\phi
\ee
Formulas (\ref{tautc}), (\ref{moma}) claim that matrix model partition function \rf{mamo},
\rf{mamoeig}, since it can be presented in the form of determinant of an
operator, exponentially depending on times or parameters of the potential
\rf{mmpot}, is a tau-function of Toda hierarchy and satisfies the corresponding
nonlinear differential equations w.r.t. these times.

To derive these equations, consider the Lax operator for
the Toda chain, which in the basis of the orthogonal polynomials (\ref{sp})
is presented in the common form of a tri-linear matrix \cite{GMMMO}
\be\label{lax}
\phi P_i(\phi) = P_{i+1}(\phi) - p_i(t)P_i(\phi) + R_i(t)P_{i-1}(\phi)
\ee
The first from the set of Lax equations
$\d_{t_k} L = \left[ L,{\cal R}\circ L^k \right]$, with $k=1$,
immediately gives rise to
\be\label{1todch}
{\partial ^2 q_i \over \partial t_1^2} = e^{q_{i+1}-q_i} - e^{q_i-q_{i-1}}
\ee
the equation of motion of exponentially interacting particles, (with co-ordinates
$q_i$ and momenta $p_i$),
or the first equation in the Toda-chain hierarchy.

Indeed,
from (\ref{sp}), (\ref{lax}) one can easily establish the relations
among the functions $q_i(t)$ and the matrix elements
$R_i(t)$ and $p_i(t)$ of the Lax operator. First,
\be
\langle i|\phi|i-1 \rangle = e^{ q_i (t)} = R_i(t) e^{ q_{i-1}(t)}
\ee
gives rise to
\be\label{rfi}
R_i(t) = e^{q_i(t)- q_{i-1}(t)}
\ee
Differentiating (\ref{sp}) for $i=j$ one gets
\be
\label{pi}
{\partial \over \partial t_1} \langle i|i \rangle = e^{ q_i}
{\partial  q_i \over \partial t_1} = \int d\phi e^{-\sum t_k \phi^k}
\left(-\phi P^2_i + 2P_i {\partial P_i \over \partial t_1}\right) =
p_ie^{ q_i}
\ee
where the second term in brackets disappears due to orthogonality
condition and condition (\ref{defpol}).
Now, from (\ref{pi}) it follows that
\be
\label{moment}
p_i(t) = {\partial  q_i (t) \over \partial t_1}
\ee
is just the momentum of $i$-th Toda particle with the coordinate $q_i(t)$.
Differentiating now (\ref{sp}) with $i > j$, we obtain
\be
0 = \int d\phi e^{-\sum t_k \phi^k} \left(- \phi P_i P_j + P_j
{\partial P_i \over
\partial t_1} \right)
\ee
and using (\ref{sp}) for
comparing this for $j \leq i-1$
with (\ref{lax}) one gets
\be\label{piri}
{\partial P_i \over \partial t_1} = R_i P_{i-1}
\ee
Now we are ready to differentiate (\ref{lax}):
\be\label{dlax}
\phi{\partial P_i \over \partial t_1} = {\partial P_{i+1} \over \partial
t_1} - {\partial p_i \over \partial t_1} P_i + O(\phi^{i-1})
\ee
Multiplying (\ref{dlax}) by $P_i$ integrating and using (\ref{sp}) and
(\ref{piri}) one finally obtains the equation
\be
{\partial p_i \over \partial t_1} = R_{i+1} - R_i
\ee
or, using (\ref{moment}) and (\ref{rfi}), arrives to the Toda chain
equation of motion (\ref{1todch}).

For the partition function of matrix model \rf{mamo}, \rf{mamoeig} or tau-function
of the Toda chain \rf{tautc}, equation of motion (\ref{1todch}) can be rewritten in
the form of one from the infinite set of the Hirota bilinear equations
\be\label{hirota}
 {\partial ^2 \over \partial t^2_1}\ \log\tau _N(t)  =
{\tau _{N+1}(t)\tau _{N-1}(t)\over \tau _N(t)^2}
\ee
In the two-matrix case instead of (\ref{sp}) one should introduce the
bi-orthogonal polynomials
\be\label{sp2}
\int_{\mathbb{C}} d^2z e^{-V(z,{\bar z})}P_i(z){\bar P}_j({\bar z}) =
\delta _{ij}e^{q_i(t,{\bar t})}
\ee
giving rise in a similar way (see \cite{GMMMO}) to the two-dimensional
Toda lattice hierarchy with the set of times, corresponding to the potential \rf{V2mm}.
In particular, the second time-derivative ${\d^2\over\d t_1^2}$ is substituted by
the two-dimensional
Laplacian ${\d^2\over\d t_1\d\bar t_1}$, and the first Hirota
equation (\ref{hirota}) is, correspondingly, replaced by
\be\label{hirota2}
{\partial ^2 \over \partial t_1\d {\bar t}_1}\ \log\tau _N(t,{\bar t}) =
{\tau _{N+1}(t,{\bar t})\tau _{N-1}(t,{\bar t})\over \tau _N(t,{\bar t})^2}
\ee

\subsection{Matrix integrals in Miwa variables}

An interesting phenomenon occurs, when we rewrite the matrix integral
\rf{mamo} in terms of the Miwa variables \cite{Miwa}
\be
\label{Miwa}
t_k = t_k^{(0)}
-{1\over k}\sum_{I=1}^L\mu_I^{-k}\equiv t_k^{(0)}-{1\over k}\Tr M^{-k}
\ee
or parameterize the times by eigenvalues of
auxiliary $L\times L$ matrix $M$. For the background values let us
take $t_k^{(0)}=\half\delta_{k,2}$, i.e. consider the matrix integral
around the gaussian potential \rf{gauss}.

Upon substitution of \rf{Miwa}, the eigenvalue integral \rf{mamoeig} acquires
the form \cite{GKMToda}
\be
\label{zmiko1}
Z(M,t^{(0)})={1\over N!}\int \prod_{i=1}^N
\left(d\phi_i e^{-W(\phi_i;t^{(0)})}\right)
\Delta^2(\phi)\prod_{i,I}\left(1-{\phi_i\over\mu_I}\right) =
\\
={1\over N!}\int \prod_{i=1}^N
\left(d\phi_i e^{-W(\phi_i;t^{(0)})}\right)
\Delta(\phi){\Delta(\phi,\mu)\over\prod_I\mu_I^N\Delta(\mu)}
\ee
where $\Delta(\mu)=\prod_{I<J}(\mu_I-\mu_J)$ and $\Delta(\phi,\mu)$ are
the corresponding $L\times L$ and $(N+L)\times (N+L)$ Van-der-Monde
determinants. As in previous section it is again convenient to introduce the
orthogonal polynomials \rf{sp}, \rf{defpol}, which allow to rewrite
\rf{zmiko1} as
\be
\label{zmiko2}
Z(M,t^{(0)})={1\over N!\prod_I\mu_I^N\Delta(\mu)}\int \prod_{i=1}^N
\left(d\phi_i e^{-W(\phi_i;t^{(0)})}\right)\times
\\
\times
\det_{ij} \|P_{j-1}(\phi_i)\|\
\det_{(i,I),(j,J)}\left\|\begin{array}{ccc}
  P_{j-1}(\phi_i) & \ldots & P_{j-1}(\mu_I) \\
  \vdots & \ddots & \vdots \\
  P_{J-1+N}(\phi_i) & \ldots & P_{J-1+N}(\mu_I)
\end{array}
\right\|
\ee
Expanding the determinants of $N\times N$ and $(N+L)\times (N+L)$ matrices
in \rf{zmiko2} and using, as before, the orthogonality relations one gets
\be
\label{zmiko3}
\tau_N(t) =
Z(M,t^{(0)})={\tau_N(t^{(0)})\over \prod_I\mu_I^N\Delta(\mu)}\
\det_{IJ}\left\|P_{J-1+N}(\mu_I)\right\|
\ee
or, introducing $\psi_J(\mu) \equiv \mu^{-N}P_{J+N}(\mu)$,
\be
\label{zmiko}
{\tau_N(t)\over \tau_N(t^{(0)})} =
{\det_{IJ}\left\|\psi_{J-1}(\mu_I)\right\|\over\Delta(\mu)}
\ee
which is a particular well-known representation for the Kadomtsev-Petviashvili (KP)
tau-function in
Miwa variables \rf{Miwa}. For the gaussian case $t_k^{(0)}=\half\delta_{k,2}$
the orthogonal polynomials can be identified with the Hermite polynomials
\be
\label{hermite}
\psi_J(\mu) = \mu^{-N}\int dx\ e^{-x^2+\mu x}\ x^{N+J}\
\stackreb{\mu\to\infty}{\sim}\ \mu^J+\dots
\ee
which allows to rewrite \rf{zmiko} in terms of the matrix integral in
external matrix field of general structure
\be\label{gkm}
Z(\Lambda) \propto
\int {\rm d}X\ e^{-\Tr \CV(X)+
\Tr \Lambda X}
\ee
where for particular case of the gaussian model \rf{zmiko1}, as follows
from the integral representation \rf{hermite}, $\Lambda=M$ and
$\CV(X) = \ha X^2 - N\log X$. Existence of dual matrix integral representation
\rf{gkm} follows from another integrable structure \cite{GKM}, arising for the
matrix integrals in external matrix field.

\subsection{Kontsevich integrals, free fermions and topological strings}

Let us now consider in detail the matrix model \rf{gkm},
where the integral is taken over the $L\times L$ matrices, and
the partition function is at the moment written up to a simple ($M$-dependent)
normalizing factor. Similarly to the expansion around the gaussian point in the
previous section, the matrix integral \rf{gkm} can be considered as a
(formal) expansion around its "classical
trajectory" $X=M$ , determined by the equation of motion
\be
\label{eqmoko}
\Lambda = \CV'(M)
\ee
The integral \rf{gkm} can be then expanded in
the {\em inverse} powers of matrix $M$, i.e. the (gauge-invariant) partition function
\rf{gkm} becomes the function of the variables \rf{Miwa}, and normalization is chosen in
such a way, that this expansion starts from unity.

In general case the
matrix integral (\ref{gkm}) can be written as well in the form of
determinant formula, like \rf{zmiko}, which states that it equals to
a tau-function of KP hierarchy
(more strictly for a polynomial potential $\CV(X)$ of degree $p$ it is
the $p$-th Gelfand-Dickey reduction of the KP hierarchy).
The determinant formula follows directly \cite{GKM}
from evaluation of the integral \rf{gkm}
\be
\label{int}
\int   {\rm d}X\ e^{- \Tr\left( \CV(X)
- \Tr\Lambda X\right)}
= {1\over \Delta (\lambda )}\left(  \prod _{I=1}^L
\int   dx_Ie^{- \CV(x_I)+\lambda _Ix_I} \right) \Delta (x) =
\\
= \Delta ^{-1}(\lambda)
\Delta \left({\partial \over \partial \lambda}\right)
\prod _I \int dx_I e^ {- \CV(x_I) + \lambda _I x_I }
= \Delta ^{-1}(\lambda ) \det _{IJ}\|\Psi_{I-1}(\lambda _J)\|
\ee
where the first equality is a consequence of the
Harish-Chandra-Itzykson-Zuber formula, and the elements
of the matrix in the r.h.s. are expressed through the contour integrals
(cf. with its particular case \rf{hermite})
\be
\label{entry}
\Psi_{I}(\lambda ) \equiv  \int   dx\ x^Ie^{- \CV(x)+\lambda x} =
\left({\partial \over \partial \lambda }\right)^{I}\Psi_0(\lambda ).
\ee
Putting on classical solution \rf{eqmoko} $\Lambda  = \CV'(M)$,
denoting the eigenvalues of  $M$  as  $\{\mu _I\}$ and restoring the
normalization, one gets the final result
\be\label{det}
Z(M) \stackreb{\rf{Miwa}}{\propto}\ {\tau_N(t)\over \tau_N(t^{(0)})} =
{\det_{IJ}\Phi _{I-1}(\mu _J)\over \Delta(\mu )}\
\ \ \ \ I,J =1,\dots,L
\ee
with
\be\label{entry3}
\Phi _I(\mu ) = \left(\CV''(\mu )\right)^{1/2}
e^{\CV(\mu )-\mu  \CV'(\mu )} \Psi_I(\CV'(\mu ))
\stackreb{\mu \to
\infty}{\to} \mu ^{I}\left(1 + O\left({1\over \mu}\right)\right)
\ee
which satisfies the Hirota equations of KP hierarchy, due to simple bilinear
relations for the determinants, see details in \cite{GKM}.
The determinant formula (\ref{det})
holds for any matrix size $L$ and in this sense, unlike the matrix integrals
\rf{mamo}, \rf{mamoeig}, \rf{mamocompl}, does not "feel" the $L\to\infty$ limit
of large size of the matrices we exploit intensively below.
We immediately see, that representation \rf{zmiko} is a particular case of
the formula \rf{det},
provided \rf{entry} are given by particular expressions \rf{hermite}.

The formulas of this section require few necessary comments (see \cite{GKM} for their
detailed discussion):
\begin{itemize}
\item The (dispersionful!) Hirota bilinear equations, satisfied
by the determinant formula \rf{det} are in fact equivalent to the
{\em free field} representation of the matrix elements \rf{entry3}
\be
\label{phif}
\Phi_I (\mu) \propto \langle I| \psi (\mu)\hat\Phi|0\rangle
\ee
with any operator of the form
$\hat\Phi=\exp\left(\sum_{i,j} A_{ij}\tilde\psi_i\psi_j\right)$. The determinant
formula \rf{det} then simply follows from the Wick theorem.

\item The particular functions \rf{entry3}, possessing simple integral
representation \rf{entry} are specified by {\em string equation}, which can
be written in the form ${\cal L}^{\{\CV \}}_{-1} Z=0$, for the first-order
differential operator w.r.to the times of the hierarchy
\be
\label{streqK}
{\cal L}^{\{\CV \}}_{-1} =\sum _{n\geq 1}\Tr
\left({1\over  \CV''(M)M^{n+1}}\right) {\partial \over \partial t_n}
+ {1\over 2}
\sum _{I,J}\left({1\over \CV''(\mu _I) \CV''(\mu _J)}{\CV''(\mu _I)-
\CV''(\mu _J)\over \mu _I - \mu _J}\right) - {\partial \over \partial t_1}
\ee
acquiring the more common form of a particular Virasoro operator
${\cal L}^{\{p \}}_{-1}\equiv {\cal L}^{\{\CV \}}_{-p}$
for monomial potential $\CV(X) = {X^{p+1}\over p+1}$.

\item Generally the Virasoro constraints ${\cal L}^{\{\CV \}}_{n} Z=0$ for $n\geq -1$
can be used as equations defining (together with some extra conditions) the partition
function of certain topological string theory \cite{Kontsevich},
or $(p,q)$-minimal string theory. Equivalently, one may impose only the constraint
\rf{streqK} together with the full set of bilinear
Hirota equations of ($p$-reduced) KP hierarchy \cite{FKN}. The planar or
quasiclassical contribution to the string partition function is then described by dKP (or
dispersionless KP) hierarchy, to be discussed in detail below.

A particular $(p,q)$ string model is then defined as a $q$-th critical point of
the $p$-reduced
KP hierarchy, which is characterized as solution to all above equations in terms
of the expansion near the point $t_k={p\over p+q}\delta_{k,p+q}$ in the space of
times or couplings.

\item
Matrix integrals \rf{gkm} are often interpreted as partition functions of the
topological Landau-Ginzburg theories of a single field with superpotential
$\CV'(X)$, interacting with two-dimensional topological gravity. The target-space
for these theories is a sphere with a marked point, or just a complex plane, and it
can be conveniently identified with the rational curve $y = \CV'(x)$, defined by
an analytic equation \rf{eqmoko} on two complex variables $(x,y) = (X,\Lambda)$.
This interpretation was proposed among
the first examples of defining the nonperturbative string theory in terms of a
(quasiclassical) tau-function, associated with a complex manifold endowed with certain
additional structure: two given functions, or, better, differentials $dx$ and $dy$ with
fixed (here vanishing) periods. The potential of matrix model in external field \rf{gkm}
$\CV = \int ydx$ is then an integral of
generating differential for the Landau-Ginzburg theory. We will see below, that this
geometric language is universal and effective description for the quasiclassics of
all matrix integrals.

\item Integral representation \rf{entry} can be then thought as a particular case of
the duality transformation \cite{KhMa}
\be
\label{pqdual}
\Psi_I (y) \propto \int d\mu(x) \Psi_I (x) e^{\int ydx}
\ee
between the $(q,p)$ and $(p,q)$ models. The dual wave functions are naturally expressed
as functions of dual local co-ordinates, $x$ and $y$ correspondingly. On classical level
this duality therefore looks like change of the local co-ordinate, and again, we will
see this below as general phenomenon in quasiclassical formulation of the matrix models.

\end{itemize}

\setcounter{equation}0
\section{Planar limit of the one-matrix model
\label{ss:1mamo}}

Let us now come back to the simplest matrix integral \rf{mamo} and
turn directly to the discussion of the planar limit $N\to\infty$, $\hbar\to 0$ with
the new parameter, introduced in \rf{t0}, $N\hbar = t_0$, being fixed.
In the limit $N\to\infty$ from the whole topological expansion
\rf{CLOSTR} only the first
term $F_0=\F$ with $g=0$ survives, corresponding to the summing over the planar
fat graphs, which can be drawn on sphere or plane without self-intersections.
Simultaneously in this limit, since $\hbar\to 0$, the matrix integral \rf{mamo}
can be calculated quasiclassically, i.e. by the stationary
phase method. The study of extrema directly in the integral over
matrices \rf{mamo} is senseless, due to the huge gauge degeneracy of the
extremal configurations in the space of all matrices, and therefore we turn below to the
study of quasiclassical properties of the eigenvalue integrals \rf{mamoeig} and
\rf{mamocompl}.

\subsection{The eigenvalue integral and complex curve}

\begin{figure}[tp]
\epsfysize=5.2cm
\centerline{\epsfbox{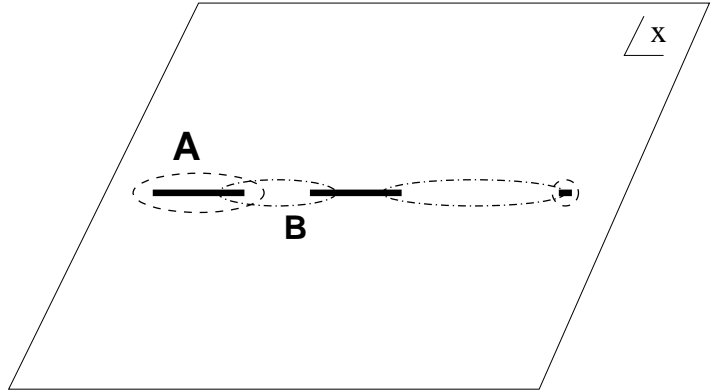}}
\caption{Cuts in the eigenvalue plane, and cycles on the $x$-sheets
of the hyperelliptic Riemann surface \rf{dvc} of the one-matrix model. The ${\bf A}$-cycles
are conventionally chosen to surround the cuts or eigenvalue supports
(on this picture supposed
to be a segments of real line $\Im x=0$), while the ${\bf B}$-cycles
connect two different eigenvalue supports.}
\label{fi:cuts}
\end{figure}

The quasiclassical value of the eigenvalue integral \rf{mamoeig} at $\hbar\to 0$,
can be found by studying the extrema of the effective potential
\be
\label{Weff}
W_{\rm eff}(\phi) = W(\phi)-2\hbar\log\Delta(\phi)
\ee
or, more strictly, solving the stationarity equation
\be
\label{mamoeq}
W'(\phi_j) = 2\hbar\sum_{k\neq j}{1\over \phi_j-\phi_k}
\ee
If $\hbar=0$ the interaction in the r.h.s. of \rf{mamoeq}
is switched off, and all eigenvalues $\phi_j$
are somehow distributed over the minima, or, in the most general setup, all extrema
where $W'(\phi)=0$. More convenient at $N\to\infty$ is to introduce
the eigenvalue density
\be
\label{mamorho}
\rho(x) = \hbar\sum_{j=1}^N\delta(x-\phi_j)
\ee
or the resolvent
\be
\label{mamoG}
G(x) = \hbar\left<\Tr{1\over x-\Phi}\right>_\Phi =
\int_{\bf C} {\rho(\phi)d\phi\over x-\phi}
\ee
defined on the $x$-plane with cutoff eigenvalue support, or, as we see below
on the double cover of some complex $x$-plane, cut along
some segments ${\bf C}=\bigcup_j C_j$, (see
fig.~\ref{fi:cuts}),
and normalized as
\be
\label{normrG}
{1\over 2\pi i}\oint_{\bf C} dxG(x)=
\int_{\bf C} d\phi\rho(\phi) = t_0
\ee
In these terms
equation \rf{mamoeq} can be rewritten as integral equation
\be
\label{inteqm}
W'(x) = 2\vpint_{\bf C}{\rho(\phi)d\phi\over x-\phi} =
G(x_+)+G(x_-), \ \ \ \
x\in\forall C_j \subset{\bf C}
\ee
where $x_\pm=x\pm i0$ are two "close" points on two different sides of
the cut -- above and below if distribution of eigenvalues is real and cut is
stretched along the real axis. This equation holds at any point
of the eigenvalue support ${\bf C}$, consisting of several disjoint pieces
for generic polynomial potentials. The formula \rf{inteqm} can
be further rewritten as an algebraic equation on the resolvent $G$, see e.g.
\cite{Migdal,David,BDE,DV}
\be
\label{mamocu}
G^2-W'(x)G=f(x)
\ee
where $f(x)$ in the r.h.s. is a polynomial of the power
one less than that of $W'(x)$. Indeed, it follows from \rf{inteqm} that
\be
\left.\left(G^2-W'G\right)\right|_{x=x_+} - \left.\left(G^2-W'G\right)\right|_{x=x_-}=
\left( G(x_+)-G(x_-)\right)\left(G(x_+)+G(x_-)-W'(x)\right) =0
\ee
i.e. the l.h.s. of \rf{mamocu} is a single-valued function $f(x)$, whose properties
(a polynomial of certain power) are determined from behavior at $x\to\infty$.

Equation \rf{mamocu} defines a hyperelliptic curve
and, as we see below, the quasiclassical free energy ${\cal F}=F_0$
(or the first term in \rf{expan}) can be entirely defined in terms of
the curve \rf{mamocu} and the generating differential
$Gdx$, determined by the resolvent \rf{mamoG} and measuring the eigenvalue
distribution. For the polynomial potentials equation
\rf{mamocu} describes an algebraic curve (of finite genus $n$ for the
potential $W$ of degree $n+1$) and defines resolvent $G$ as an
{\em algebraic function} on the curve, which is
double-cover of $x$-plane\footnote{However,
if one allows all possible long operators $\Tr\Phi^L$ for $L\to\infty$
it becomes a curve of infinite genus and nothing can be said about the
resolvent $G$ immediately.}.

Introducing new variable by
\be
\label{yG}
y=W'(x)-2G
\ee
one can rewrite \rf{mamocu} as
\be
\label{dvc}
y^2 = W'( x)^2 + 4f( x) = R( x)
\ee
where $R(x)$ is a polynomial of the degree $2n$ for the potential
$W(x)$ of degree $n+1$. For such polynomial equation \rf{dvc} is a canonical
representation for the hyperelliptic curve of genus $g=n-1$.

\subsection{Free energy: geometric definition
\label{ss:free}}

By computation of free
energy for the matrix model (\ref{mamo})
in large $N$ or quasiclassical limit,
one usually understands solution to the variational problem \rf{mamoeq} for
the functional\footnote{\label{fu:modulus}
By writing modulus in the argument of logarithms we generally imply choosing appropriate
branch of the logarithm function, it literally coincides with modulus when all eigenvalues
are supposed to be real.}
\be
\label{variF}
\F = -\left[
\sum_i W(\phi_i) - \hbar\sum_{i\neq j} \log\left|\phi_i-\phi_j\right|
\right]_{N=\sum_\alpha N_\alpha} =
\\
= -\left[\int dx\rho(x) W(x)-
\int dx_1\int dx_2\rho(x_1)\log\left|x_1-x_2\right|\rho(x_2)
+ \sum_\alpha \Pi_\alpha \left( \int_{C_\alpha}dx\rho(x)\right)-S_\alpha
\right]_{{\delta \F\over\delta\rho}=0}
\ee
In addition to formal definition \rf{mamoeig} we have assumed here
some fixed distribution of all $N\to\infty$ eigenvalues into certain (finite number
of) groups with the fractions $\sum_\alpha N_\alpha = N$, implying an extra linear
condition imposed on new quasiclassical variables
$\sum_\alpha S_\alpha = \hbar N = t_0$. Let us discuss, first, the nature
and meaning of these groups and corresponding fractions.

Intuitively it is clear, that in quasiclassical
limit the eigenvalues form some condensates, which can be
located around any minimum (or even near any extremum $W'(\phi)=0$ as we
determine immediately).
If one turns off the Coulomb repulsion, coming from the
Van-der-Monde determinant, by $\hbar\to 0$ the eigenvalues will be located exactly
at these extrema.
Therefore, for some small eigenvalue fractions this picture is evident,
for the large eigenvalue fractions it can be more complicated (the condensates
can grow collide) and we will discuss some details of this below.
The jumps between different condensates are exponentially suppressed, since
they are of the order $e^{-{\rm const}/\hbar}\sim e^{-{\rm const}\cdot N}$, and
therefore, it is natural to introduce extra (to the coefficients of the potential
\rf{mmpot}) variables, or to fix the fractions of eigenvalues
\be
\label{fracS}
S_\alpha=\hbar N_\alpha = \int_{C_\alpha}dx\rho(x) = {1\over 2\pi i}\oint_{C_\alpha} Gdx
\ee
at each extremum, when solving the variational problem for the functional \rf{variF},
say, introducing the corresponding Lagrange multipliers $\Pi_\alpha$.

Moreover, it is clear, that at switched off Coulomb interaction, the stationary
configurations, in general position, consist of separated points -- the
extrema $W'(\phi)=0$ of the potential, for large $N$ each
taken with certain multiplicity. However, the interaction smoothes the
picture -- instead of separate points one now gets the set of one-dimensional
lines, to be possibly thought of as analogs of mechanical trajectories or
Ising strings. For analytic continuation of the partition function in the space
of parameters and deformation of the integration contours, this leads to appearance
of the complex curves of statistical matrix ensembles.
\begin{figure}[tp]
\centerline{\epsfig{file=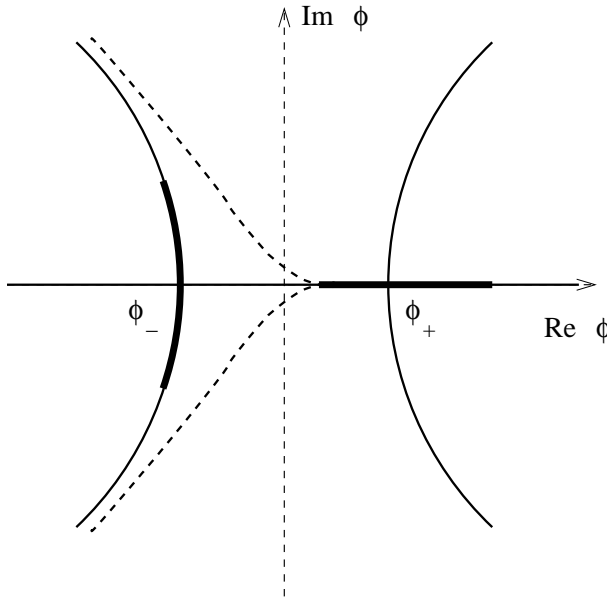,width=80mm}}
\caption{Complex geometry of the potential
$W(\phi) = {\phi^3\over 3} - t\phi$. The level lines, passing through the
critical points $\phi_\pm=\pm\sqrt{t}$ are $\Im\phi=0$ and
$(\Re\phi)^2 - {(\Im \phi)^2\over 3} - t =0$. The Coulomb repulsion
stretches the eigenvalues along the segments of the level lines
near the critical points.}
\label{fi:w3}
\end{figure}

To understand this better, one should remember that as any integral, the
eigenvalue integral \rf{mamoeig} depends on the choice of the
integration contours. For the potential of degree $n+1$ the basis in linear
space of all possible contours, when integral over each eigenvalue
converges, is $n$-dimensional. It is easy to see, that each "steepest descent"
contour can be drawn in the complex plane
exactly passing through one of the $n$ extrema points $W'(\phi)=0$, the recent
detailed discussion of this issue can be found in \cite{Feld}.

For the simplest non gaussian potential $W(\phi) = {\phi^3\over 3} - t\phi$
this geometry is illustrated at fig.~\ref{fi:w3}. Two critical points of
potential are at $\phi_\pm =
\pm\sqrt{t}$ (at the figure we have chosen $t$ to be real and positive).
The steepest descent lines of $\Re W$, passing through the
critical points $\phi_\pm$, are the level lines
of $\Im W = \Im\phi \left((\Re\phi)^2 - {(\Im \phi)^2\over 3} -
t\right) =0$.
The integration contours, for which the eigenvalue integrals converge should
start and end at one of the $n+1$ sectors of the complex $\phi$-plane, where
$\Re W >0$ ($n=2$ for fig.~\ref{fi:w3}, i.e. for this case there are three such
sectors). Since integrands are holomorphic and the value of the integral does not
depend on the local deformations of the contour, the steepest descent method
"chooses" the integration contour from one allowed sector to another,
passing through a critical point (two such possible contours are
shown by dashed lines at fig.~\ref{fi:w3}).
As already mentioned above,
the contribution of the Van-der-Monde determinant to the effective potential \rf{Weff}
leads to repulsion
of the eigenvalues, so that from their locations exactly at the extrema $W'(x)=0$
they stretch along the segments of the allowed contours, see
fig.~\ref{fi:w3}.

The quasiclassical free energy \rf{variF} is calculated
at critical densities, i.e. when
\be
\label{vareq}
{\delta \F\over\delta \rho(x)} =
-W(x) + 2\int_{\bf C}
d\phi \rho(\phi)
\log\left|x-\phi\right| - \Pi_\alpha  = 0, \ \ \ \ \ x\in C_\alpha
\ee
The right equality can used for computation of the Lagrange multipliers
$\Pi_\alpha$, which will be intensively used below, since they are equal to
the partial derivatives of the free energy \rf{variF} w.r.t.
the eigenvalue fractions \rf{fracS}
\be
\label{dFPi}
{\d \F\over\d S_\alpha} = \Pi_\alpha + \int_{\bf C} dx {\delta\F\over\delta\rho(x)}
{\d\rho(x)\over\d S_\alpha} = \Pi_\alpha
\ee
since the variational derivative vanishes due to \rf{vareq}, which
simultaneously determines the Lagrange multiplier itself
\footnote{Note also, that the Lagrange multipliers in
(\ref{variF}) are also necessary for satisfying
dispersionless Hirota equations and their analogs, i.e. they play the
same role as careful normalization in the measure (\ref{measure1mm}) dictated by
(\ref{hirota}), (\ref{hirota2}).}.

Formulas (\ref{variF}) can be now accepted as a separate
geometric definition of the planar free energy $\F$. To make it strict,
the solution to (\ref{vareq}), (\ref{variF})
should be supplemented by boundary condition -- a choice of support
${\bf C}$ for the nonvanishing density $\rho(x)\neq 0$ or non-analyticity of the
resolvent $G(x)dx$ \rf{mamoG}. This support (the set of "cuts" for the
one-matrix model or "drops" in two-dimensional case of \rf{mamocompl},
see fig.~\ref{fi:cuts} and fig.~\ref{fi:w3}) is introduced "by hands",
but its length and
form (or size and shape in two-dimensional case, to be discussed in detail below)
is then determined dynamically.
The density of eigenvalues in one-dimensional domain
(union of "cuts") is expressed by \rf{mamocu}
through the parameters of the potential and coefficients of auxiliary
function
\be
\label{f}
f(x) = \sum_{k=0}^{n-1}f_k x^k
\ee
related to the filling fractions. Note immediately that ambiguity in the choice of
extra free parameters of the polynomial \rf{f}, arising in the r.h.s. of \rf{mamocu}
is exactly "eaten" introducing extra parameters - the filling fractions \rf{fracS}.

Below we are going to show that the geometric definition of free energy \rf{dFPi}
coincides with a particular case of so called prepotential or quasiclasical tau-function,
which can be defined in principle for generic complex manifold endowed with some extra
structure, usually encoded into a meromorphic differential form. For the
planar limit of one matrix-model discussed in this section the complex manifold can be
identified with the hyperelliptic curve \rf{mamocu}, \rf{dvc}.

\paragraph{Example: the gaussian case}\

Let us now solve explicitly the above equation \rf{inteqm} for the simplest gaussian
potential $W = \ha x^2$. The r.h.s. of \rf{mamocu} or the polynomial \rf{f}
in this case ($n=1$) is some constant, and the solution to \rf{mamocu} reads
\be\label{Ggau}
G = {x\over 2} + {1\over 2}\sqrt{x^2+4f} \equiv {x\over 2} + {1\over 2}\sqrt{x^2-a^2}
\ee
i.e. it has a single cut in the eigenvalue plane,
with the only parameter $a^2=-4f$, identified with the squared half-length of a
symmetric around $x=0$ cut in the $x$-plane. It can be
determined from the normalization \rf{normrG}, which gives rise to $a^2=4t_0$.
In this case the matrix model curve \rf{mamocu}, \rf{dvc} is rational or has genus
$g=0$, and the density equals
\be
\label{rhogau}
\rho(x) = {1\over 2\pi}\sqrt{a^2-x^2}
\ee
and one can check that the partition function for this case is expressed by
the following formula
\be
\label{Fgau}
\left.\F\right|_{t_1=0} = \left.\left({t_0^2\over 2}\left(\log t_0 - {3\over 2}\right)
+ \half t_0t_1^2\right)\right|_{t_1=0}= {t_0^2\over 2}\left(\log t_0 - {3\over 2}\right)
\ee
coinciding with the first term of the quasiclassical expansion of \rf{gauss} at
$t_1=0$, corresponding to symmetrically located cut ${\bf C}=[-a,a]$ in the $x$-plane
and the Wigner eigenvalue density \rf{rhogau}. Formula \rf{Fgau} follows from the
only in the gaussian case relation \rf{dFPi} expressing derivative of free energy
w.r.t. the total number of eigenvalues $S=t_0$
\be
\label{dFPigau}
{\d\F\over\d t_0} = -W(x^\ast) +
2\int_{\bf C} d\phi \rho(\phi) \log\left|x^\ast-\phi\right|\
\stackreb{x^\ast=0}{=}\ {2\over\pi}\int_0^a\sqrt{a^2-x^2}\log x dx \
\stackreb{4a^2=t_0}{=}\ t_0\left(\log t_0-1\right)
\ee
Integration of \rf{dFPigau} leads to \rf{Fgau}, where the dependence on $t_1$ is trivially
restored by the shift argument (e.g. from \rf{gauss} it immediately follows, that
$Z_{\rm gauss} = e^{-{t_0t_1^2\over 2\hbar^2}}\left.Z_{\rm gauss}\right|_{t_1=0}$.

\setcounter{equation}0
\section{Quasiclassical tau-function
\label{ss:taufun}}

The quasiclassical free energy of the matrix model, defined by formulas \rf{dFPi}
is a particular case of so called prepotentials, or more strictly, quasiclassical
tau-functions, introduced in \cite{KriW}. In order to continue discussion of the
geometric properties of matrix models we need to learn more about the
properties of quasiclassical tau functions and to remind, first, some necessary
facts from geometry of complex curves.

\subsection{Riemann surfaces, holomorphic and meromorphic differentials}

\begin{figure}[tp]
\epsfysize=3.2cm
\centerline{\epsfbox{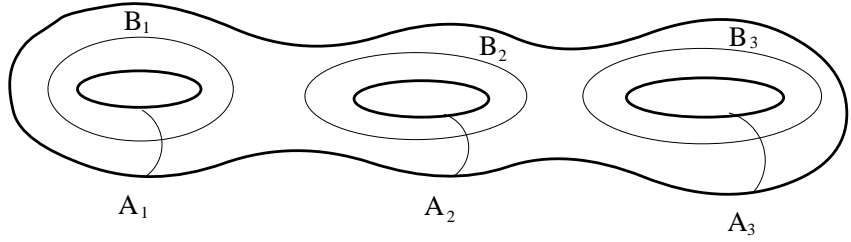}}
\caption{Compact Riemann surface of genus $g=3$ with chosen
basis of $A$ and $B$ cycles, with the intersection
form $A_\alpha\circ B_\beta = \delta _{\alpha\beta}$. Relabelling of $A$- and
$B$-cycles is called a {\em duality} transformation.}
\label{fi:riemann}
\end{figure}

Let us start with reviewing some well-known facts
from the theory of Riemann surfaces or one-dimensional complex manifolds -
the complex curves. The
topology of compact oriented Riemann surface is characterized by a single
non-negative integer - genus $g$, generally seen as number of "handles", attached
to sphere. The simplest examples are sphere itself with no handles or $g=0$
and torus with $g=1$, an example of Riemann surface with $g=3$ can be found at
fig.~\ref{fi:riemann}.

\begin{figure}[tp]
\centerline{\epsfig{file=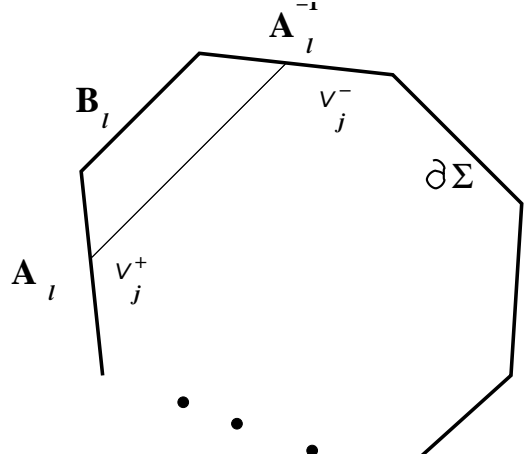,width=60mm,angle=-90}}
\caption{Cut Riemann surface with the boundary $\d\Sigma$. The integral over the
boundary can be divided into several contributions (see formula (\ref{derdpm})). In
the process of computation we use the fact that the boundary values of
Abelian integrals
$v_\alpha^\pm$ on two boundaries of the cut differ from each other
by the period integral of
the corresponding differential $d\omega_\alpha$ over the dual cycle.}
\label{fi:cut}
\end{figure}

On genus $g$ Riemann surface $\Sigma _g$ there
exists $2g$ independent noncontractable contours
which can be split into pairs, usually called by
$A\equiv\{ A_\alpha\}$ and $B\equiv\{ B_\alpha\}$ cycles, (see
fig.~\ref{fi:riemann})
$\alpha = 1,\dots,g$, with the
intersection form $A_\alpha\circ B_\beta = \delta _{\alpha\beta}$.
The basis in the space of of one-forms is
naturally dual to the basis of one-dimensional cycles, e.g.
the {\em holomorphic} (or the first kind Abelian)
differentials $\bar\d (d\omega_\alpha)=0$
are canonically chosen to be normalized to the $A$-cycles
\be\label{normA}
\oint _{A_\beta}d\omega _\alpha = \delta _{\alpha\beta}
\ee
Their integrals along the $B$-cycles form then the {\em period matrix}
\be\label{pemat}
\oint _{B_\beta}d\omega _\alpha =  T_{\alpha\beta}
\ee
(a symmetric $g\times g$ matrix with the positive definite imaginary part),
which is one of the most important characteristics of complex curve.
The symmetricity of period matrix follows from the
Riemann bilinear relations for holomorphic differentials
\be
\label{sypema}
0=\int_{\Sigma} d\omega_\beta\wedge d\omega_\gamma=
\sum_\alpha\left( \oint_{A_\alpha}d\omega_\beta\oint_{B_\alpha}
d\omega_\gamma-\oint_{A_\alpha}
d\omega_\gamma\oint_{B_\alpha}d\omega_\beta \right)
= T_{\beta\gamma} - T_{\gamma\beta}
\ee
coming from the Stokes theorem on the cut Riemann surface, (see
fig.~\ref{fi:cut}),
and vanishing of any $(2,0)$-form, in particular $d\omega_\beta\wedge
d\omega_\gamma$, on a one-dimensional complex manifold. Analogously, the imaginary part
of the period matrix is positive-definite, since $\Im T_{\alpha\beta}\propto
\int_{\Sigma}d\omega_\alpha\wedge d\bar\omega_\beta$.

In contrast to holomorphic, the
meromorphic differentials $d\Omega$ are analytic everywhere, except for the finite number
of points $P$, where they can have poles and $\bar\d d\Omega\propto \delta(P)$.
Their canonical form is fixed not only by half of
the periods (as for the holomorphic ones, see \rf{normA}), but as well
by their behavior ("main part") at singular points. The common classification for the
meromorphic differentials includes the second kind Abelian differentials, with a
single singularity at some point $P_0$ of the form
$d\Omega_k \sim
{d\zeta\over\zeta^{k+1}}+\dots$, $k\geq 1$,
where by dots we denoted a
nonsingular part, and $\zeta$ is a local co-ordinate near the point $P_0$:
$\zeta(P_0)=0$. Note, that this set starts from differentials with the second order poles,
since there is no meromorphic differential with a single first-order pole,
due to vanishing of the total residue. Instead, for taking into account
the first-order poles,
it is convenient to introduce the third-kind Abelian differentials
$d\Omega_\pm$, with {\em two} first order poles at some points $P_\pm$ and,
opposite by sign, unit residues there.

Note now, that to each so defined second and/or third-kind Abelian differential one
can add any linear combination of the holomorphic ones, without changing
their behavior at the singular points. To fix this ambiguity we should
impose $g$ constraints (the dimension of the space of holomorphic differentials)
to their periods, the canonical choice of such constraint is
vanishing of all their $A$-periods $\oint_{\bf A} d\Omega = 0$. This
means, that as a strict definitions of the Abelian differentials one should
write
\be
\label{2kA}
d\Omega_k\ \stackreb{P\to P_0}{\sim}\
{d\zeta\over\zeta^{k+1}}+\dots,\ \ \ \ \oint_{\bf A} d\Omega_k = 0,
\ \ \ \ \ k\geq 1
\ee
for the second kind, and
\be
\label{3kA}
d\Omega_0\ \stackreb{P\to P_\pm}{\sim}\
\pm{d\zeta_\pm\over\zeta_\pm}+\dots,\ \ \ \ \oint_{\bf A} d\Omega_0 = 0,
\ee
for the third kind, where $\zeta_\pm(P_\pm)=0$.
The bipole differential (\ref{3kA}) can be also presented as
\be
\label{bp}
d\Omega_{0} = d\log
{E(P,P_+)\over E(P,P_-)}
\ee
where $E(P,P')$ is the so called prime form on $\Sigma\times\Sigma$, the
$-\half$-bidifferential with the only zero at coinciding arguments, see
details and its properties in the book \cite{Fay}.

In the same way as with \rf{sypema}, one can write down the Riemann bilinear
relations for generally meromorphic Abelian differentials.
For example, for the first and third kind Abelian differentials one gets
\be
\label{syToda}
0=\int_{\Sigma} d\omega_\beta\wedge d\Omega_0=
\sum_\alpha\left( \oint_{A_\alpha}d\omega_\beta\oint_{B_\alpha}
d\Omega_0-\oint_{A_\alpha}
d\Omega_0\oint_{B_\alpha}d\omega_\beta \right) + \\ +
\res_\infty(d\omega_\beta)\int_{\infty_-}^{\infty_+}d\Omega_0
- \res_\infty(d\Omega_0)\int_{\infty_-}^{\infty_+}d\omega_\beta
= \oint_{B_\beta}d\Omega_0 - \int_{\infty_-}^{\infty_+}d\omega_\beta
\ee
which is widely used the theory of Toda integrable systems, appearing
rather naturally in the context of matrix models.

In practice, complex manifolds can be effectively described by systems of
polynomial equations in multi-dimensional complex Euclidean spaces $\mathbb{C}^K$,
to get a complex curve one should have $K-1$ equations for $K$
variables. The coefficients of these equations play the role of {\em
moduli} of the complex structures on a given Riemann surface, for a genus $g>1$
surface the complex dimension of moduli space is $\dim_\mathbb{C}{\cal
M}_g=3g-3$, that can be found from the Riemann-Roch theorem.
The Riemann surfaces with different complex structures cannot be
holomorphically mapped to each other, i.e. are different complex manifolds,
though can be isomorphic as real surfaces. Note also, that it means that matrix
elements of the period matrices \rf{pemat} are not functionally independent, since
there are totally $g(g+1)/2$ functions of only $3g-3$ moduli parameters,
except for the cases
$g=2,3$ being solutions to the quadratic equation $g(g+1)/2=3g-3$ and exceptional
case $g=1$ with the only modulus of a complex torus.

\subsection{Definition of quasiclassical tau-function
\label{ss:KriW}}

Let $\Sigma$ be a Riemann surface of some finite genus $g$, and suppose we have
defined on $\Sigma$ two meromorphic differentials, say denoted $dx$ and
$dy$, with the fixed periods. As we already discussed above, without loosing
of generality one can always
take the canonically normalized meromorphic differentials, with, say,
vanishing $A$-periods. However, if we simultaneously fix
the rest $B$-periods, this gives already nontrivial constraints on the moduli of
$\Sigma$. In practice this restricts us onto some subspace in the
total moduli space, and the dimension of this subspace can be estimated in
the following way:
the differentials $dx $ and $dy$ are only defined up to multiples, and fixing
the $B$-periods of $dx $ gives $g-1$ constraints (taking account of
the scaling freedom), while the $B$-periods of $dy$ give a further
$g-2$ constraints.
Altogether we come to a system with $(3g-3)-(g-1)-(g-2)=g$ parameters, i.e. the
number of moduli equal to the genus of $\Sigma$.

A complex curve endowed with two meromorphic differentials with the fixed
periods, or, briefly, the $g$-parametric family of curves defines an
{\em integrable system}\footnote{\label{fu:is}
In the most general spirit of the Liouville theorem,
one can say that on $g$-dimensional subspace of moduli space of $\Sigma$ one can always
choose
$g$ independent functions - Hamiltonians or actions, while the co-ordinates on Jacobian
{\bf Jac} of $\Sigma$, a $g$-dimensional complex torus, play
the role of complexified angle variables.}.
In practice, one considers a $g$-parametric family
of Riemann surfaces $\Sigma$, endowed with a {\em generating} meromorphic
differential, given, up to a constant, by the formula
\be
\label{dS}
dS \propto ydx, \ \ \ \ \delta_{\rm moduli} dS = {\rm holomorphic}
\ee
where $y(P)=\int^P dy$, $P\in\Sigma$ is an Abelian integral on $\Sigma$,
and by variation in moduli we mean any local variation of the rest $g$ parameters
of the family, which "survives" after one fixed all periods of $dx$ and $dy$.

A general definition for the {\em prepotential} of an integrable system for
a $g$-dimensional family of smooth curves
$\Sigma_g$ of genus $g$ endowed with meromorphic one-forms \rf{dS}, is
of the following form
\be
\label{prep}
{\bf S} = \oint_{\bf A} dS,
\\
{\d\F\over\d {\bf S}} = 2\pi i\oint_{\bf B} dS
\ee
where $A$ and $B$ are dual cycles in the homology basis. The prepotential
is defined by \rf{prep} locally on the moduli space of $\Sigma$, or better, on
the Teichm\"uller space with the fixed basis in $H^1(\Sigma,\mathbb{Z})$.
For complex curves, where it is natural to add to \rf{prep} the variables associated
with the degenerate cycles or marked points
(as in the simplest example of gaussian matrix model \rf{Fgau}
we have considered above),
the exact definition of so generalized prepotential
was given in \cite{KriW}, where it was called
the (logarithm of) quasiclassical tau-function or
tau-function of generalized Whitham hierarchy.
The consistency of \rf{prep} is guaranteed by integrability condition
\be
\label{period}
{\d^2\F\over \d S_\alpha\d S_\beta} = 2\pi iT_{\alpha\beta}
\ee
following directly from \rf{dS} and \rf{sypema} or
symmetricity of the period matrix $T_{\alpha\beta}$ of the Riemann surface $\Sigma_g$.

For degenerate curves or even the curves with the marked points,
the formulas \rf{prep} should be treated more
carefully and cannot be applied directly. A degenerate $A$-cycle turns into
a pair of marked points $P_\pm$ (a solitonic limit of a handle) and for such
degenerate handle formulas \rf{prep} turn into
\be
\label{t0res}
t_0 = {1\over 2\pi i}\res_{P_+} dS = - {1\over 2\pi i}\res_{P_-} dS
\\
{\d\F\over\d t_0} = 4\pi i\int_{P_-}^{P_+} dS
\ee
where the last expression is naively divergent and should be applied only after
extra careful definition, where the divergent integral is replaced by
appropriate finite quantity, like it happened in the simplest gaussian example
\rf{dFPigau}.

In most general setup, one should
complete the definition \rf{prep} by the time-variables associated with the
second-kind Abelian differentials with singularities at a point $P_0$
\be
\label{tP}
t_k = {1\over w\pi ik}\res_{P_0} \xi^{-k}dS,\ \ \ k>0
\\
{\d\F\over \d t_k} = {1\over 2\pi i}\res_{P_0} \xi^{k}dS,\ \ \ k>0
\ee
where $\xi$ is an {\em inverse} local co-ordinate at $P_0$:
$\xi(P_0)=\infty$.
The consistency condition for \rf{tP} is ensured by
\be
\label{sysi}
{\d^2\F\over \d t_n\d t_k} = {1\over 2\pi i}\res_{P_0} (\xi^k d\Omega_n)
\ee
and symmetricity of \rf{sysi} is provided by $\Omega_n= \xi^n_+$, where
operation $+$ leaves only the main, singular at $P_0$, part of the function $\xi^n$.
In other words, the differential $d\Omega$ is a meromorphic on $\Sigma_g$ with
the only singularity at $P_0$, i.e. is the second-kind Abelian differential
\rf{2kA}.

Altogether, the basis of "flat" times $({\bf S}, {\bf t}, t_0)$ exactly
corresponds to the basis of the first, second and third kind Abelian
differentials. For the generating one-form \rf{dS} one has
\be
\label{shol}
{\d dS\over \d S_\alpha} = d\omega_\alpha, \ \ \ \alpha=1,\dots,g
\ee
together with
\be
\label{s3}
{\d dS\over \d t_0} = d\Omega_0
\ee
and
\be
\label{smer}
{\d dS\over \d t_k} = d\Omega_k, \ \ \ k\geq 1
\ee
which are "dual" formulas to the first lines of \rf{prep}, \rf{t0res} and \rf{tP},
and derivative over moduli is taken at constant local co-ordinate $\xi$ (that corresponds
to choice of connection on moduli space of $\Sigma$).
For the rational curve $\Sigma_0$ and a single marked point $P_0$ the variables
\rf{prep} and \rf{t0res} are absent, and formulas
\rf{tP} define the tau-function of dispersionless dKP hierarchy. In this
case the singular at $P_0$ part $\xi^n_+$
is a $n$-th degree polynomial in some global uniformizing co-ordinate $\lambda$.

\subsection{Prepotential beyond one complex dimension}

The quasiclassical tau-function \rf{prep} can be considered as a particular case of
so called prepotentials of complex manifolds, which exist as well for
higher complex dimensions, like in the case of the Calabi-Yau three-folds, being
one of the most interesting examples for the purposes of string theory.
A general definition of prepotential contains:
\begin{itemize}
    \item A complex manifold $\Sigma$, homology basis with symplectic
structure $A_\alpha\circ B_\beta=\delta_{\alpha\beta}$, moduli space ${\cal M}$
of complex structures on $\Sigma$;

\item Generating differential form $\Omega$;

\item The set of period variables $S_\alpha = \oint_{A_\alpha}\Omega$;

\item Dual variables - the dual periods $\Pi_\alpha = \oint_{B_\alpha}\Omega$;

\item Integrability condition ${\d \Pi_\alpha\over\d S_\beta} = {\d \Pi_\beta
\over\d S_\alpha}$,
following from bilinear relations $\int_{\Sigma}\delta\Omega\wedge\delta\Omega=0$;
\end{itemize}

If so, like in the one-dimensional case considered in previous section, one can
argue, that there exists a prepotential
${\cal F}$: $\Pi_\alpha = {\d\F\over\d S_\alpha}$.
As in the one-dimensional case, the whole picture is defined only locally on the
Teichm\"uller space, but on the moduli space of complex structures it
is consistent with the duality
transformations: $A \leftrightarrow B$,
$S \leftrightarrow \Pi$, in the sense of the Legendre transform of the prepotential
\be
\label{duality}
\F \leftrightarrow \F + \sum_\alpha S_\alpha\Pi_\alpha
\ee
or, generally, the whole "electro-magnetic" duality of $Sp(2h,\mathbb{Z})$, where $2h$ is
the dimension of the homology group \cite{dWM}.

However, there are few very important distinctions to be necessarily commented.
For the moduli spaces in the case of
Calabi-Yau three-folds one has literally $h=\dim {\cal M} = h^{(2,1)}+1 =
\# {\rm (A-cycles)}$ coincidence of dimension of moduli space with the number of
$A$-cycles, unlike the case of
complex curves, where $\dim {\cal M} = 3g-3$, while $\# {\rm (A-cycles)}=g$ and
there is discrepancy with naive counting. Therefore, in the latter case
one has necessarily to restrict the family of $\Sigma$
to some subspace of moduli space, the most invariant way to do this is, as
already mentioned above,
to consider a pair of meromorphic one-forms with the fixed periods, i.e., if
on the curve $\Sigma$
$$
\exists\ \Omega_{1,2}=\{d\lambda,\ dz\}:\ \ \ \oint \Omega_{1,2}={\rm const}
$$
this leaves exactly $g$ moduli from $3g-3$ and this is the most
"rough" definition of an integrable system - at the level of counting, see
footnote~\ref{fu:is}.
However, for higher dimensional complex manifolds
the number of cycles usually exceeds the number of moduli.

For generating differential forms one gets the following picture
\begin{itemize}
    \item For the Calabi-Yau 3-folds $\Omega \in H^{(3,0)}$ is a unique (up to
    multiplication by a factor) {\em holomorphic} $(3,0)$-form and
    relation $\int_\Sigma \delta\Omega\wedge\delta\Omega=0$
follows from the decomposition $\delta\Omega\in H^{(3,0)}\oplus H^{(2,1)}$;
\item For the complex curves $\Omega=dS$ is a {\em meromorphic} $(1,0)$-form,
in addition it may have nontrivial residues $\res (dS)$, $\res (\xi^{-k}dS)$,
which play the role of the "generalized" periods \rf{t0res}, \rf{tP}.
\end{itemize}
Hence, in the last case we restore the picture of the previous paragraph.

\subsection{Tau-function of the one-matrix model}

Now let us go back and continue discussion of the one-matrix model free energy,
using the new notions introduced above\footnote{The geometric formulation for
the free energy of one-matrix model was first proposed in \cite{David}
and was discussed later in \cite{BDE,DV,ChM}.}. The functions $G$ and $y$
(see \rf{mamocu} and \rf{dvc}) are defined on the double cover of
the complex $x$-plane and on each sheet of this double cover
they acquire discontinuities along the segments or cuts, proportional to
the eigenvalue density $\rho( x)$, vanishing outside the cuts (see
fig.~\ref{fi:cuts}). Therefore, the eigenvalue
distribution \rf{fracS} can be described by the periods
of generating differential \rf{dS}
\be
\label{DVper}
S_\alpha = \oint_{A_\alpha}dS
\ee
where the contour integrals are taken around the eigenvalue supports ${\bf C}_\alpha$,
to be identified (except for one of the supports, i.e. $k=1,\dots,n-1$) with the
set of canonical $A$-cycles on the curve \rf{dvc} (see
fig.~\ref{fi:cuts}). The generating differential $Gdx$ with the resolvent
\rf{mamoG} for the one-matrix model can be equally chosen as
\be
\label{dvds}
dS = {i\over 4\pi}\ yd x
\ee
where now $x$ and $y$ are the co-ordinates on
$\Sigma$ determined by \rf{dvc}, since $\oint Gdx = -\half\oint ydx$, see
\rf{yG}\footnote{The question about residues \rf{t0res}, \rf{tP} is only slighty
more delicate. For the symmetric choice of generating differential \rf{dvds} one
can compute them at any of two infinities $x=\infty$, while in the asymmetric case
\rf{mamoG} they should be computed only at infinity on unphysical sheet, where
the resolvent \rf{mamoG} is singular.}.
Both $x$ and $y$ are globally defined meromorphic functions on \rf{dvc}, so that
the integrals $\oint dx =0$ and $\oint dy =0$ obviously vanish for any
choice of the closed contours. The relations \rf{shol} are certainly valid
for the differential \rf{dvds},
when the derivatives are taken at fixed residues and coefficients $\{ t_l\}$
of the potential (\ref{mmpot}).

Now, an important fact is that the Lagrangian
multipliers (see (\ref{vareq})) can be also rewritten as period
integrals of the same
generating differential (\ref{dvds}) over the dual $B$-contours (see
again fig.~\ref{fi:cuts})
\be
\label{DVF0}
\Pi _\alpha = -W(x^*_\alpha) + 2\int_{\bf C} dx \rho(x) \log\left(x-x^*_\alpha\right)
\ee
where the points $x^*_\alpha\in {\bf C}_\alpha$ can be conveniently
chosen, each on $\alpha$-th piece of the
support; then the modulus can be "forgotten" (see footnote~\ref{fu:modulus})
and $\Pi_\alpha$ become holomorphic quantities.

Formula \rf{DVF0} can be transformed easily to the contour integral
(for simplicity, we consider explicitly the case of only two cuts,
as on fig.~\ref{fi:w3}, the generalization is straightforward).
In this case we have two fractions $S_1+S_2=t_0$ and
it is more convenient to choose the independent variables $t_0$ and
$S=S_2$. Then, $\sum_\alpha \Pi_\alpha S_\alpha = \left(\Pi_2-\Pi_1\right)S+\Pi_1t_0$
and variation of the free energy \rf{variF}
w.r.t. $S$ gives
\be
\label{dFS}
{\d\F\over \d S} = W(x^*_1)-W(x^*_2) +
2\sum_{\alpha=1,2} \int_{{\bf C}_\alpha} dx\rho(x)\log{x-x^*_2\over x-x^*_1}
\ee
Since the integral over the contour around total support vanishes
$\oint_{{\bf C}} dxG(x)\log{x-x^*_1\over x-x^*_2} = 0$
(this contour can be deformed to the contour around the point $x=\infty$ where the
integrand has no residue), the last expression can be transformed to
\begin{figure}[tp]
\epsfysize=4cm
\centerline{\epsfbox{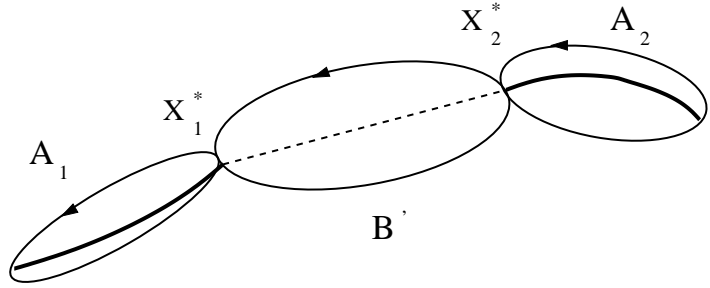}}
\caption{The calculation of the integral in \rf{int1mm}. The integral over the
cuts ${\bf C}_k$ for $k=1,2$ (fat lines), surrounded by the $A_k$-cycles, are
replaced, first, by the integral over the $B'$-cycle around the logarithmic cut
(the dashed line). The latter is transformed to the integral along the contour
between the points $x_1^*$ and $x_2^*$ (without logarithm in the integrand), which
is half of the period taken along the $B$-cycle, going one way between the cuts
${\bf C}_1$ and ${\bf C}_2$ on one sheet and vice versa on the other.}
\label{fi:logcut1}
\end{figure}
(see fig.~\ref{fi:logcut1})
\be
\label{int1mm}
\sum_{\alpha=1,2} \int_{{\bf C}_\alpha} dx\rho(x)\log{x-x^*_1\over x-x^*_2} =
\sum_{\alpha=1,2}{1\over 2\pi i}\oint_{{\bf C}_\alpha} dxG(x)\log{x-x^*_1\over x-x^*_2} =
\\ =
- {1\over 2\pi i}\oint_{B'} dxG(x)\log{x-x^*_1\over x-x^*_2} =
\int_{x^*_2}^{x^*_1} dx G(x)
\ee
Altogether, this leads for \rf{dFS} to the formula
\be
{\d\F\over \d S} = W(x^*_1)-W(x^*_2) -
2\int_{x^*_2}^{x^*_1} dx G(x) = \int_{x^*_2}^{x^*_1} ydx =
-\half\oint_{B} ydx = 2\pi i\oint_B dS
\ee
where the contour $B$ goes between $x_1^*$ and $x_2^*$ and back on the other
sheet of the curve (do not mix with the $B'$-contour around the points $x_1^*$ and
$x_2^*$!).
For generic number of cuts one should similarly replace $\sum_{\alpha=1}^n
\Pi_\alpha S_\alpha = \left(\Pi_1-\Pi_n\right)S_1+\dots+
\left(\Pi_{n-1}-\Pi_n\right)S_{n-1}+\Pi_nt_0$ and proceed to the trick with
the logarithmic cuts. This leads to identifying free energy of the
one-matrix model with the prepotetial \rf{prep}.

To the set of parameters (\ref{DVper}) one should also
add\footnote{By the $\infty$-point in what follows we call for short the point
$\infty_+$ or $ x=\infty$ on the ``upper" sheet of hyperelliptic Riemann
surface (\ref{dvc}) corresponding to the positive sign of the square root,
i.e. to $y=+\sqrt{W'( x)^2+4f( x)}$.} the
total number of eigenvalues \rf{normrG} 
\be
\label{t01mm}
t_0 = {1\over 2\pi i}\res_\infty\left( dS\right) = {2f_{n-1}\over (n+1)t_{n+1}}
\ee
which is exactly the zero time \rf{t0res},
and the parameters of the potential (\ref{mmpot}),
rewritten in the form of the times \rf{tP}, i.e. as
\be
\label{LGtimes}
t_k = {1\over 2\pi ik}\res_\infty\left( x^{-k}dS\right),
\ \ \ \
k=1,\dots,n
\ee
since $x$ plays the role of the inverse local co-ordinate at the
infinity point $x=\infty$.
Then, explicitly
\be
\label{bipole}
d\Omega_0 =
\left.{\d dS \over \d t_0}\right|_{{\bf S},{\bf t}}
=(n+1)t_{n+1}{ x^{n-1} d x\over y} +
2\sum_{k=0}^{n-2}{\d f_k\over\d t_0}{ x^k d x\over y}
\ee
and the dependence of $\{ f_k\}$ with $k=0,1,\dots,n-2$ on $t_0$ is fixed
by the period integrals
\be
\oint_{A_\alpha}\left((n+1)t_{n+1}{ x^{n-1} d x\over y} +
2\sum_{k=0}^{n-2}{\d f_k\over\d t_0}{ x^k d x\over y}\right)=0
\ee
which for $\alpha=1,\dots,n-1$ gives exactly $n-1$ relations on the derivatives
of $f_0,f_1,\dots,f_{n-2}$ w.r.t. $t_0$.

For the derivatives w.r.t. parameters of the potential
(\ref{LGtimes}), one gets
\be
\label{omes}
d\Omega_k = \left.{\d dS \over \d t_k}\right|_{{\bf S},t_0} =
{W'( x)k x^{k-1} d x\over y} +
2\sum_{j=0}^{n-2}{\d f_j\over\d t_k}{ x^k d x\over y}
\ee
analogously obeying
\be
\label{peom}
\oint_{A_\alpha}d\Omega_k =
\oint_{A_\alpha}{W'( x)k x^{k-1} d x\over y} +
2\sum_{j=0}^{n-2}{\d f_j\over\d t_k}\oint_{A_\alpha}
{ x^k d x\over y} = 0
\ee
and this is again a system of linear equations, resolved for ${\d f_j\over\d t_k}$.
To complete the setup
one should also add to (\ref{DVF0})
\be\label{v}
{\d\F\over\d t_k}
= {1\over 2\pi i}\res_\infty\left( x^k dS\right), \ \ \ \ k>0
\ee
and the following formula\footnote{\label{reg}
Naively understood the integral in (\ref{dfdt0}) is divergent
and should be supplemented by a proper regularization.
This subtlety can be just ignored, if one needs only
the residue formulas for the third derivatives, (to be considered in the next section),
say for the WDVV equations \cite{DVWDVV}. The
simplest way to avoid these complications is to think of the pair of marked
points $\infty$ and $\infty_-$ as of degenerate handle; then the residue
(\ref{t01mm}) comes from degeneration of the extra $A$-period, while the
integral (\ref{dfdt0}) from degeneration of the extra $B$-period, see
fig.~\ref{fi:cuts}. For practical purposes this divergent period
can be always replaced by a finite
quantity, see e.g. the gaussian example \rf{dFPigau}.}
\be
\label{dfdt0}
{\d \F \over \d S_n} =
\Pi_n = \int_{B'_n} dS
\ee
with the appropriately chosen contour
$B'_n$ passing from $\infty_-$ on the lower sheet to
$\infty_+$ on the upper sheet through the $n$-th cut
(we again remind that, instead of $t_0$, the parameter $S_n =
t_0 - \sum_{i=1}^{n-1}S_\alpha $ can be used equivalently).
In the case of differential \rf{dvds} for the one-matrix model these
definitions requires some care due to divergences, however, in
the geometric context of the AdS/CFT correspondence \cite{KMMZ} one considers
an example of the generating differential
with regular behavior in $\infty_\pm$, and the formulas like \rf{dfdt0}
can be literally applied.

\subsection{Residue formula
\label{ss:residue}}

Now let us discuss one of the most universal and nice formulas for the quasiclassical
tau-function - the residue formula for its third derivatives
\be
\label{residue}
{\d^3 \F\over \d T_I\d T_J\d T_K} =
{1\over 2\pi i}\res_{dx=0}\left(dH_IdH_JdH_K\over dx dy\right)
\ee
which is universal for dependence of $\F$ upon any
variables $\{ T_I\}$, with $\{ dH_I\}$ being
the corresponding differentials. Existence of simple formula \rf{residue} for the third
derivatives together with absence if similar expressions for the higher derivatives
has definitely the string theory origin, coming from the world sheet theory, where the
three-point correlation functions on sphere play a distinguished role.

\subsubsection{Holomorphic differentials
\label{ss:reshol}}

Let us derive the formulas for the third derivatives of prepotential
$\F$,
following the way proposed by Krichever in \cite{KriW}, and presented
explicitly in \cite{DVWDVV}.
We first note, that the derivatives of the elements $T_{ij}$
of the period matrix \rf{period}
can be expressed through the integral over the boundary
$\d\Sigma$ of the cut Riemann surface $\Sigma$ (see fig.~\ref{fi:cut})
\be
\label{dpm}
{\d T_{\alpha\beta}\over \d S_\gamma}\equiv \d_\gamma T_{\alpha\beta} =
\int_{B_\beta}\d_\gamma d\omega_\alpha =
-\int_{\d\Sigma}\omega_\beta\d_\gamma d\omega_\alpha
\ee
where $\omega_\beta = \int^P d\omega_\beta$ are the Abelian integrals, whose values
on two copies of cycles on the cut Riemann surface (see fig.~\ref{fi:cut})
are denoted below as $\omega_\beta^\pm$.
Indeed, the computation of the integral in the r.h.s. of (\ref{dpm}) gives
\be
\label{derdpm}
\int_{\d\Sigma}\omega_\beta\d_\gamma d\omega_\alpha =
\sum_\rho\left(\int_{B_\rho}\omega_\beta^+\d_\gamma d\omega_\alpha -
\int_{B_\rho}\omega_\beta^-\d_\gamma d\omega_\alpha\right) -
\sum_\rho\left(\int_{A_\rho}\omega_\beta^+\d_kd\omega_\alpha -
\int_{A_\rho}\omega_\beta^-\d_\gamma d\omega_\alpha\right) = \\ =
\sum_\rho\oint_{B_\rho}\left(\oint_{A_\rho}d\omega_\beta\right)\d_\gamma d\omega_\alpha  -
\sum_\rho\oint_{A_\rho}\left(\oint_{B_\rho}d\omega_\beta\right)\d_\gamma d\omega_\alpha =
\\ =
\sum_\rho\left(\oint_{A_\rho}d\omega_\beta\right)\oint_{B_\rho}\d_\gamma d\omega_\alpha  -
\sum_\rho\left(\oint_{B_\rho}d\omega_\beta\right)\oint_{A_\rho}\d_\gamma d\omega_\alpha\
\stackreb{\oint_{A_\alpha}d\omega_\beta=\delta_{\alpha\beta}}{=}\ -\d_\gamma  T_{\alpha\beta}
\ee
One can now rewrite formula (\ref{dpm}) as
\be
\label{dpmres}
\d_\gamma  T_{\alpha\beta} = -\int_{\d\Sigma}\omega_\beta\d_\gamma  d\omega_\alpha =
\int_{\d\Sigma}\d_\gamma  \omega_\beta d\omega_\alpha = \sum\res_{d\lambda = 0}
\left(\d_\gamma  \omega_\beta d\omega_\alpha\right)
\ee
where the sum is taken over all residues of the integrand, i.e.
over all residues of $\d_\gamma  \omega_\beta$ since the differentials $d\omega_\alpha$ are
holomorphic. In order to investigate these singularities and
clarify the last equality in (\ref{dpmres}), we
discuss first the strick sense of the derivatives $\d_\gamma $ w.r.t. moduli, or
introduce the corresponding connection on moduli space of $\Sigma$.

To this end, let us introduce a
covariantly constant function of moduli, say the Abelian integral $x=\int^P dx$,
i.e. choose such a connection on moduli space, that $\d_\gamma  x=0$. Roughly
speaking, the role of covariantly constant function can be played
by one of co-ordinates,
in the simplest possible description of complex curve by a
single equation on two complex variables - any of two functions constrained by the equation
$F(x,y)=0$. Then, using this
equation, one may express the other co-ordinate $y$ as a function of
$x$ and moduli (the coefficients of the equation). Any Abelian integral $\omega_\beta$
can be then, in principle, expressed in terms of $x$,
and in the vicinity of critical points $\{x_a\}$
where $dx=0$ (in general position) we get an expansion
\be
\label{vexp}
\omega_\beta(x)\ \stackreb{x\to x_\alpha}{=}\ \omega_{\beta a} +
c_{\beta a}\sqrt{x-x_a}+ \dots
\ee
whose derivatives
\be
\label{dervj}
\d_\gamma \omega_\beta \equiv \left.\d_\gamma \omega_\beta\right|_{x= {\rm const}} =
- {c_{\beta a}\over
2\sqrt{x-x_a}}\d_\gamma x_a + {\rm regular}
\ee
possess the first order poles at $x=x_a$, and they are written up to
regular terms which do not contribute to the expression
(\ref{dpmres}). The exact coefficient in (\ref{dervj}) can be
computed relating the function $x$ with
the generating differential $dS = ydx$. Using
\be
\label{yl}
y(x)\ \stackreb{x\to x_a}{=}\
\Gamma_a\sqrt{x-x_a}+ \dots
\ee
where $\Gamma_a =
\sqrt{\prod_{b\neq a}(x_a-x_b)}$ or
\be
\label{deryl}
{\d\over\d S_\gamma} y(x) = - {\Gamma_a\over
2\sqrt{x-x_a}}{\d x_a\over\d S_\gamma} + {\rm regular}
\ee
together with
\be
\label{difyv}
dy = {\Gamma_a \over
2\sqrt{x-x_a}}\ dx + {\rm regular}
\ee
and
\be
\label{difzv}
d\omega_\beta = {c_{\beta a}\over
2\sqrt{x-x_a}}\ dx + \dots
\ee
and, following from (\ref{dvds}) and (\ref{shol}) expansion
\be
d\omega_\gamma  = \d_\gamma  dS = - {\Gamma_a\d_\gamma x_a\over
2\sqrt{x-x_a}}\ dx + {\rm regular}
\ee
one finally gets for (\ref{dpmres})
\be
\res\left(\d_\gamma  \omega_\beta d\omega_\alpha\right) = \sum_a
\res\left( {c_{\beta a}\d_\gamma x_a\over
2\sqrt{x-x_a}}d\omega_\alpha\right)
= \sum_a
\res\left( {d\omega_\beta\over
dx}d\omega_\alpha\d_\gamma x_a\right)=
\sum_a
\res\left( {d\omega_\alpha d\omega_\beta d\omega_\gamma \over
dx dy}\right)
\ee
In the case with hyperelliptic curves \rf{dvc} for one-matrix model, this
general derivation of the residue formula can be replaces by based on
the formula \cite{Fay}
\be
\label{Fay}
\frac{\partial
T_{\alpha\beta}}{\partial x_a} = \hat
\omega_\alpha(x_a)\hat\omega_\beta(x_a)
\label{derram}
\ee
where $\hat\omega_\alpha(x_a) =
\left.{d\omega_\alpha(x)\over
d\sqrt{x-x_a}}\right|_{x=x_a}$ is
the "value" of canonical differential at a critical point.

\subsubsection{Meromorphic differentials
\label{ss:resmero}}

Almost in the same way the residue formula can be derived for the
meromorphic differentials.
One gets
\be
{\d\F \over\d t_k} = {1\over 2\pi i}\res_\infty \left(x^k dS \right), \ \ \ \ k>0
\ee
therefore
\be
{\d^2\F \over\d t_k\d t_n} =
{1\over 2\pi i}\res_\infty \left(x^k d\Omega_n\right) =
{1\over 2\pi i}\res_\infty \left((\Omega_k)_+ d\Omega_n\right)
\ee
where $\left(\Omega_k\right)_+$ is the singular part of the integrated
one-form $d\Omega_k$. Further
\be
{\d\over\d t_m}\res_\infty \left(x^k d\Omega_n\right) =
\res_\infty \left(x^k {\d d\Omega_n\over\d t_m}\right) =
 -\res_\infty \left((d\Omega_k)_+ {\d \Omega_n\over \d t_m}\right) =
-\res_\infty \left(d\Omega_k {\d \Omega_n\over \d t_m}\right)
\ee
The last expression can be rewritten as
\be
\label{3mero}
-\res_\infty \left(d\Omega_k {\d \Omega_n\over \d t_m}\right) =
\oint_{\d\Sigma}\left(d\Omega_k {\d \Omega_n\over \d t_m}\right) +
\sum \res_{x_a} \left(d\Omega_k {\d \Omega_n\over \d t_m}\right)
= \sum \res_{x_a} \left(d\Omega_k {\d \Omega_n\over \d t_m}\right)
\ee
since $\oint_{\d\Sigma}\left(d\Omega_k {\d \Omega_n\over \d t_m}\right) =0$
due to $\oint_{A_\alpha}d\Omega_n = 0$, (cf. with (\ref{derdpm})):
\be
\label{derome}
\int_{\d\Sigma}\Omega_j{\d\over\d t_k} d\Omega_i =
\sum_\alpha\left(\int_{B_\alpha}\Omega_j^+{\d\over\d t_k}d\Omega_i -
\int_{B_\alpha}\Omega_j^-{\d\over\d t_k}d\Omega_i\right) -
\sum_\alpha\left(\int_{A_\alpha}\Omega_j^+{\d\over\d t_k}d\Omega_i -
\int_{A_\alpha}\Omega_j^-{\d\over\d t_k}d\Omega_i\right) = \\ =
\sum_\alpha\oint_{B_\alpha}\left(\oint_{A_\alpha}d\Omega_j\right){\d\over\d t_k}d\Omega_i
-
\sum_\alpha\oint_{A_\alpha}\left(\oint_{B_\alpha}d\Omega_j\right){\d\over\d t_k}d\Omega_i =
\\ =
\sum_\alpha\left(\oint_{A_\alpha}d\Omega_j\right)\oint_{B_\alpha}{\d\over\d t_k}d\Omega_i
-
\sum_\alpha\left(\oint_{B_\alpha}d\Omega_j\right)\oint_{A_\alpha}{\d\over\d t_k}d\Omega_i\
\stackreb{\oint_{A_\alpha}d\Omega_j=0}{=}\ 0
\ee
Now, as in the holomorphic case, one takes the initial terms of the expansion
\be
\label{oexp}
\Omega_n(x)\ \stackreb{x\to x_a}{=}\
\Omega_{na} +
\gamma_{na}\sqrt{x-x_a}+ \dots
\ee
and, therefore
\be
\label{deroj}
{\d\over\d t_k}\Omega_j \equiv \left.{\d\over\d t_k}\Omega_j\right|_{x=const} =
- {\gamma_{ja}\over
2\sqrt{x-x_a}}{\d x_a\over\d t_k} + {\rm regular}
\ee
Then, using (\ref{yl}), (\ref{deryl}) and (\ref{difyv}) together with
\be
d\Omega_j = {\gamma_{ja}\over
2\sqrt{x-x_a}}\ dx + \dots
\ee
and the relation, following from (\ref{smer}), (\ref{deryl})
\be
d\Omega_k = {\d\over\d t_k}dS  = - {\Gamma_a dx\over
2\sqrt{x-x_a}}\ {\d x_a\over\d t_k}
+ {\rm regular}
\ee
one gets for (\ref{3mero})
\be
{\d^3\F\over\d t_k\d t_n\d t_m} =
{1\over 2\pi i}\sum \res_{x_a} \left(d\Omega_k {\d \Omega_n\over \d t_m}\right)
= - {1\over 2\pi i}\sum \res_{x_a} \left(d\Omega_k{\gamma_{na}\over
2\sqrt{x-x_a}}{\d x_a\over\d t_m}\right) =
\\\
= - {1\over 2\pi i}\sum \res_{x_a} \left(d\Omega_k{d\Omega_n\over dx}
{\d x_a\over\d t_m}\right) =
{1\over 2\pi i}\sum \res_{x_a} \left({d\Omega_kd\Omega_nd\Omega_m
\over dx dy}\right)
\ee
The derivation of the residue
formula for the set of parameters including $t_0$ corresponding to the
third-kind Abelian differential (\ref{bipole}) can be performed almost
identically, and in a similar way, one proves the residue formula for the mixed derivatives.
Thus, we finally conclude
\be
\label{resgen}
{\d^3 \F\over \d T_I\d T_J\d T_K} = {1\over 2\pi i}\sum_{x_a}
\res_{x_a}\left(dH_IdH_JdH_K\over dx dy\right) =\\=
{1\over 2\pi i}\sum_{x_a}\res_{x_a}
\left({\phi_I\phi_J\phi_K\over {dx /dy}}dy\right) =
\sum_{x_a}\Gamma_a^2
\phi_I(x_a)\phi_J(x_a)
\phi_K(x_a)=\sum_{x_{a}}{\hat H_I(x_{a})
\hat H_J(x_{a})\hat H_K(x_{a})\over
\prod_{b\ne a}(x_{a}-x_{b})^2}
\ee
for the whole set of variables $\{ T_I \} = \{ t_k, t_0, S_i \}$ and
corresponding to them one-forms
$\{ dH_I \} = \{ d\Omega_k, d\Omega_0, d\omega_i \}$. In the second line of
(\ref{resgen}) we have introduced the meromorphic functions
\be
\label{phi}
\phi_I(x) = {dH_I\over dy} = {\hat H_I(x)\over R'(x)}
\ee
for any (meromorphic or holomorphic) differential on hyperelliptic curve \rf{dvc}
$dH_I = \hat H_I(x){dx\over y}$.

\setcounter{equation}0
\section{Complex curve of the two-matrix model
\label{ss:cc2m}}

Now let us turn to the case of two-matrix model \rf{mamocompl}. As for
the one-matrix model we first discuss the auxiliary complex manifold
$\Sigma$, and then turn to the quasiclassical tau-function.

\subsection{Quasiclassics of the two-matrix model}

\begin{figure}[tp]
\epsfysize=5.2cm
\centerline{\epsfbox{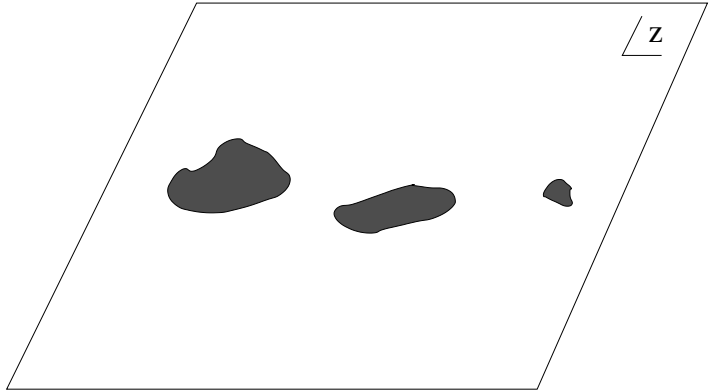}}
\caption{Drops in the eigenvalue plane of
the real problem in two-matrix model. In what follows this set of drops will be
referred to as a domain ${\sf D}$.}
\label{fi:drops}
\end{figure}

Consider the free energy of the two-matrix model
\rf{V2mm}, \rf{mamocompl} in the planar limit.
Again, as in the one-matrix case
in the planar or quasiclassical limit
one can replace the direct computation of the eigenvalue integral
\rf{mamocompl} by solution of the corresponding variational problem.
At least for the real problem in two-matrix model, with the mutually
complex conjugated eigenvalues\footnote{Sometimes it is also
called, in a rather misleading way, the
"normal" matrix model. We stress here, that studying the partition function
\rf{mamocompl} with the particular "cross-term" $\Tr\left(\Phi^\dagger\Phi\right)$
in the potential \rf{V2mm} the only essential thing is the choice of "real section"
or the particular class of integration contours over each $z_i$ and ${\bar z}_i$
integration variable. The difference between "normal" (with the commuting matrices $\Phi$
and $\Phi^\dagger$) and usual two-matrix
model arises only when studying multitrace correlators, which are beyond the scope
of geometric description discussed here.}
the analysis becomes easier due to the basic property of the complex logarithm
\be
\label{log}
\Delta \log\left|z-z'\right| = 2\pi\delta^{(2)}(z-z')
\ee
i.e. the two-dimensional Coulomb repulsion of the complex-conjugated eigenvalues is now
governed by a kernel of the operator inverse to the two-dimensional Laplacian
$\Delta =4\p_z \p_{\bar z}$. Therefore for the potential \rf{V2mm} the eigenvalue
density is some constant inside the support or (two-dimensional) eigenvalue domain and
vanishing outside, i.e.
\be
\label{rho2mm}
\rho (z,\bar z) \propto \left\{
\begin{array}{c}
  1, \ \ \ \ \ \ \  z\in D \\
  0, \ \ \ \ \ \ \  z\nin D
\end{array}\right.
\ee
and the corresponding real problem reduces to finding the potential energy of
some domain (a set of "drops") in complex $z$-plane, filled by charged liquid with
constant density of charge, see fig.~\ref{fi:drops}\footnote{This problem has
attracted recently lots of attention in the context of so called Laplacian growth,
see, for example \cite{Lapgro} and references therein.}. In contrast to
one-dimensional case, now the eigenvalue density \rf{rho2mm} is constant and
the nontrivial information about their distribution is encoded into the area and shape
of domain, expressed through the coefficients of matrix model
potential \rf{V2mm} (more strictly of its harmonic part).

The stationarity equation
\be
\label{sta2mm}
\bar z_i = W'(z_i) - \hbar\sum_{j\neq i}{1\over z_i-z_j}
\ee
following from the two-dimensional
free-energy functional, corresponding to the effective potential
\be
\label{variF2}
V_{\rm eff}(z, \bar z) = \sum_i \left({\bar z}_iz_i - W(z_i) -
{\tilde W}(\bar z_i)\right) + \hbar\sum_{i<j}\log \left|z_i-z_j\right|^2
\ee
for the two-matrix case \rf{mamocompl}
can be re-written in the form
\be
\label{2mmeq}
{\bar z} = W'(z) - G(z)
\ee
(together with the complex conjugated formula)
after introducing, like in the one-matrix case \rf{mamorho}, \rf{mamoG}, the density
\rf{rho2mm} and the following function
\be
\label{2mamoG}
G(z) = \hbar\left<\Tr{1\over z-\Phi}\right>_{\Phi,\Phi^\dagger}
= {1\over\pi}\int_{\mathbb{C}} {\rho(\zeta,\bar\zeta)d^2\zeta\over z-\zeta}\
\stackreb{\rf{rho2mm}}{=}\ {1\over\pi}\int_{\sf D} {d^2\zeta\over z-\zeta}
\ee
or the (holomorphic) resolvent for the two-matrix model.
When interaction in r.h.s. of \rf{sta2mm} is switched off at $\hbar\to 0$,
the resolvent vanishes, and \rf{2mmeq} can be symmetrically written as
\be
\label{CUCLA}
\left({\bar z} - W'(z)\right)\left(z - {\bar W}'({\bar z})\right) = 0
\ee
Generally equation \rf{CUCLA} has $n\tilde n$ solutions - the points in
complex plane $\mathbb{C}$, or in complexified situation
$(z,\bar z)\to (z,\tilde z)\in \mathbb{C}^2$ in the two-dimensional complex euclidean space.
Equation \rf{CUCLA}, like in the one-matrix
case, can be then treated as a degenerate classical curve.
When the Coulomb interaction
between the eigenvalues is switched on (in \rf{variF2} this is
literally a two-dimensional Coulomb interaction, justifying the terminology we
have formally introduced even in the one-matrix model case), this set of points turns
into a set of one-dimensional "trajectories", or into a real section of some
complex curve. In what follows, we are going first to study the
structure of this curve or Riemann surface in general position, corresponding to the complex problem
in two-matrix model and return to a real problem later.

\subsection{Equation and genus of the curve in the complex problem}

For general polynomial potentials $W$ and $\tilde W$ of
powers $n+1$ and $\tilde n+1$ correspondingly, the highest degree terms
of the analytic equation of the classical curve \rf{CUCLA} have the form
$z^n{\tilde z}^{\tilde n} + a_{\tilde n}z^{n+1} +
{\tilde a}_{n}{\bar z}^{\tilde n+1}$ with the coefficients
\be
\label{aa}
a_{\tilde n} =
-\left((\tilde n+1)\tilde t_{\tilde n+1}\right)^{-1},
\ \ \ \
\tilde a_{n} =
-\left((n+1) t_{n+1}\right)^{-1}
\ee
This means that the generic
desingularization of the classical curve \rf{CUCLA} acquires the following
form \cite{KM} (see also \cite{E2MM})
\be
\label{NONSY}
F(z, {\tilde z}) = \left((n+1)(\tilde n+1) t_{n+1}\tilde t_{\tilde n+1}\right)^{-1}
\left({\tilde z}-W_n'(z)\right)
\left(z-{\tilde W}_{\tilde n}'({\tilde z})\right)+\dots = \\ =
z^n{\tilde z}^{\tilde n} + a_{\tilde n}z^{n+1} +
{\tilde a}_n{\bar z}^{\tilde n+1} +
\sum_{(i,j)\in (N.P.)_+} f_{ij}z^i{\tilde z}^j
\ee
where we have already used complexified notations $(z,\tilde z)\in \mathbb{C}^2$
for the
co-ordinates embedding the curve \rf{NONSY} into the two-dimensional complex
space, instead of mutually complex-conjugated co-ordinates $(z,\bar z)\in\mathbb{C}$
of a point in the eigenvalue plane of fig.~\ref{fi:drops}. For the future purposes
we will precise the form of algebraic equation of the curve \rf{NONSY} for the
real potential \rf{V2mm} with complex-conjugated coefficients and equal
powers $n=\tilde n$
\be\label{complcu}
F(z, {\tilde z}) = z^n{\tilde z}^n + a_nz^{n+1} + {\tilde a}_n{\tilde
z}^{n+1} + \sum_{i,j\in (N.P.)_+} f_{ij}z^i{\tilde z}^j = 0
\ee
%
\begin{figure}[tp]
\centerline{\epsfig{file=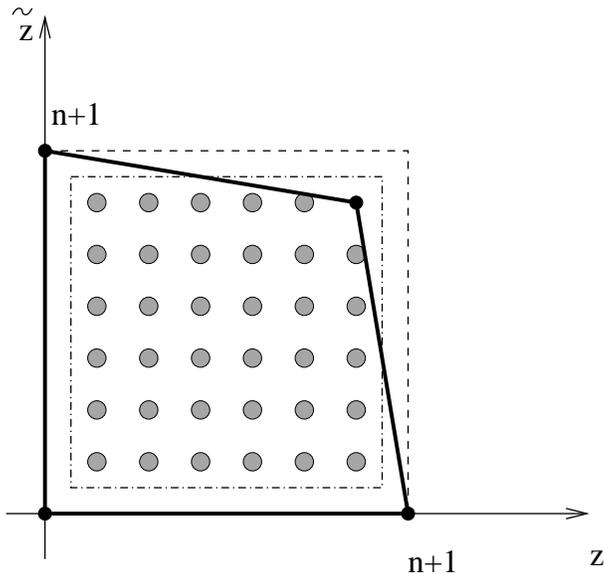,width=80mm}}
\caption{ The Newton polygon for the curve \rf{complcu}.
The highest degree terms in \rf{complcu} determine the shape of the
polygon and the integer dots inside it count the number of independent holomorphic
differentials, or genus of the curve. Clearly this number is equal to
the area of "dual" square except for one (black) point, so that
$g=n\tilde n-1=n^2-1$.}
\label{fi:newton}
\end{figure}
The properties of the curves (\ref{NONSY}), \rf{complcu} can be easily established via
the Newton polygon (see fig.~\ref{fi:newton} for the case $n=\tilde n$,
for general values of $n$ and $\tilde n$
the $(n+1)\times(n+1)$ square on fig.~\ref{fi:newton} should be replaced by the
rectangle of the size $(n+1)\times({\tilde n}+1)$ with all other
elements of the construction remaining intact).
The first three terms in \rf{NONSY} correspond to the three points on the
boundary lines of the Newton polygon,
while the sum over $(N.P.)_+$ in the last term stays for the sum over all
points inside the Newton polygon (including the points on both axis
not marked on fig.~\ref{fi:newton}).

The simplest basis for the holomorphic differentials on the curves \rf{NONSY}, \rf{complcu}
can be chosen as
\be\label{vij}
  dv_{ij} = z^i{\tilde z}^j {d{\tilde z}\over F_z} =
-z^i{\tilde z}^j {dz\over F_{\tilde z}},
\ee
with the degrees $i=i'-1$ and $j=j'-1$, where $(i',j')\in N.P.$ are
coordinates of the points strictly inside the Newton polygon, without
the boundary points (see fig.~\ref{fi:newton})\footnote{For example,
for $n=2$ there are three points inside the
polygon: $i',j'>0$ and $i'+j'\leq 2$, then the holomorphic differentials
are labelled by $i,j\geq 0$ and $i+j\leq 1$.}.
Counting the number of integer points inside the polygon one finds
that the number of linear independent holomorphic differentials, or genus of the
curve \rf{NONSY}, equals to
\be
\label{genus}
g = n\tilde n-1
\ee
For ${\tilde n}=1$ the gaussian integration over the matrix $\Phi^\dagger$
returns us to the one-matrix model with the hyperelliptic curve of genus $n-1$,
considered in sect.~\ref{ss:1mamo}, for generic $n$ and $\tilde n$ the
curves \rf{NONSY}, \rf{complcu} are certainly not hyperelliptic.

\begin{figure}[tp]
\centerline{\epsfig{file=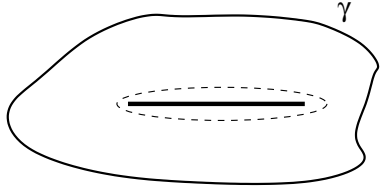,width=50mm}}
\caption{ The boundary of the drop $\bgamma$ and a cut of
a multi-valued function $\tilde z(z)$ inside the drop. On $\bgamma$ one has
an equality $\bar z = \tilde z(z)$ but this is, certainly, not true on the cut.}
\label{fi:cutspo}
\end{figure}

\begin{figure}[tp]
\centerline{\epsfig{file=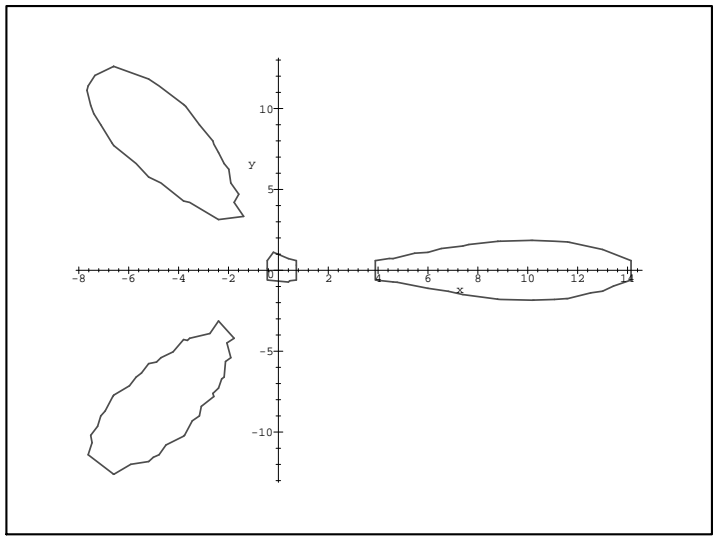,width=80mm,angle=-90}}
\caption{Curve for $0.11z^2{\bar z}^2 - z^3-{\bar z}^3 +
6z{\bar z}-6.5  =0$ by MAPLE computation.}
\label{fi:curve4h}
\end{figure}

The generating differential, measuring the constant density of eigenvalues \rf{rho2mm},
has the form
\be
\label{diff2mm}
dS = {1\over 2\pi i}\ {\tilde z}dz
\ee
The origin of this formula is demonstrated on fig.~\ref{fi:cutspo}, looking
at this figure one immediately comes to the following simple
relations for the two-dimensional and contour integrals
\be\label{RHODS}
\int_{\rm drop} dz\wedge d{\bar z} = \oint_\bgamma \bar z dz =
\oint_\bgamma \tilde z dz = \oint_{\rm cut} \tilde z dz
\ee
meaning that one can write the eigenvalue fractions as the period integrals
\be
\label{pertmm}
S_i = {1\over 2\pi i} \int_{i-\rm th\ drop} dz\wedge d{\bar z}
= {1\over 2\pi i}\oint_{A_i}{\tilde z}dz
\ee
The
relations \rf{RHODS} allow to endow the complex curve \rf{complcu}
(or \rf{NONSY} in the asymmetric case) with a
meromorphic generating differential \rf{diff2mm}, and, therefore to formulate
the planar two-matrix model in the geometric way we have discussed above.
They also clarify the relation between the complex and real problems;
we are now going to discuss in detail the first one, intensively using the
algebraic curve \rf{complcu}, and postpone discussion of the real
problem till next section, where it will be formulated, following \cite{KMZ},
in terms of "doubling" fig.~\ref{fi:drops} or the so called Schottky
double\footnote{
As a demonstration of non-trivial relation between the complex and real
problems, we present at fig.~\ref{fi:curve4h} the real section of the curve,
described by particular cubic equation. Clearly, formula \rf{genus}
for the complex problem gives $g=3$, i.e. four independent variables
\rf{pertmm}, corresponding to all extrema. Moreover, even the real
section of complex curve in this example consists of four disjoint real contours.
However, since cubic potential
possesses a single minimum, only one of these contours can be located on physical
sheet for a real problem. Nevertheless all extrema are indistinguishable
from the point of view of complex problem and "holomorphic" data \rf{complcu},
\rf{diff2mm}.
The "unphysical" drops give also essential
contribution to the
full non-perturbative partition function, but the detailed discussion of this
topic is beyond the scope of this paper.}.

The derivatives of generating differential \rf{diff2mm} w.r.t. coefficients
of the equation (\ref{NONSY}) can be
computed in a standard way. Choosing $z$ as a covariantly constant function,
when taking derivatives over moduli of the curve, one writes for \rf{complcu}, (\ref{NONSY})
\be
\label{deltaF}
F_{\tilde z}\delta{\tilde z} + \delta F = 0
\ee
where $\delta F \equiv \sum \delta f_{ij}z^i{\bar z}^j$ is variation
of the coefficients in equations \rf{complcu}, (\ref{NONSY}).
Then, for the generating differential (\ref{diff2mm}) one gets
\be
\label{vardS}
\delta\left({\tilde z} dz\right) = - \delta F {dz\over F_{\tilde z}} =
- \sum \delta f_{ij} {\tilde z}^j {z^idz\over F_{\tilde z}}
\ee
Expression (\ref{vardS}) contains decomposition of variation of the
meromorphic differential (\ref{diff2mm}) over the basis of Abelian (meromorphic and
holomorphic) differentials on the curve \rf{complcu}, (\ref{NONSY}). It is easy to check
that the coefficients $f_{ij}$, corresponding to the meromorphic
Abelian differentials of the second kind, can be expressed through the
parameters of potential \rf{V2mm} of the two-matrix model, i.e. in terms of
coefficients $t$ and $\bar t$ of its harmonic part \cite{KM}.
This follows immediately from substitution into \rf{complcu}, (\ref{NONSY}) of the
asymptotic expansion of the following branch of the function $\tilde z$
\be
\label{branch}
{\tilde z} = W'(z) + O\left( z^{-1}\right) =
\sum_{k=1}^{n+1}kt_kz^{k-1} + O\left( z^{-1}\right)
\ee
which gives rise the formulas \rf{aa} etc. Using \rf{branch} and
\rf{diff2mm} one can immediately conclude that
\be
\label{times2mm}
t_k = {1\over 2\pi ik} \res_{z=\infty}\left( z^{-k}{\tilde z}dz\right)
\ee
The rest of the expansion \rf{vardS} consists of the coefficients,
corresponding to the linear combination of the third kind Abeian
differential and the holomorphic differentials (\ref{vij}).

The dependence of the two-matrix model planar free energy
upon the filling numbers \rf{pertmm} and parameters of the potential
\rf{V2mm}, as in the one-matrix case, can be geometrically formulated by
\be
\label{dptmm}
{\p {\cal F}\over\d S_i} = {1\over 2\pi i}\oint_{B_i} {\tilde z}dz
\ee
where $\{ B_i \}$ are the canonical dual cycles $A_i\circ
B_j=\delta_{ij}$ (see fig.~\ref{fi:riemann}) on the curve \rf{NONSY}, \rf{complcu}, and
\be
\label{dptmm2}
{\d{\cal F}\over\d t_k} =
{1\over 2\pi i}\ {\rm res}_{z=\infty} \left( z^k {\tilde z} dz\right)
\ee
The integrability of \rf{dptmm} follows from the
symmetry of the period matrix of the curve \rf{complcu} and the integrability of
\rf{dptmm2} from the Riemann bilinear relations analogous to \rf{syToda}.

\subsection{The structure of the two-matrix model complex curve
\label{ss:strcu2mm}}

To understand better the structure of the curve \rf{NONSY}, \rf{complcu}, consider
first the cubic example. Writing equation \rf{complcu} for $n=\tilde n=3$ with
some arbitrary coefficients
\be\label{cuthree}
F(z,{\tilde z})= z^2{\tilde z}^2 + az^3 + {\bar a}{\tilde z}^3 +
bz^2{\tilde z} + {\bar b}z{\tilde z}^2 + cz^2 + {\bar c}{\tilde z}^2
+ fz{\tilde z} + qz + {\bar q}{\tilde z} + h = 0
\ee
one has to make it consistent with the asymptotic \rf{branch}
\be\label{asyone}
{\tilde z} = W'(z) - G(z) = \sum_{k=1}^{3}kt_kz^{k-1} +
O\left( z^{-1}\right)
\ee
Substituting \rf{asyone} into the eq.~\rf{cuthree} and collecting the
coefficients of the terms $z^6$, $z^5$ and $z^4$ one gets
\be\label{coeff}
a =  -{1\over 3t_3},
\ \ \ \
b = {2\tilde t_2\over 3\tilde t_3},
\ \ \ \
c =  {\tilde t_1\over 3\tilde t_3} -
{2t_2\over 9t_3\tilde t_3} + {8\tilde t_2^2\over 9\tilde t_3^2}
\ee
together with their complex-conjugated counterparts, i.e. the coefficients at
higher powers of equation \rf{cuthree} are indeed completely fixed by parameters of the
potential \rf{V2mm}. Four lower coefficients $f$, $q$, ${\bar q}$
and $h$ correspond to the bipole differential \rf{bp} and three holomorphic
differentials and their values depend also on the filling fractions \rf{RHODS},
i.e. of the periods \rf{pertmm} of the differential \rf{diff2mm}\footnote{Note,
that the equations of the curve \rf{complcu}, \rf{cuthree} is often
written implying some reality condition onto the coefficients, but as
usual, the deformations of these coefficients even in this case should be considered as
independent complex variables.}. The classical "expectation values"
of these coefficients (at vanishing filling fractions) can be extracted
from eq.~\rf{CUCLA}.

\begin{figure}[tp]
\centerline{\epsfig{file=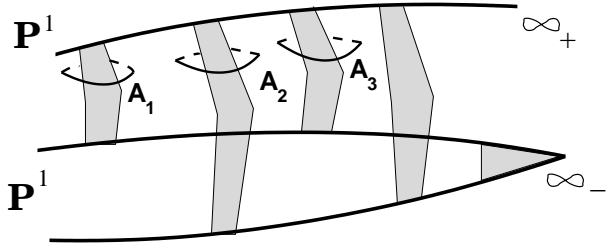,width=80mm}}
\caption{ Cubic curve as a cover of
$z$-plane. }
\label{fi:cutre}
\end{figure}
Let us now  think of the curve \rf{cuthree} as of the Riemann surface of
multi-valued function ${\tilde z}(z)$, being then
a three-sheet cover of the complex $z$-plane. On the first, physical
sheet, there are no branching at $z\to \infty$, as it follows from
the asymptotic \rf{branch}, \rf{asyone}. However, the "complex-conjugated" asymptotic
\be\label{asytwo}
z = {\overline W}'({\tilde z}) + O\left({\tilde z}^{-1}\right)
\ee
on two unphysical sheets ${\tilde z} \propto \sqrt{z}$ stays, that their two
infinities are glued, being an end-point of a cut, see fig.~\ref{fi:cutre}.

The branch points at $z$-plane are determined by zeroes of the
differential $dz$, or by $F_{\tilde z}=0$. Considering the simplest
non-degenerate case of the curve \rf{cuthree}
\be\label{CUSIMP}
z^2{\tilde z}^2 + az^3 + {\bar a}{\tilde z}^3
+  h = 0
\ee
it is easy to see that there are nine branch points in the $z$-plane
without infinity $z=\infty$
(of course, one comes to the same conclusion looking at the Cardano
formula, or from the index theorem, see below).

The structure of the curve is depicted at fig.~\ref{fi:cutre}.
The curve can be seen as two copies
of $\mathbb{P}^1$, glued by four cuts, i.e. in general position it has
genus \rf{genus} $g=3$. There are two "infinities" $z=\infty$, ${\tilde
z}=\infty$, one of them being a branch point. We have shown schematically
on fig.~\ref{fi:cutre} the possible cuts, and the corresponding choice of
the canonical $A$-cycles. In classical situation \rf{CUCLA} the degenerate
curve can be seen as two parabolas
intersecting at four points, and under the "quantum resolution" of singularities
these points turn
into four cuts connecting two spheres $\mathbb{P}^1$ as at fig.~\ref{fi:cutre}.

\begin{figure}[tp]
\centerline{\epsfig{file=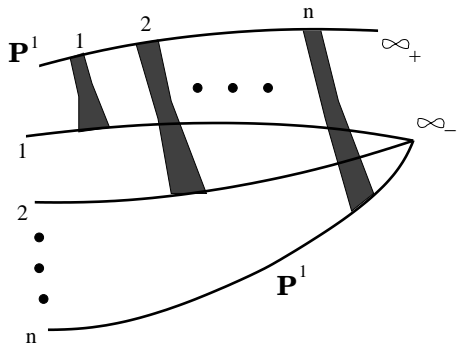,width=60mm}}
\caption{ Generic curve of the two matrix model with symmetric potential as
a cover of the $z$-plane. In contrast to fig.~\ref{fi:cutre} each fat line consists
of a stack of $n$ cuts.}
\label{fi:cun}
\end{figure}
Now it is now almost clear, how the curves \rf{NONSY} and
\rf{complcu} look for generic polynomial potentials.  Say, the curve \rf{complcu}
of degree $n$, i.e. when
$W'(z)\sim z^n + \dots$ (see fig.~\ref{fi:cun}), can be again presented
as two spheres $\mathbb{P}^1$ glued by $n$ stacks of cuts. One of these
$\mathbb{P}^1$'s corresponds to the "physical" $z$-sheet, the other one is glued at
the $\infty_-$ from $n$ copies of "unphysical" $z$-sheets. Each stack
consists of $n$ cuts, so their total number is $n^2$ among which one
can choose $n^2-1$ independent, in the sense of surrounding them cycles,
whose number is equal to the genus \rf{genus} of
this Riemann surface.

The differential $dz$ has always a pole of the second order at $\infty_+$ on
the upper, or "physical" sheet, and  a pole of the order $n+1$ at
$\infty_-$ since $\left. z\right|_{\infty_-} \propto {\tilde z}^n + \dots$. It
gives altogether $n+3$ poles, and from the Riemann-Roch theorem one concludes
that the number of branching points, or zeroes of $dz$ is equal to
\be\label{ZERDZ}
\# (dz=0) = n+3 + 2(n^2-1)-2 = 2n^2 + n -1
\ee
reproducing nine for $n=2$. In general position this gives
exactly $2n^2$ branch points, being the end-points of $n^2$ simple cuts,
and $n-1$ ramification
points, connected by cuts with $\infty_-$.

\subsection{Degenerations of the maximal genus curve}

Up to now we have considered the complex curve of two matrix model of its maximal
possible genus \rf{genus}, corresponding to the situation when all extrema
are filled in by nonvanishing fractions of eigenvalues \rf{RHODS}. However, in many cases,
as for example for the real problems to be discussed below, some part
of extrema, corresponding to unstable
configurations, can remain empty, what from the point of view of the curves \rf{NONSY},
\rf{complcu} corresponds to their degenerations. Before passing to detailed discussion
of how the real problem can be formulated in terms of the Schottky double,
let us consider, following \cite{KM}, how the
curve \rf{complcu} can be in principle degenerated.

The genus $g=n^2-1$ of the curve \rf{complcu} decreases if there exists
nontrivial solution to the following system of equations
\be\label{DEGENR}
F(z,{\tilde z}) = 0, \ \ \ \ \ dF = \d_zF dz + \d_{\tilde z}F d{\tilde z} = 0
\ee
This system imposes constraints to the coefficients of $f_{ij}$ of the
equation \rf{complcu}, which can be found, computing the resultant of
the equations \rf{DEGENR}, or the discriminant of the curve. However,
these constraints cannot be effectively resolved in general position.

To get an idea how the curve \rf{complcu} can be degenerated consider, first,
the cubic case \rf{cuthree} and let, in addition, all coefficients of this equation
be real. Then, it is easy to see that equation \rf{cuthree} can be rewritten in the form
\be\label{ELLW}
Y^2 + aX^3 + cX^2 + qX + h - {1\over 4}\left((3a-b)X + 2c-f\right)^2
\equiv Y^2+P(X) = 0
\ee
%
\begin{figure}[tp]
\centerline{\epsfig{file=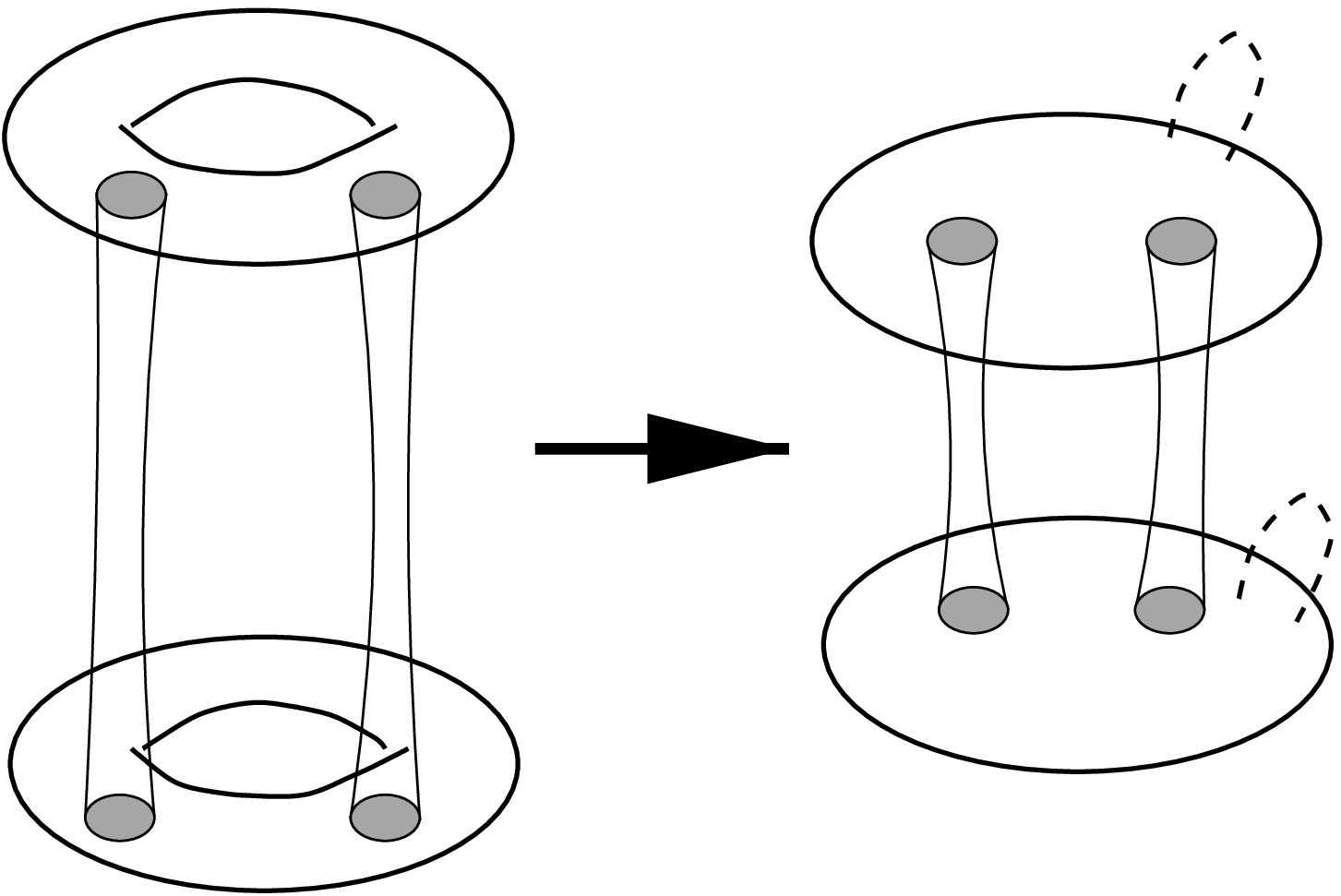,width=80mm}}
\caption{The curve \rf{cuthree} as double
cover of the torus. When the torus \rf{ELLW} degenerates, the genus
$g=n^2-1=3$ curve \rf{cuthree} falls down to $g_{\rm
red}=n-1=1$.}
\label{fi:cutrto}
\end{figure}
where\footnote{One may also "tune" for simplicity the coefficients of the potential
\rf{coeff} to get $3a=b$ and $2c=f$.}
\be\label{COVER}
X = z + {\tilde z}, \ \ \ \ \
Y = z{\tilde z} - {1\over 2}\left((3a-b)X + 2c-f\right)
\ee
The formulas \rf{COVER} show that our
curve \rf{complcu} can be presented as a double cover of the torus
\rf{ELLW} with four branch points (where the transformation
\rf{COVER} becomes singular) being solutions to equation \rf{cuthree}
under the substitution ${\tilde z}=z$. Hence, the curve \rf{cuthree} can be also presented
(in addition to the picture of fig.~\ref{fi:cutre}) as two tori glued by
two cuts (see fig.~\ref{fi:cutrto}).

Now it becomes clear, how this picture can be degenerated. Rewriting
equations \rf{DEGENR} as
\be\label{FDF}
F_{\tilde z} = zF_{Y} + P'(X)
\\
F_{\tilde z} = {\tilde z}F_{Y} + P'(X)
\ee
one immediately finds that they lead either to $z={\tilde z}$ or to
$F_Y=0$ and, hence to $P'(X)=0$. In the second case the torus \rf{ELLW}
degenerates, while $z={\tilde z}$ leads to degeneration of the cover
of this torus.
When the torus degenerates into a rational curve, one gets the degeneration of the
Riemann surface
\rf{cuthree} presented as a double cover of sphere with two cuts,
i.e. as an elliptic curve of genus $g=1$ with an extra two
pairs of the singular points (see fig.~\ref{fi:cutrto}).

\begin{figure}[tp]
\centerline{\epsfig{file=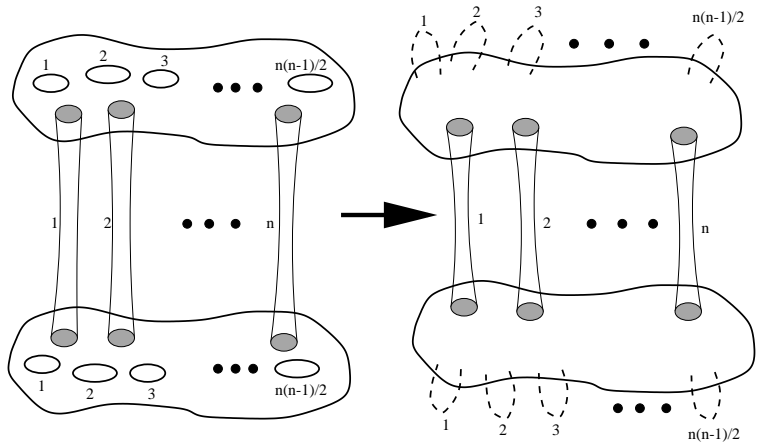,width=100mm}}
\caption{The general curve \rf{complcu} as a double
cover of the curve \rf{QMM} with a genus $g_\ast = {n(n-1)\over 2}$.
Similarly to fig.~\ref{fi:cutrto}, when the curve \rf{QMM} completely degenerates
into a rational curve, the Riemann surface of the two-matrix model
\rf{complcu} degenerates into the curve of genus $g_{\rm red}=n-1$.}
\label{fi:cucomm}
\end{figure}

Now, in the general case  \rf{complcu} with real coefficients,
the substitution analogous to \rf{COVER} brings
it to the form
\be\label{QMM}
Y^n + X^{n+1} + \dots = 0
\ee
%
\begin{figure}[tp]
\centerline{\epsfig{file= 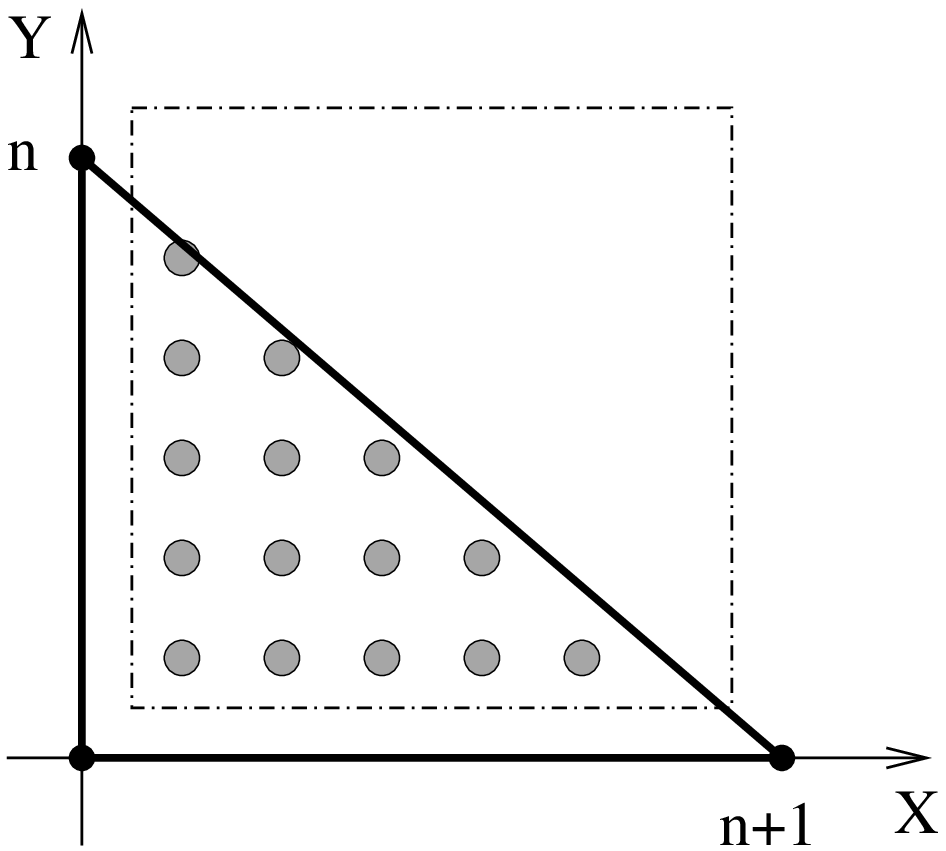,width=80mm}}
\caption{The Newton polygon for the curve
\rf{QMM} gives the genus $g_\ast = {n(n-1)\over 2}$.}
\label{fi:newtxy}
\end{figure}
where by dots we denoted monomials of lower powers in $X$ and $Y$,
and there are no "mixed" terms\footnote{The curves of this
type will be considered in the second part of this paper in the context of
matrix model solutions to minimal string theory, or the $(p,q)$ critical points
of two-dimensional gravity with $|p-q|=1$.}.
The genus of the curve \rf{QMM} can be again easily computed by the Newton polygon
(see fig.~\ref{fi:newtxy}), which gives
\be\label{GQMM}
g_\ast = {n(n-1)\over 2}
\ee
In the same way one may present the generic curve of the two matrix model
\rf{complcu} as a double cover of the Riemann surface \rf{QMM} with $2n$ branch points.
Indeed, the Riemann-Hurwitz formula
\be\label{RH}
2-2g = \#\ S\cdot(2-2g_0) - \#\ B.P.
\ee
where $\#\ S$ is number of sheets of the cover and $\#\ B.P.$ is number of branch
points, gives for $g=n^2-1$ and $g_0=g_\ast$ exactly $\#\ B.P.=2n$. It
means that the generic curve of the two matrix model \rf{complcu} can be
presented as a double cover of the curve \rf{QMM} with $n$ cuts, and when
the curve \rf{QMM} degenerates into a rational one, the genus of the curve \rf{complcu}
falls down to
\be\label{GRED}
g_{\rm red} = n-1
\ee
growing already linearly with the highest power of potential,
like in the case of one-matrix model, see fig.~\ref{fi:cucomm}.
%
Finally in this section, let us say few words about the  rational degenerations of
\rf{complcu}, i.e. when its smooth genus completely vanishes. A particular example of
such total degeneration is given by the "classical" curve
\rf{CUCLA}, but the rational case can be easily studied for the generic
values of coefficients in \rf{complcu}, i.e. without any reality
restriction.

In such situation equation of the curve \rf{complcu} can be resolved via the generalized
rational conformal map
\be\label{COMAP}
z = rw + \sum_{k=0}^n {u_k\over w^k},
\ \ \ \ \ \
\tilde z = {r\over w} + \sum_{k=0}^n \bar u_k w^k
\ee
with uniformizing parameter $w$,
and the substitution of \rf{COMAP} into \rf{complcu} gives a system of
equations, expressing all coefficients $f_{ij}$ in terms of
parameters of the conformal map \rf{COMAP}, and their explicit form can be found
in Appendix~\ref{ap:rat}.

\setcounter{equation}0
\section{Two-matrix model and the Dirichlet problem}

In this section we continue to study the partition function of the two-matrix model
\rf{mamocompl}.
It turns out, that for the real problem, when the maximal genus
\rf{genus} of the smooth curve
cannot be achieved by filling only the minima of the potential \rf{V2mm},
the most effective way to rewrite the free energy
as a quasiclassical tau-function is based on studying
the Dirichlet problem for the boundary of the eigenvalue
domain, shown at fig.~\ref{fi:drops}. For the simply-connected domains this problem
is solved in terms of the conformal maps \rf{COMAP}, but for many eigenvalue drops
or the multiply-connected domains the solution to the Dirichlet problem requires the
whole machinery introduced in sect.~\ref{ss:taufun}.

\subsection{The Dirichlet boundary problem and integrability
\label{ss:simply}}

Let us, first, reformulate some results of
\cite{MWZ,KMZ} in the form closely connected with the
quasiclassical solution of the two-matrix model. As we already mentioned above,
the eigenvalue distribution for the real problem in two-matrix model \rf{mamocompl}
acquires the form of some drops in the complex plane (see fig.~\ref{fi:drops}),
with the constant density \rf{rho2mm} for the potential \rf{V2mm}.
The density remains constant
and the quasiclassical hierarchy describes now the deformation of the shape of these
drops, which can be encoded in the solution to corresponding
Dirichlet boundary problem.

\begin{figure}[tb]
\epsfysize=4cm
\centerline{\epsfbox{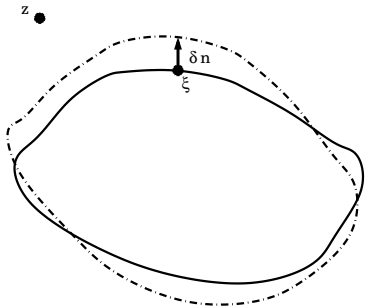}}
\caption{\sl The deformation of domain with normal displacement
${\bf n}(\xi)$.}
\label{fi:adam}
\end{figure}
We start with a single drop solution, corresponding to a
connected domain ${\sf D}$, bounded by a simple smooth curve.
Following \cite{MWZ,KMZ}, consider the exterior Dirichlet problem in
${\sf D^c}=\mathbb{C}\setminus {\sf D}$
\be\label{Dirih}
u(z)=
- \frac{1}{2\pi}\oint_{\d {\sf D}}
u_0(\xi )\p_{n} G(z,\xi ) |d\xi |
\ee
where $G(z,z')$ is the Dirichlet Green function, i.e. satisfying the boundary conditions
$\left.G(z,z')\right|_{z\in\d{\sf D}}=\left.G(z,z')\right|_{z'\in\d{\sf D}}=0$,
and assume that ${\sf D}$ contains the point $z=0$, then
${\sf D^c}$ contains the point where $z=\infty$.
Our main tool in this section
is the Hadamard variational formula, expressing
variation of the Dirichlet Green function $G(z,z')$ under small deformations
of domain ${\sf D}$ in terms of the Green function itself:
\be\label{Hadam}
\delta G(z, z')=\frac{1}{2\pi}\oint_{\d {\sf D}}
\p_{n}G(z, \xi)\p_{n}G(z',\xi)\delta n(\xi)|d\xi |.
\ee
Here $\delta n(\xi)$ is the normal displacement
at the boundary point $\xi$ (see fig.~\ref{fi:adam}), and $\d_n$ in \rf{Dirih}
and \rf{Hadam} is the corresponding normal derivative.

For a one-drop solution,
the solution to the Dirichlet problem is equivalent \cite{MWZ} to finding
a conformal map from
${\sf D^c}$ onto the complement to unit disk $|w|>1$,
or any other reference domain, where the Green function
is known explicitly. Such conformal map $w(z)$ exists due
to the Riemann mapping theorem, then
\be\label{Gconf}
G(z, z')=\log \left |
\frac{w(z)-w(z')}{w(z)\overline{w(z')} -1} \right |
\ee
where bar means complex conjugation.
An example of such map is given by
inverse rational function \rf{COMAP}, and this basically solves the problem
for a single-drop solution of matrix model \rf{mamocompl} with a polynomial potential.

Let $t_k$ the be moments of the domain ${\sf D^c}=\mathbb{C}\setminus {\sf D}$
defined w.r.t. harmonic functions $\{ {z^{-k}/ k}\}$:
\be\label{momt}
t_k= \, - \, \frac{1}{\pi k}
\int_{{\sf D^c}}z^{-k} \,d^2 z\,, \,\;\;\;\;\;k=1,2,\ldots
\ee
and $\{{\bar t}_k\}$ be the complex conjugate moments, i.e.
$ {\bar t}_k = -{1\over\pi k}\int_{{\sf D^c}}d^2 z
\bar z^{-k}$; the coincidence of notations with the harmonic parameters of
the matrix model potential
\rf{V2mm} is certainly not accidental.
The Stokes formula represents them as contour integrals, e.g.
\be\label{momtcont}
t_{k}=\frac{1}{2\pi i k}\oint_{\d {\sf D}}
z^{-k} \bar z dz = \frac{1}{2\pi i k}\oint z^{-k} \tilde z dz
\ee
providing, in particular, a regularization of possible divergencies in
(\ref{momt}) and directly relating them with the matrix model
times \rf{times2mm}, if $\tilde z(z)$ is analytic continuation of the
function $\bar z(z)$ from ${\d {\sf D}}$, see fig.~\ref{fi:cutspo}.
Besides, we denote by $t_0$ the area of domain ${\sf D}$:
\be\label{t0area}
t_0 =\frac{1}{\pi}\int_{{\sf D}} d^2z
\ee
or the total number of eigenvalues distributed in the domain with the constant
density \rf{rho2mm}.

The harmonic moments of ${\sf D^c}$ \rf{momt} are coefficients of
the Taylor expansion of the potential
\be\label{dd1}
\Phi (z,\bar z) =-\frac{2}{\pi}\int_{{\sf D}} \log |z-z'|d^2z'
\ee
induced by the domain ${\sf D}$, and directly related
to the potential \rf{V2mm} of the two-matrix model
\be\label{deftk}
V(z,\bar z) = \Phi (0)-\Phi (z,\bar z) = z\bar z - W(z) - \bar W(\bar z)
=|z|^2 -\sum_{k\geq 1}
\left (t_k z^k +\bar t_k \bar z^k \right )
\ee
The derivative of the potential, or resolvent \rf{2mamoG}
\be
\label{fG}
\d_z \Phi (z) =-\frac{1}{\pi}\int_{{\sf D}}
\frac{d^2 z'}{z-z'} = - {1\over \pi}G(z)
\ee
is continuous across the boundary and holomorphic for
$z\in {\sf D^c}$ while for $z \in {\sf D}$ the function
$\d_z \Phi +\bar z$ is holomorphic.
If the boundary is an analytic {\em real} curve,
both these functions can be analytically continued,
and there indeed exists a function $\tilde z(z)$,
analytic at least in some strip-like neighborhood
of the boundary contour and being $\tilde z(z)=\bar z$ on the contour.
In other words, the analytic continuation $\tilde z(z)$ of
the function $\bar z$ away from the boundary contour,
directly related to the generating differential \rf{diff2mm},
completely determines the shape of
the boundary and is called, in this context, the Schwarz function.

The basic fact of the theory of deformations of
closed smooth curves is that the complex moments \rf{momt}, \rf{momtcont},
supplemented by real variable \rf{t0area} form a set of good local coordinates in
the ``moduli space" of such curves. Moreover, it is not
an overcomplete set, as follows from the explicit construction
of corresponding vector fields, see \cite{KMZ} and references
therein for details. The proof of this statement is based on the observation,
that the difference of the boundary values
$\d_tC^{\pm}(\zeta)d\zeta$
of the derivative of the Cauchy
integral
\be\label{C}
C(z)dz=
{dz\over 2\pi i}\oint_{\d {\sf D}} {\bar \zeta d\zeta
\over \zeta -z}
\ee
is purely imaginary differential on the boundary of ${\sf D}$:
\be\label{C1}
\d_tC(z)dz=
{dz\over 2\pi i}\oint_{\d {\sf D}} \left({\bar \zeta_t \zeta_{\sigma}+
\bar \zeta \zeta_{t,\,\sigma}
\over \zeta -z}-{\bar \zeta \zeta_{\sigma}\zeta_t
\over (\zeta -z)^2}\right)d\sigma
={dz\over 2\pi i}\oint_{\d {\sf D}} \left({\bar \zeta_t \zeta_{\sigma}-
\bar \zeta_{\sigma} \zeta_{t}
\over \zeta -z}\right)d\sigma
\ee
where $\zeta(\sigma ;t)$ is
parametrization of the real curve $\d {\sf D}(t)$. Hence,
\be
\label{imC}
\left(\d_tC^{+}(\zeta)-\d_tC^{-}(\zeta)\right)d\zeta
=\d_t\bar \zeta d\zeta -\d_t\zeta d\bar \zeta =
2i \Im\left(\d_t\bar \zeta d\zeta \right)
\ee
is indeed purely imaginary, and if a $t$-deformation preserves all the
moments $t_k$, $k\geq 0$, the differential\\
$\d_t \bar\zeta d\zeta-\d_t\zeta d\bar\zeta$ extends to a holomorphic
differential in ${\sf D^c}$. Indeed,
if $|z|<|\zeta |$ for all $ \zeta \in \d {\sf D}$,
one can expand:
\be\label{7}
\d_tC^+(z)dz=\frac{\d}{\d t}
\left ( {dz\over 2\pi i}\sum_{k=0}^{\infty}
z^k \oint_{\d{\sf D}} \zeta^{-k-1}\bar \zeta d\zeta \right ) =
\sum_{k=1}^{\infty} k\left(\d_t t_{k}\right)z^{k-1}dz=0
\ee
and, since $C^+$ is analytic in ${\sf D}$,
we conclude that $\d_t C^+ \equiv 0$.
The expression
$\d_t \bar\zeta d\zeta-\d_t\zeta d\bar\zeta$ is the boundary value
of the differential
$- \d_t C^-(z)dz$ which has at most simple pole at the infinity
and holomorphic everywhere else in ${\sf D^c}$. The equality
\be
\d_t t_0 =\frac{1}{2\pi i}\oint_{\d{\sf D}}
(\d_t \bar\zeta d\zeta  - \d_t \zeta d\bar \zeta ) = 0
\ee
then implies that the
residue at $z=\infty$ vanishes, therefore $\d_t C^-(z)dz$ is holomorphic.
\begin{figure}[tb]
\epsfysize=4cm
\centerline{\epsfbox{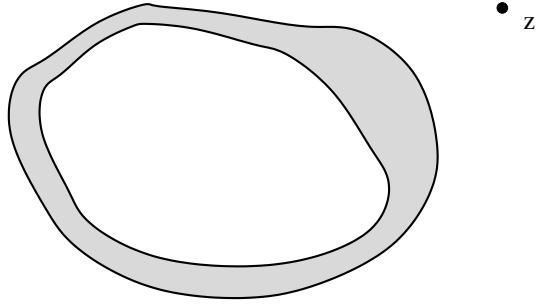}}
\caption{\sl The elementary deformation
with the base point $z$.}
\label{fi:bump}
\end{figure}
Further, any holomorphic differential, purely imaginary along the boundary of a
simply-connected
domain, must vanish in this domain due to the Schwarz symmetry principle
on the Schottky double,
obtained by attaching to ${\sf D^c}$ its
{\em complex-conjugated} copy along the boundary
(from the Schwarz symmetry principle,
$\d_tC^-dz$ extends to a
globally defined holomorphic differential on sphere $\mathbb{P}^1$, i.e.
must vanish due to the Riemann-Roch theorem).

\begin{figure}[tb]
\epsfysize=4cm
\centerline{\epsfbox{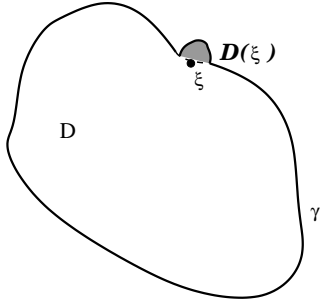}}
\caption{\sl The bump deformation at the boundary point $\xi$.}
\label{fi:bump1}
\end{figure}

For a fixed point $z\in {\sf D^c}$ one can consider
a special infinitesimal
deformation of the domain such that the normal displacement
of the boundary is proportional to the gradient of the
Green function $G(z,\xi)$ (see fig.~\ref{fi:bump})
\be\label{small}
\delta_z n(\xi)
=-\frac{\epsilon}{2}\d_n G(z, \xi)
\ee
to be called, following \cite{KMZ}, the
elementary deformations
with the base point $z$.
It is easy to see that
\be
\label{spe}
\delta_z t_0 =
\frac{1}{\pi}\oint_{\d{\sf D}}
\delta n(\xi) |d\xi|=
-\frac{\epsilon}{2\pi }\oint
\d_n G(z, \xi ) |d\xi|= \epsilon
\\
\delta_z t_k =
\frac{1}{\pi k}\oint_{\d{\sf D}} \xi^{-k}
\delta n(\xi) |d\xi|=
-\frac{\epsilon}{2\pi k}\oint \xi^{-k}
\d_n G(z, \xi ) |d\xi|= {\epsilon \over k} z^{-k}
\ee
as a direct consequence of the Dirichlet formula
\rf{Dirih}\footnote{Note that the elementary deformation with the base
point at $\infty$ keeps all moments except $t_0$ fixed.
Therefore, the deformation which changes
only $t_0$ is given
by $\delta n(\xi )=-\frac{\epsilon}{2} \d_n G(\infty , \xi )$.}.
Consider now the variation $\delta_z X$ of
any functional $X=X({\bf t})$ under the
elementary deformation with the base point $z$ in
the leading order in $\epsilon$, i.e.
\be
\label{D4}
\delta_z X =
\sum_k \frac{\d X}{\d t_k}\, \delta_z t_k =
\epsilon \nabla (z)X
\ee
where the differential operator $\nabla (z)$
is defined as
\be\label{D2}
\nabla (z)
=\d_{t_0} +\sum_{k\geq 1} \left (
\frac{z^{-k}}{k} \d_{t_k} +
\frac{\bar z^{-k}}{k}\d_{\bar t_k}\right )
\ee
Note that the elementary deformation can be intuitively understood
\cite{MWZ} as a ``bump" on the boundary, continued harmonically into ${\sf D^c}$,
see fig.~\ref{fi:bump1}.
Indeed, these two are related by
$\delta_z \propto \oint |d\xi| \d_n G(z,\xi)
\delta^{\rm bump}(\xi)$, and the ``bump" deformation
should be understood as
a (carefully taken) limit of $\delta_z$ when the point $z$ tends
to the boundary $\d{\sf D}$.

Fix now
three points $z_1,z_2,z_3\in\mathbb{C}\setminus{\sf D}$ and
compute $\delta_{z_i}G(z_j,z_k)$ by means of the
Hadamard formula (\ref{Hadam}).
Using (\ref{D4}), one can identify the result
with the action of the vector field $\nabla (z_i)$
onto the Green function:
\be\label{Th0}
\nabla (z_3)G(z_1,z_2) =
-\, \frac{1}{4\pi }\oint_{\d {\sf D}}
\d_n G(z_1 , \xi) \d_n G(z_2 , \xi)
\d_{n} G(z_3 , \xi) |d\xi|
\ee
Remarkably, the r.h.s. of (\ref{Th0})
is {\em symmetric} in all three arguments, i.e.
\be\label{symha}
\nabla  (z_1)G(z_2,z_3)
=\nabla (z_2)G(z_3,z_1)
=\nabla (z_3)G(z_1,z_2)
\ee
This is the key relation which allows to represent the deformation
of the Dirichlet problem \rf{Dirih}
as an integrable hierarchy of non-linear
differential equations \cite{MWZ}, with (\ref{symha}) being
the integrability condition of the hierarchy.
It follows from (\ref{symha}) that there exists a function $\F_{\sf D}=\F_{\sf D}({\bf t})$ of the
moments \rf{momt}, \rf{t0area} such that
\footnote{Such formula was first conjectured
by L.~Takhtajan, see \cite{Taht} for and discussion.}
\be
\label{gf}
G(z,z') = \log\left|{1\over z} - {1\over z'}\right|
+ \ha \nabla (z)\nabla (z')\F_{\sf D}
\ee
i.e. the Green function is a double $\nabla(z)$-derivative, up to a
time-independent part, which is determined from the boundary conditions at
the coinciding points $z=z'$ and when $z,z'\to\infty$.

Eq.\,(\ref{gf}) allows one to obtain a representation
of the function $\F_{\sf D}=\F_{\sf D}({\bf t})$ as a double integral over the domain ${\sf D}$.
Set $\tilde \Phi (z)=\nabla (z)\F_{\sf D}$,
this function is determined by its variation under the elementary
deformation
\be\label{dual1a}
\delta_\zeta \tilde \Phi (z) =
-2\epsilon \log \left |\zeta^{-1}-z^{-1}\right | +
2\epsilon G(\zeta,z)
\ee
which is read from eq.\,(\ref{gf}) by virtue of
(\ref{D4}).
This allows one to identify $\tilde \Phi$
with the ``modified potential of domain
$\tilde \Phi (z)=\Phi (z)-\Phi (0) +t_0 \log |z|^2$,
where $\Phi$ is given by (\ref{dd1}).
Thus, one can write
\be
\label{modpot}
\nabla (z)\F_{\sf D} =\tilde \Phi (z)=
-\frac{2}{\pi} \int_{{\sf D}} \log |z^{-1}-\zeta^{-1}| d^2 \zeta
= v_0 + 2\Re\sum_{k>0}\frac{v_k}{k}z^{-k}
\ee
The last equality should be understood as
the Taylor expansion around infinity, and
the coefficients $v_k$ are the moments of domain ${\sf D}$
(the dual to \rf{momt} harmonic moments), defined as
\be
\label{vk}
v_k= \frac{1}{\pi }\int_{{\sf D}}z^{k}\,d^2 z=\p_{t_k}\F_{\sf D} \ \ \ (k>0)\,,
\;\;\;\;\;
v_0 =-\Phi(0)=\frac{2}{\pi}\int_{{\sf D}} \log |z|d^2 z=\p_{t_0}\F_{\sf D}
\ee
i.e. the moments of the complementary domain ${\sf D}$ are
completely determined by the function $\F_{\sf D}$ of
harmonic moments of ${\sf D^c}$. Formulas \rf{vk}, after rewriting them as contour
integrals, allow to identify the
function $\F_{\sf D}\equiv\F$ with the quasiclassical tau-function corresponding
to the one-drop solution of the planar matrix model \rf{mamocompl}.

In a similar manner, one arrives at the integral representation of the
tau-function itself. Using the fact, that the elementary deformation $\delta_\xi$
or the operator $\nabla(\xi)$ applied at the boundary point $\xi\in\p {\sf D}$
(where $G(z,\xi)=0$) is the bump deformation (see fig.~\ref{fi:bump1}) or attaching
a ``small piece" to the integral over the domain ${\sf D}$, and interpreting
(\ref{modpot}) as a variation $\delta_z \F_{\sf D}$ we arrive at the
following double-integral representation of the tau-function
\be\label{F}
\F_{\sf D}=\frac{1}{2\pi}\int_{{\sf D}}
\tilde \Phi (z) d^2 z=-\frac{1}{\pi^2}\int_{{\sf D}} \! \int_{{\sf D}}
\log |z^{-1} -\zeta^{-1}| d^2 z d^2 \zeta
\ee
This is nothing, but a continuous version of the effective potential \rf{variF2},
calculated on its extremal value, and
this formula remains intact in the multi-support (or multiply-connected) case below,
though an identification with the free energy for the multi-support solution
of the two-matrix model \rf{mamocompl} requires slightly more care.

\subsection{The multiply-connected case and the Schottky double
\label{ss:taumu}}

\begin{figure}[tb]
\epsfysize=7.5cm
\centerline{\epsfbox{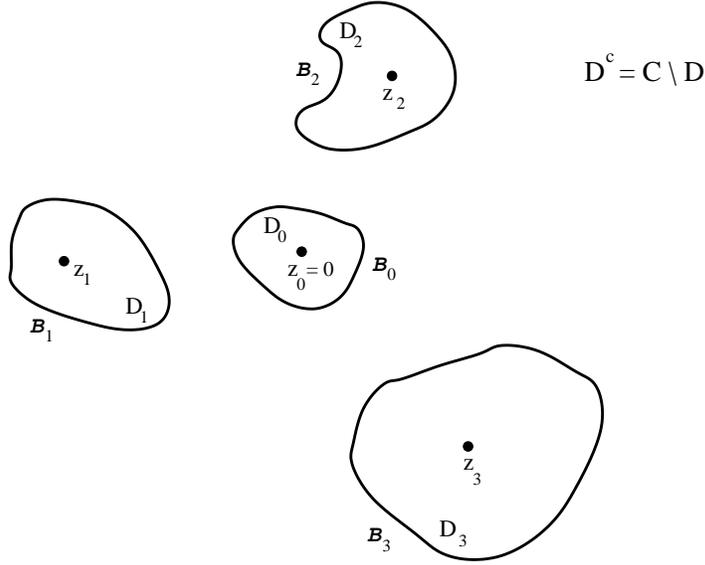}}
\caption{\sl A multiply-connected domain
${\sf D^c}=\mathbb{C}\setminus {\sf D}$ for $g=3$.
The domain ${\sf D} = \bigcup_{\alpha=0}^3 {\sf D}_\alpha$
consists of $g+1=4$ disconnected
parts ${\sf D}_\alpha$ with the boundaries $\d{\sf D}_\alpha$.
To define the complete set of harmonic
moments, we also need the auxiliary points
$z_\alpha\in {\sf D}_\alpha$ which should be
always located inside the corresponding domains.}
\label{fi:multid}
\end{figure}

Now we can turn to generic multi-support solution.
Let ${\sf D}_{\alpha}$, $\alpha =0, 1, \ldots , g$, be
a collection of $g+1$ non-intersecting bounded
connected domains in the complex plane
with smooth boundaries $\d {\sf D}_{\alpha}$ so that
${\sf D}= \cup _{\alpha =0}^{g} {\sf D}_{\alpha}$,
and the complement ${\sf D^c} = \mathbb{C}\setminus {\sf D}$ becomes
a multiply-connected domain in the complex plane
(see fig.~\ref{fi:multid}).
Let $B_{\alpha}$ be the homology classes of the boundary curves $\d {\sf D}_{\alpha}$,
assumed to be positively oriented as boundaries of ${\sf D^c}$, so that
$\cup _{\alpha =0}^{g} B_{\alpha} \simeq -\d {\sf D}$, or each
$B_{\alpha}$ has the clockwise orientation.

Comparing to the simply-connected case, nothing
is changed in posing the standard Dirichlet problem.
The definition of the Green function
and the formula (\ref{Dirih}) for the solution
of the Dirichlet problem through the Green
function remain intact.
A difference is, however, in the set of
harmonic functions: any harmonic function
is still the real part of an analytic function
but in the multiply-connected case
these analytic functions are not necessarily single-valued
(only their real parts have to be single-valued).
In other words, the harmonic
functions may have non-zero ``periods" over non-trivial
cycles\footnote{By ``periods"
of a harmonic function $f$ we mean the integrals
$\oint \d_n f \, dl$ over the non-trivial cycles.} - here over
the boundary contours or $B_{\alpha}$.
In general, the Green function has non-zero ``periods"
over all boundary contours, hence in the multiply-connected case
it is natural to introduce new objects, related to the periods
of the Green function.

First, the {\em harmonic measure}
$\varpi_{\alpha}(z)$ of the boundary component $\d{\sf D}_{\alpha}$
is the harmonic function in ${\sf D^c}$
such that it is equal to unity on $\d{\sf D}_{\alpha}$
and vanishes on the other boundary curves.
In other words, the harmonic measure solves
the particular Dirichlet problem, which from the formula (\ref{Dirih})
looks as
\be
\label{periodG}
\varpi_{\alpha}(z)=
\frac{1}{2\pi}\oint_{\d{\sf D}_{\alpha}}
\d_n G(z, \zeta )|d\zeta |,
\ \ \ \ \ \ \alpha=1,\ldots,g
\ee
so the harmonic measure is the period of the
Green function w.r.t. one of its arguments. From the maximum
principle for harmonic functions it
follows that $0< \varpi_{\alpha}(z)<1$ in internal points, and
moreover, it is obvious, that $\sum_{\alpha =0}^{g}\varpi_{\alpha}(z)=1$.
In what follows we consider the
linear independent functions
$\varpi_{\alpha} (z)$ with $\alpha =1, \ldots , g$.

Further, taking ``periods", we define
\be\label{periodG2}
T_{\alpha \beta} = \frac{i}{2}
\oint_{\d{\sf D}_{\beta}}\d_n \varpi_{\alpha} (\zeta )|d\zeta |,
\ \ \ \ \ \ \alpha , \beta =1,\ldots,g
\ee
which is a symmetric,
non-degenerate and positively-definite (imaginary) matrix.
It will be clear below that
the matrix \rf{periodG2}
can be identified with the matrix of periods \rf{pemat} of holomorphic differentials
on the Schottky double of the domain ${\sf D^c}$.

For the harmonic measure and the period matrix
there are variational formulas similar to the
Hadamard formula (\ref{Hadam}). They can be derived
either by a direct variation of (\ref{periodG}) and
(\ref{periodG2}) using the Hadamard formula or
by a ``pictorial'' argument (see fig.~\ref{fi:bump},~\ref{fi:bump1}),
and the formulas themselves look as
\be\label{varomega}
\delta \varpi_{\alpha}(z)=
\frac{1}{2\pi}\oint_{\d {\sf D}}
\p_n G(z, \xi)\, \p_n \varpi_{\alpha}(\xi)\,
\delta n(\xi) \, |d\xi |
\ee
\be\label{varperiod}
\delta T_{\alpha \beta}=
\frac{i}{2}\oint_{\d {\sf D}}
\p_n \varpi_{\alpha}(\xi)\, \p_n \varpi_{\beta}(\xi)
\, \delta n(\xi) \, |d\xi |
\ee
With a planar multiply-connected domain one can associate its
Schottky double -- a compact Riemann surface $\Sigma$,
endowed with antiholomorpic involution, the boundary
of the initial domain being the set of the fixed points
of the involution.
The Schottky double of the domain ${\sf D^c}$ can be
thought of as two copies of ${\sf D^c}$ (``upper" and ``lower" sheets
of the double) glued along their boundaries
$\d {\sf D^c}$, with two infinities added
($\infty$ and $\bar \infty$).
In this set-up the holomorphic coordinate on
the upper sheet is $z$ inherited from ${\sf D^c}$,
while the holomorphic
coordinate\footnote{More precisely,
the proper
coordinates should be $1/z$ and $1/\bar z$, which have
first order zeros instead of poles at $z=\infty$ (and
$\bar z=\bar\infty$).}
on the other sheet is $\bar z$.
The Schottky double of the multiply-connected domain
${\sf D^c}$ is a Riemann surface $\Sigma$ of genus
$g = \#\{ {\sf D}_\alpha\}-1$.
A meromorphic function on the double is a pair
of meromorphic functions $f,\tilde f$ on ${\sf D^c}$
such that $f(z)=\tilde f (\bar z)$ on the boundary $\d{\sf D^c}$,
similarly, a meromorphic differential on the double
is a pair of meromorphic differentials
$f(z)dz$ and $\tilde f (\bar z)d\bar z$ such that
$f(z)dz =\tilde f(\bar z)d\bar z$ along the boundary
curves.

\begin{figure}[tb]
\epsfysize=7.5cm
\centerline{\epsfbox{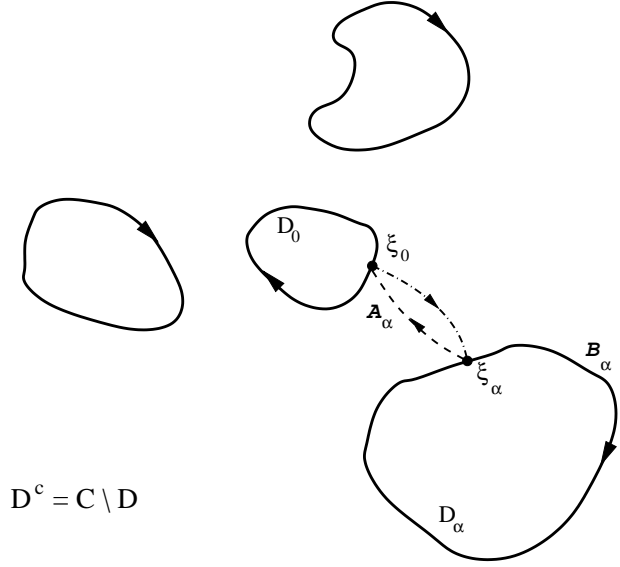}}
\caption{\sl The domain ${\sf D^c}$ with
the $A_\alpha$-cycle, going one way along the ``upper
sheet" and back along the ``lower sheet" of the
Schottky double of ${\sf D^c}$.
For such choice one clearly gets the intersection form
$A_\alpha\circ B_\beta = \delta_{\alpha\beta}$ for
$\alpha,\beta=1,\dots,g$.}
\label{fi:Dcut}
\end{figure}

We fix on the double a canonical basis
of cycles: the $B_\alpha$-cycles
to be homologically equivalent to the boundaries of the
holes $-\d{\sf D}_{\alpha}$ for $\alpha =1, \ldots , g$, and
the $A_{\alpha}$-cycle connects $\alpha$-th hole
with the distinguished 0-th one. To be more precise, when computing integrals,
fix points $\xi_{\alpha}$ on the boundaries, then
the $A_{\alpha}$-cycle starts from $\xi_{0}$,
goes to $\xi_{\alpha}$ on the ``upper'' (holomorphic)
sheet of the double and goes back
the same way on the ``lower'' sheet, where
the holomorphic coordinate is
$\bar z$, see fig.~\ref{fi:Dcut}.
Harmonic $\varpi_{\alpha}$ can be represented
as real parts
$\varpi_{\alpha}(z)=\omega_{\alpha}(z)+\overline{\omega_{\alpha}(z)}$
of the holomorphic multivalued functions in ${\sf D^c}$. The differentials
$d\omega_{\alpha}$
are single-valued holomorphic in ${\sf D^c}$ and purely imaginary
on all boundary contours, so they
can be extended holomorphically to the lower sheet
as $-d\overline{\omega_{\alpha}(z)}$.
In fact this is the canonically normalized basis
of holomorphic differentials \rf{normA} on the double $\Sigma$ with chosen canonic basis
of the cycles, since
\be
\label{omdoub}
\oint_{A_{\alpha}}\! d\omega_{\beta} =
\int_{\xi_0}^{\xi_{\alpha}}
d\omega_{\beta}(z)+
\int_{\xi_{\alpha}}^{\xi_0}
\left ( - d\overline{\omega_{\beta}(z)}\right )=
2 {\rm Re} \int_{\xi_0}^{\xi_{\alpha}}
d\omega_{\beta}(z)=
\varpi_{\beta}(\xi_{\alpha})\!-\!
\varpi_{\beta}(\xi_0) =\delta_{\alpha \beta}
\\
\oint_{B_{\alpha}}d\omega_{\beta} =
\frac{i}{2}\oint_{\d{\sf D}_{\alpha}}\p_n \varpi_{\beta}(\zeta )|d\zeta |
= T_{\alpha \beta}
\ee
where the last equality gives the period matrix of the
Schottky double (\ref{periodG2}).

One can still use harmonic moments to characterize
the shape of a multiply-connected domain.
However, the set of harmonic
functions should be now extended by adding
functions
with poles in any hole (not only in ${\sf D}_0$
as before) together with the functions,
whose holomorphic parts are not
single-valued. To specify this set,
let us mark
points $z_{\alpha}\in {\sf D}_{\alpha}$,
one in each hole (see fig.~\ref{fi:multid},
without loss of generality, it is convenient to put
$z_0 =0$), and consider single-valued
analytic functions in ${\sf D^c}$
of the form
$(z-z_{\alpha})^{-k}$ and harmonic functions
$\log \left |1-{z_{\alpha}\over z}\right |^2$ with the
multi-valued holomorphic part.
The arguments almost identical to the ones used in the simply-connected
case show \cite{KMZ}, that the parameters
\be\label{n1}
M_{n,\, \alpha}=
- {1\over \pi}\int_{{\sf D^c}} (z-z_{\alpha})^{-n}d^2 z,
\;\;\;\;\; \alpha =0,1, \ldots , g, \;\; n\geq 1
\ee
together with their complex conjugate,
\be\label{n0}
\phi_{\alpha}=-{1\over \pi }\int_{{\sf D^c}}
\log \left | 1-\frac{z_{\alpha}}{z}\right |^2
d^2z = \Phi (0) - \Phi (z_{\alpha})-|z_{\alpha}|^2,
\;\;\;\;\;
\alpha =1, \ldots , g
\ee
and $t_0 \propto\mbox{Area}({\sf D})$
still given by (\ref{t0area}),
uniquely define ${\sf D^c}$, i.e. any deformation
preserving these parameters is trivial.
A crucial step is the change of
variables from $M_{n,\alpha}$ to the variables $\tau_k$ \cite{KMZ}, which
are {\em finite} linear combinations of the $M_{n,\alpha}$'s or the moments
\be\label{t1}
\tau_0=t_0,\ \ \
{\tau}_{k}={1\over 2\pi i }
\oint_{\p {\sf D}} {\sf a}_{k}(z)\bar zdz =
-{1\over\pi}\int_{{\sf D}^c}d^2z\ {\sf a}_k(z),\ \ \ k> 0
\ee
with respect to a kind of Krichever-Novikov basis of functions
(the indices $\alpha$ and $\beta$ are understood modulo $g+1$):
\be\label{a}
{\sf a}_{m(g+1)+\alpha}(z)=
\prod_{\beta=0}^g (z-z_{\beta})^{-m}\prod_{\beta=0}^{\alpha-1}
(z-z_{\beta})^{-1}
\\
{\sf a}_k(z)\ \stackreb{z\to\infty}{=}\ z^{-k}+O(z^{-k-1})
\ee
Any analytic function in ${\sf D^c}$ vanishing at infinity
can be represented as a linear combination
of ${\sf a}_k$ which is convergent in domains such that
$|\prod_{\beta=0}^g (z-z_{\beta})| > {\rm const}$\footnote{In the
case of a single hole $g=0$ formulas (\ref{a})
give the basis used in the previous section ${\sf a}_{k}=z^{-k}$;
note also, that always ${\sf a}_0 =1$, ${\sf a}_1 =1/z$, therefore $\tau_0 = t_0$ and
$\tau_1 = M_{1,0}= t_1$.}.

The parameters
$\tau_k$, $\phi_{\alpha}$ can be treated as local coordinates
in the space of multiply-connected domains, analogously to the simply-connected
case, the details of the proof can be found in \cite{KMZ}, see also Appendix~\ref{app:co}.
Instead of $\phi_{\alpha}$ one can use
the already known fractions of eigenvalues, or the areas of the holes
\be\label{dual4}
S_{\alpha}
=\frac{\mbox{Area}({\sf D}_{\alpha})}{\pi}
=\frac{1}{\pi}\int_{{\sf D}_{\alpha}}d^2z =
\frac{1}{2\pi i}\oint_{\p{\sf D}_{\alpha}}
\bar z dz=
\frac{1}{2\pi i}\oint_{B_{\alpha}}\tilde z dz\,,
\;\;\;\;\; \alpha =1, \ldots , g
\ee
where we have again introduced the Schwarz function
$\tilde z(z)\stackreb{z\in\d{\sf D}}{=}\bar z$
in some strip-like neighborhoods of all boundaries.
Note, that in the notations of this section the $S$-variables are
expressed naturally
through the $B$-periods of the generating differential, in contrast to the  more
common choice
\rf{pertmm} used before. This is related to the fact, for the Schottky
double it is convenient to choose the canonical differentials \rf{omdoub} related to
harmonic measure and normalized to the $A$-cycles, as their are chosen on
fig.~\ref{fi:Dcut}. For these different choices the tau-functions are related by the
duality transformation, already mentioned above and to be discussed in detail in
sect.~\ref{ss:dual}.

At the same time, the variables $\phi_{\alpha}$ do not have any geometric sense
on the Schottky double,
instead, one can introduce
the variables $\Pi_{\alpha}$ (again the coincidence of notations
with the previously introduced Lagrangian multipliers is not accidental),
which were referred to in \cite{KMZ} as {\em virtual} $A$-periods
of the differential $\tilde z(z)dz$ on the Schottky
double, i.e. coinciding with these $A$-periods always, when the latter can be
rigorously defined.
Following \cite{KMZ}, consider the basis of differentials
$d{\sf b}_k$, satisfying the ``orthonormality" relations\footnote{Some
properties of the functions \rf{a} and their dual differentials \rf{o} are
collected in Appendix~\ref{app:co}.}
\be\label{n3}
\frac{1}{2\pi i}
\oint_{\p {\sf D}}{\sf a}_kd{\sf b}_{k'}=\delta_{k,k'}
\ee
for all integer $k,k'\in\mathbb{Z}$. Explicitly they are given by:
\be\label{o}
d{\sf b}_{m(g+1)+\alpha}={dz \over z-z_g}\prod_{\beta=0}^g (z-z_{\beta})^{m}
\prod_{\beta=0}^{\alpha-1} (z-z_{\beta-1})
\ee
where we identify $z_{-1}\equiv z_g$.
Now one can introduce the meromorphic differential on $\Sigma$ with the
only pole at $\infty$ on the upper sheet, where it has the form
\be\label{n25a}
d\tilde\Omega_k(z)=d{\sf b}_{k}(z)+O(z^{-2})dz
\ee
and vanishing $A$-periods
\be\label{n26a}
\oint_{A_{\alpha}}d\tilde\Omega_k=0
\ee
i.e. it is a canonically normalized meromorphic differential with a singular part
\rf{n25a}.
The normal displacements of the boundary given by real and imaginary parts of the
normal derivative $\p_n \Omega_k$ define a complex tangent vector field (the
partial derivatives at constant $\Pi_\alpha$)
\be\label{n33}
\p_{\tau_k}^{\Pi}=\p_{\tau_k}^{\phi}-
\sum_{\alpha}{\sf b}_k(z_{\alpha})\p^{\phi}_{\alpha}
\ee
to the space of multiply-connected domains,
where
\be\label{n32a}
{\sf b}_k(z)=\int_0^zd{\sf b}_{k}
\ee
is a polynomial of degree $k$.
These vector fields keep fixed the formal variable
\be\label{n34}
\Pi_{\alpha}=\phi_{\alpha}+2{\rm Re}\sum_{k}{\sf b}_k(z_{\alpha})\tau_k
\ee
The sum in the r.h.s. generally does not converge, but
in the case when the Schwarz function has
a meromorphic extension to the double $\Sigma$, the sum converges
and the corresponding quantity coincides with the
$A_{\alpha}$-period of the extension of the differential $\tilde z dz$,
see formula \rf{period33} below.

As in the simply-connected case, we introduce the elementary deformations
\be
\label{dispmu}
\delta_z
\;\;\;\mbox{with}\;\;\;\;
\delta n (\xi )=-\, \frac{\epsilon}{2}
\p_n G(z, \xi )\,,
\;\;\;\; z \, \in \, {\sf D^c}
\\
\delta^{(\alpha )}
\;\;\mbox{with}\;\;\;\;
\delta n (\xi )=-\, \frac{\epsilon}{2}
\p_n \varpi_{\alpha}(\xi)\,,
\;\;\;\; \alpha =1, \ldots , g
\ee
where $\varpi_{\alpha}(z)$ is the harmonic measure of the boundary
component $\d{\sf D}_{\alpha}$ (see (\ref{periodG})).
Then the variations of the local
coordinates under elementary deformations are:
\be\label{elem1}
\delta_z \tau_k =\epsilon {\sf a}_k(z),
\;\;\;\;
\delta_z \phi_{\alpha}=
\epsilon \log \left |1\! -\! \frac{z_{\alpha}}{z}\right |^2,
\;\;\;\;
\delta^{(\alpha )}\tau_k =0,
\;\;\;\;
\delta^{(\alpha )}\phi_{\beta} =-2\epsilon
\delta_{\alpha \beta}
\ee
since
\be
\delta_z \int_{{\sf D^c}}  f(\zeta )d^2 \zeta
=\frac{\epsilon}{2} \oint_{\p {\sf D^c}} f(\zeta )
\p_n G(z, \zeta )|d\zeta | =-\epsilon \pi f(z)
\ee
for any harmonic function $f$ in ${\sf D^c}$, and
\be
\delta^{(\alpha )}
\int_{{\sf D^c}}  f(\zeta )d^2 \zeta
=\frac{\epsilon}{2} \oint_{\p {\sf D^c}} f
\p_n \varpi_{\alpha} \, |d\zeta |=
-\frac{\epsilon}{2} \oint_{\p {\sf D}_{\alpha}}
\p_n f \, |d\zeta |=-i\epsilon
\oint_{\p {\sf D}_{\alpha}}
\p_{\zeta}f \, d\zeta
\ee
leading to (\ref{elem1}) for particular choices of $f$.
Variations of the variables $\Pi_{\alpha}$
(in the case when they are well-defined) then read
\be\label{elem1aaa}
\delta_z \Pi_{\alpha}=0,
\;\;\;\;
\delta^{(\alpha )}\Pi_{\beta} =-2\epsilon
\delta_{\alpha \beta}
\ee
and for any functional $X$
on the space of the multiply-connected domains
the following equations hold
\be\label{m6}
\delta_z X=\epsilon\nabla(z)X
\\
\delta^{(\alpha)}X =-2\epsilon \p_{\alpha}^{\phi}X=-2\epsilon
\p_{\alpha}^{\Pi}X
\ee
The differential operator $\nabla(z)$ is defined now
by the formula
\be\label{D2m}
\nabla (z)
=\p_{\tau_0}^{\Pi} +\sum_{k\geq 1}
\left( {\sf a}_k(z)\p_{\tau_k}^{\Pi} +
\overline{{\sf a}_k(z)}\p_{\bar \tau_k}^{\Pi}\right )
\ee
The functional $X$ can be regarded as a function
$X=X^\phi(\phi_{\alpha},\tau_k)$
on the space with the local coordinates
$\phi_{\alpha},\tau_k$, or as a function $X=X^\Pi(\Pi_{\alpha},\tau_k)$
on the space with the local coordinates $\Pi_{\alpha},\tau_k$
(we stress again, that although the variables $\Pi_{\alpha}$ are
generally formal, their variations under elementary deformations
and the vector-fields $\p_{\tau_k}^{\Pi}$ are always well-defined).

Let $K(z, \zeta)d\zeta$
be a unique meromorphic Abelian differential of the third kind \rf{3kA}, \rf{bp}
on the Schottky double $\Sigma$ with the simple poles at $z$ and $\infty$
on the upper sheet with residues $\pm 1$.
Then
\be\label{m5}
2\p_{\zeta}G(z,\zeta)d\zeta - 2\p_{\zeta}G(\infty,\zeta)d\zeta
=K(z,\zeta)d\zeta +K(\bar z,\zeta)d\zeta
\ee
and the differential $d\tilde\Omega_k(\zeta)$ \rf{n25a}, \rf{n26a} can be
represented in the form
\be\label{m4}
d\tilde\Omega_k(\zeta)={d\zeta \over 2\pi i}
\oint_{\infty} K(u,\zeta) d{\sf b}_k(u)
\ee
where the $u$-integration goes along a big circle
around infinity.
Using summation formulas (\ref{n4}) (see Appendix~\ref{app:co}) we obtain that
\be\label{m41}
-\sum_{k\geq 1} {\sf a}_k(z)d\tilde\Omega_k(\zeta)=
{d\zeta \over 2\pi i}\oint_{\infty}
{K(u,\zeta)du\over u-z}=K(z,\zeta)d\zeta
\ee
Therefore, for the Green function one gets an expansion
\be\label{expder}
2\p_{\zeta}G(z, \zeta )d\zeta =d\Omega_0 (\zeta)-
\sum_{k\geq 1}\left (
{\sf a}_k (z)d\tilde\Omega_k (\zeta)+c.c.\right )
\ee
where the complex conjugated $\overline{d\tilde \Omega_k}$ is a unique meromorphic
differential on $\Sigma$ with the only pole
at infinity $\bar\infty$ on the lower sheet with the principal
part $-d\overline{{\sf b}_k(z)}$ and
vanishing $A$-periods.

\subsection{Quasiclassical tau-function of the multi-support domain}

Applying the formulas (\ref{Hadam}),
(\ref{varomega}), (\ref{varperiod}), one
finds the variations of the Green function, harmonic measure
and period matrix under the elementary deformations:
\be\label{elem3}
\delta_{z_1} G({z_2},{z_3}) =\delta_{z_2} G({z_3},{z_1}) =\delta_{z_3} G({z_1},{z_2})
\\
\delta_{z_1} \varpi_{\alpha}({z_2})=\delta^{(\alpha )}G({z_1},{z_2})
=\delta_{z_2} \varpi_{\alpha}({z_1})
\\
\delta^{(\alpha)}\varpi_{\beta}(z)=
\delta^{(\beta)}\varpi_{\alpha}(z)
\\
\delta_z T_{\alpha \beta}=
i\pi\delta^{(\alpha)}\varpi_{\beta}(z)
\\
\delta^{(\alpha)}T_{\beta \gamma}=
\delta^{(\beta)}T_{\gamma \alpha}=
\delta^{(\gamma)}T_{\alpha \beta}
\ee
From (\ref{m6})
it follows that formulas (\ref{elem3}) can be rewritten in terms
of the differential operators $\nabla (z)$ and
$\p_{\alpha}=\p / \p\phi_{\alpha}=\p/\p\Pi_{\alpha}$:
\be\label{elem4}
\nabla ({z_1}) G({z_2},{z_3}) =\nabla ({z_2}) G({z_3},{z_1}) =\nabla ({z_3}) G({z_1},{z_2})
\\
\nabla ({z_1})\varpi_{\alpha}({z_2})=-2\p_{\alpha} G({z_1},{z_2})
\\
\p_{\alpha}\varpi_{\beta}(z)=
\p_{\beta} \varpi_{\alpha}(z)
\\
\nabla (z) T_{\alpha \beta}=
-2\pi i\p_{\alpha} \varpi_{\beta}(z)
\\
\p_{\alpha}T_{\beta \gamma}=
\p_{\beta} T_{\gamma \alpha}=
\p_{\gamma} T_{\alpha \beta}
\ee
These integrability relations generalize
formulas (\ref{symha}) to the multiply-connected case;
the first line just coincides with (\ref{symha}),
while the other ones extend the symmetricity of the
derivatives
to the harmonic measure and the period matrix.

Again, (\ref{elem4}) can be regarded
as a set of compatibility
conditions of an infinite hierarchy of
differential equations.
They imply that there exists a function
$\F_{\sf D} = \F_{\sf D}(\Pi_{\alpha}, \mathbf{\tau})$ such that
\be\label{elem5}
G({z_1},{z_2})=\log \left |{z_1}^{-1}-{z_2}^{-1}\right |+
\frac{1}{2}\nabla ({z_1})\nabla ({z_2})\F_{\sf D}
\\
\varpi_{\alpha}(z)=-\, \p_{\alpha}\,\nabla (z)\F_{\sf D}
\\
T_{\alpha \beta}
=2\pi i \,
\p_{\alpha}\p_{\beta} \F_{\sf D}
\ee
The function $\F_{\sf D}$ is the (logarithm of the)
tau-function of multiply-connected domains and will be related
below in sect.~\ref{ss:dual} with the free energy of multi-support solutions
to the matrix model \rf{mamocompl} after duality transformation.

Set again $\tilde \Phi (z)=\nabla (z)\F_{\sf D}$,
equations (\ref{elem5}) then determine $\tilde \Phi (z)$
for $z\in {\sf D^c}$ via its variations under the elementary
deformations
\be\label{dual1}
\delta_\zeta \tilde \Phi (z) =-2\epsilon
\log \left |\zeta^{-1}-z^{-1}\right | +2\epsilon G(\zeta,z)
\\
\delta^{(\alpha)}\tilde \Phi (z)=
2\epsilon \varpi_{\alpha}(z)
\ee
Indeed, using (\ref{dispmu}), for the variation of
\be\label{modpot1}
\tilde \Phi (z)=-\frac{2}{\pi}\int_{{\sf D}}
\log |z^{-1}-\zeta^{-1}|d^2 \zeta =
\Phi (z)-\Phi (0)+\tau_0 \log |z|^2
\ee
just coinciding with (\ref{modpot}),
if ${\sf D}$ is understood as union of all ${\sf D}_{\alpha}$'s,
one gets
\be
\delta_\zeta \left(-\frac{2}{\pi}\int_{{\sf D}}
\log |z^{-1}-z'^{-1}|d^2 z'\right) = {\epsilon\over\pi}
\oint_{\p{\sf D^c}} |d\xi|\p_n G(\zeta,\xi) \log |z^{-1}-\xi^{-1}| =
\\
={\epsilon\over\pi}
\oint_{\p{\sf D^c}} |d\xi|\p_n G(\zeta,\xi)\left(\log |z^{-1}-\xi^{-1}|
-G(z,\xi)\right) = -2\epsilon
\log \left |\zeta^{-1}-z^{-1}\right | +2\epsilon G(\zeta,z)
\ee
Similarly, for $z \in {\sf D^c}$ one obtains:
\be
\delta^{(\alpha)} \left(-\frac{2}{\pi}\int_{{\sf D}}
\log |z^{-1}-\zeta^{-1}|d^2 \zeta\right) = {\epsilon\over\pi}
\oint_{\p{\sf D^c}} |d\xi|\p_n
\varpi_\alpha(\xi) \log |z^{-1}-\xi^{-1}| =
\\
={\epsilon\over\pi}
\oint_{\p{\sf D^c}} |d\xi|\p_n \varpi_\alpha(\xi)
\left(\log |z^{-1}-\xi^{-1}| -G(z,\xi)\right) =
\\
=-{\epsilon\over\pi}
\oint_{\d{\sf D}_\alpha} |d\xi|\p_n \left(\log |z^{-1}-\xi^{-1}|
-G(z,\xi)\right) =
2\epsilon \varpi_\alpha(z)
\ee
The same calculation for $z\in {\sf D}$ yields
\be\label{zind}
\delta^{(\alpha )}\tilde \Phi (z)=
\left \{
\begin{array}{c}
0  \;\;\mbox{if $z\in {\sf D}_0$}
\\
2\epsilon \delta_{\alpha \beta}
\;\;\mbox{if $z\in {\sf D}_{\beta}$}\, ,
\;\;\beta =1, \ldots , g
\end{array}\right.
\ee
The coefficients of an expansion of $\tilde \Phi$ at infinity
define the dual moments $\nu_k$:
\be\label{n50}
\nabla (z)\F_{\sf D} =\tilde \Phi (z)=
-\frac{2}{\pi} \int_{{\sf D}} \log |z^{-1}-\zeta^{-1}| d^2 \zeta
= v_0 + 2\Re\sum_{k>0}\nu_k {\sf a}_k(z)
\ee
which are moments of the
union of the interior domains with respect to the
dual basis
\be\label{n51}
\nu_k={1\over \pi}\int_{{\sf D}} {\sf b}_k(z)d^2z
\ee
From (\ref{n50}) it follows that
\be\label{n52}
\nu_k=\p_{\tau_k}^\Pi \F_{\sf D}
\ee
and the same arguments show that the derivatives
\be\label{dual2}
S_{\alpha}=-\, \p_{\alpha} \F_{\sf D}
\ee
are just areas of the holes (\ref{dual4}). Indeed,
\be\label{dual5}
\delta^{(\alpha)}\F_{\sf D} = \frac{1}{2\pi}
\delta^{(\alpha)} \left(\int_{{\sf D}}
\tilde\Phi(z) d^2z\right) = -{\epsilon\over 4\pi}
\oint_{\p{\sf D^c}} |d\xi|\p_n \varpi_\alpha(\xi)\tilde\Phi(\xi)
+\frac{1}{2\pi} \int_{{\sf D}} \delta^{(\alpha )}\tilde
\Phi (\zeta ) \, d^2 \zeta = \\
\stackreb{(\ref{zind})}{=}\
{\epsilon\over 4\pi}
\oint_{\d{\sf D}_\alpha} |d\xi|\p_n\tilde\Phi(\xi)
+\frac{\epsilon}{\pi}\int_{{\sf D}_{\alpha}}d^2 \zeta
=-\frac{\epsilon}{4\pi}\int_{{\sf D}_{\alpha}}\Delta
\tilde \Phi \, d^2 \zeta +\epsilon S_{\alpha}=2\epsilon S_{\alpha}
\ee
The integral representation of $\F_{\sf D}$ is found
similarly through its variations, read from (\ref{n50}), and the result is given
by the same formula (\ref{F}) as in the
simply-connected case provided by
${\sf D}=\cup_{\alpha=0}^g{\sf D}_{\alpha}$ is now
the union of all ${\sf D}_{\alpha}$'s.

\subsection{Polynomial potentials and algebraic domains}

The domains are called algebraic
if the Schwarz function $\tilde z(z)$ has a meromorphic
extension to a Riemann surface $\Sigma$ with antiholomorphic
involution,
then $\Sigma$ can be naturally
divided in two ``halves'' (say upper and lower sheets)
exchanged by this involution.
The domain ${\sf D^c}$ is algebraic if and only if
the Cauchy integrals
\be
C^{\alpha}(z)=\frac{1}{2\pi i}
\oint_{\p {\sf D}}\frac{\bar \zeta d\zeta}{\zeta -z}
\;\;\;\mbox{for $z\in {\sf D}_{\alpha}$}
\ee
are extendable to a rational meromorphic function $\CS(z)$ (the same for all
$\alpha$!) on the whole complex plane $\mathbb{C}$ with a marked point at
infinity. Equality $\tilde z(z)=\CS(z)-C^{-}(z)$,
valid by definition for $z\in \partial {\sf D^c}$, can be used for
analytic extension of the Schwarz function.
Since $C^{-}(z)$ is analytic in
${\sf D^c}$, $\CS(z)$ and $\tilde z(z)$ have the same singular parts
at their poles in ${\sf D^c}$.
One may treat $\tilde z(z)$ as a function on the
Schottky double extending it to the lower sheet as
$\bar z$.

It is also convenient to introduce the Abelian integral
\be\label{alg2}
{\sf b}(z)=\int_{0}^{z}\CS(z)dz
\ee
which is multi-valued if $\CS(z)$ has simple poles
(to fix a single-valued branch, one has to make cuts from
$\infty$ to all simple poles of $\CS(z)$).
In neighborhoods of the points $z_{\alpha}\in {\sf D}_\alpha$, see \rf{a}, \rf{o},
one has
\be
\label{JV}
\CS(z)dz=\sum_{k\geq 1}\tau_k d{\sf b}_k(z),
\;\;\;\;\;
{\sf b}(z)=\sum_{k\geq 1}\tau_k {\sf b}_k(z)
\ee
Formula (\ref{n34})
shows that for the algebraic domains the variables
$\Pi_{\alpha}$, introduced in the general case as formal
quantities, are well-defined and  equal to the $A$-periods
of the differential $\tilde z(z)dz$ on the Schottky double $\Sigma$.
Indeed, using the fact that $C^0(z)$ and $C^{\alpha}(z)$
represent restrictions of the {\em same} function $\CS(z)$,
one can use representation (\ref{n201}) from Appendix~\ref{app:co}
and rewrite it in the form
\be
\phi_{\alpha}=-2\, {\rm Re}\left(
\int_{0}^{z_{\alpha}}\CS(z) dz
+\int_{\xi_0}^{\xi_{\alpha}}(\bar z +C^{-} (z)-\CS(z))dz \right)
\ee
Combining this equality with the definition of $\Pi_{\alpha}$
(\ref{n34}), we obtain:
\be\label{period33}
\Pi_{\alpha}=2\, {\rm Re}
\int_{\xi_0}^{\xi_{\alpha}} (\tilde z(z)-\bar z)dz =
\int_{\xi_0}^{\xi_{\alpha}}
\left (\tilde z(z)dz -\bar z d \overline{\tilde z(z)} \right )=
\oint_{A_{\alpha}} \tilde zdz
\ee
As an example of algebraic domains,
it is instructive to consider
the case with only a finite number of the non-vanishing moments $\tau_k$, i.e.
$\tau_k=0$ for $k>n+1$, which corresponds to the two-matrix models with polynomial
potentials \rf{V2mm}. Then $\tilde z(z)$ extends to a meromorphic function on $\Sigma$
with a pole
of order $n$ at $\infty$ and a simple pole at $\bar{\infty}$.
The function $z$ extended to the lower sheet of the Schottky double
as $\overline{\tilde z(z)}$ has a simple pole at $\infty$ and a pole
of order $n$ at $\bar{\infty}$.
For such a domain ${\sf D^c}$
the moments with respect to the Laurent
basis (\ref{momt}), as in the simply-connected case,
coincide with the coefficients of the expansion of the Schwarz
function near $\infty$:
\be\label{n70}
\tilde z(z)=\sum_{k=1}^{n+1}kt_kz^{k-1}+O(z^{-1}),\ \ \ z\to\infty
\ee
literally coinciding with formula \rf{branch}.
The normal displacement of
the boundary of an algebraic domain, which changes the variable $t_k$
keeping all the other moments (and $\Pi_{\alpha}$) fixed
is defined by normal derivative of the function
$2\, {\rm Re}\, \int^z d \Omega_k$, where
$d\Omega_k$ is a second kind Abelian normalized
meromorphic differential \rf{2kA} on $\Sigma$  with the only
pole at $\infty$
\be\label{n60}
d\Omega_k=d(z^k+O(z^{-1})),\ \ \ \ \
\oint_{A_{\alpha}}d\Omega_k=0
\ee
Let also $d\Omega_0$ be a third-kind Abelian differential \rf{3kA}, \rf{bp} on $\Sigma$
with simple poles at two infinities $\infty$ and $\bar\infty$, its Abelian integral
\be\label{n71}
\log w(z)=\int^z_{\xi_0} d\Omega_0
\ee
defines in the neighborhood of $\infty$ a function $w(z)$ which
has a simple pole at infinity. The dependence of the inverse function
$z(w)$ on the variables $t_k$ is described by the Whitham
equations for the
two-dimensional Toda lattice hierarchy of the form
\be\label{n80}
\p_{t_k}^{\Pi} z(w)=\{\Omega_k(w),z(w)\}=
{d\Omega_k(w)\over d\log w}\p_{t_0}z(w)-
\p_{t_0}\Omega_k(w){dz\over d\log w}
\ee
In this case one can also write a quasi-homogeneity condition \cite{KriW}
for the quasiclassical tau-function, which acquires the form
\be\label{quasihom}
2\F_{\sf D}=-\frac{1}{2}\tau_{0}^{2} +\tau_0 v_0
+\frac{1}{2} \sum_{k \geq 1}(2-k)
(\tau_k \nu_k +\bar \tau_k \bar \nu_k )-
\sum_{\alpha =1}^{g}\Pi_{\alpha} S_{\alpha}
\ee
Algebraic domains of a more general form
correspond to the universal Whitham hierarchy \cite{KriW}, the discussion of
this issue can be found in \cite{KMZ}.

\subsection{The duality transformation and free energy of two-matrix model
\label{ss:dual}}

Passing from $\Pi_{\alpha}$ to $S_{\alpha}$ is a particular case of
duality transformation \rf{duality} which is equivalent to
the interchanging of the $A$ and $B$ cycles
on the Riemann surface $\Sigma$, see fig.~\ref{fi:riemann}.
For the quasiclassical tau-function, depending also upon some extra
$\mathbb{\tau}$-variables this is achieved by the
partial Legendre transform
$\F_{\sf D}(\Pi_{\alpha}, \mathbb{\tau} )
\rightarrow
\F(S_{\alpha}, \mathbb{\tau} )$,
where
\be\label{p1}
\F_{\sf D} =\F +\sum_{\alpha =1}^{g} \Pi_{\alpha}S_{\alpha}
\ee
The function $\F$ is dual prepotential to the tau-function of the Dirichlet problem,
it solves the modified Dirichlet problem and can be identified
with the free energy of two-matrix model in the planar
large $N$ limit in the case when the support of eigenvalues
consists of a few disconnected drops.

The main properties of $\F$ follow from those of
$\F_{\sf D}$. According to (\ref{n52}), (\ref{dual2}) we have
$d\F_{\sf D} = -\sum_{\alpha} S_{\alpha}d\Pi_{\alpha}
+\sum_k \nu_k d\tau_k$
(for brevity, $k$ is assumed to run over all integer values,
$\tau_{-k}\equiv \bar \tau_k$), so
$d\F = \sum_{\alpha} \Pi_{\alpha}dS_{\alpha}
+\sum_k \nu_k d\tau_k$.
This gives the first order derivatives:
\be\label{f4a}
\Pi_{\alpha}=  \frac{\d \F}{\d S_{\alpha}}\,,
\;\;\;\;
\nu_{k}= \frac{\d \F}{\d \tau_k}
\ee
The second order derivatives are transformed as follows
(see e.g.\,\cite{dWM}): set
\be
{\F_{\sf D}}_{\alpha \beta}=\frac{\d^2 \F_{\sf D}}{\d \Pi_{\alpha} \d \Pi_{\beta}}\,,
\;\;\;
{\F_{\sf D}}_{\alpha k}=\frac{\d^2 \F_{\sf D}}{\d \Pi_{\alpha} \d \tau_k}\,,
\;\;\;
{\F_{\sf D}}_{i k}=\frac{\d^2 \F_{\sf D}}{\d \tau_i \d \tau_k}
\ee
and similarly for $\F$, then
\be\label{p2}
{\F_{\sf D}}_{\alpha \beta}=-(\F^{-1})_{\alpha \beta}
\\
{\F_{\sf D}}_{\alpha k}=\sum_{\gamma =1}^{g}(\F^{-1})_{\alpha \gamma}
\F_{\gamma k}
\\
{\F_{\sf D}}_{i k}=\F_{i k} -
\sum_{\gamma , \gamma ' =1}^{g}
\F_{i \gamma }
(\F^{-1})_{\gamma \gamma '}
\F_{\gamma ' k}
\ee
where $(\F^{-1})_{\alpha \beta}$ means the matrix element
of the matrix inverse to the $g\times g$ matrix
$\F_{\alpha \beta}$.

Using these formulas,
it is easy to see that the main properties
(\ref{elem5}) of the tau-function
are translated to the dual tau-function as follows:
\be\label{p3}
\tilde G(z, \zeta )
=\log |z^{-1}-\zeta^{-1}|
+\frac{1}{2} \nabla  (z) \nabla  (\zeta)\F
\\
2\pi i \, \tilde \varpi_{\alpha}(z)=
-\, \p_{S_{\alpha}}\! \nabla (z) \F
\\
2\pi i \, \tilde T_{\alpha \beta}=
\frac{\p^2 \F}{\p S_{\alpha} \p S_{\beta} }
\ee
where
$\tau_k$-derivatives in $\nabla (z)$ are taken at
fixed $S_{\alpha}$. The objects in the left hand sides
of these relations are:
\be\label{modified}
\tilde G(z,\zeta)=G(z,\zeta)+i\pi
\sum_{\alpha, \beta =1}^{g}\omega_{\alpha}(z)
\tilde T_{\alpha \beta} \, \omega_{\beta}(\zeta)
\\
\tilde \omega_{\alpha}(z)=
\sum_{\beta =1}^{g}\tilde T_{\alpha \beta}\,
\omega_{\beta}(z)
\\
\tilde T=-T^{-1}
\ee
The function $\tilde G$ is the Green function of the modified
Dirichlet problem, see \cite{KMZ}.
The matrix $\tilde T$ is the matrix of $A$-periods of the
holomorphic differentials $d\tilde \omega_\alpha$
on the double $\Sigma$ (so that $\tilde\varpi_\alpha(z) = \tilde \omega_\alpha(z)
+ \overline{\tilde \omega_\alpha(z)}$), normalized with respect to
the $B$-cycles
$\oint_{B_\alpha}d\tilde \omega_\beta = -\delta_{\alpha\beta}$,
$\oint_{A_\alpha}d\tilde \omega_\beta = \tilde T_{\alpha\beta}$,
i.e. more precisely, the change of cycles under duality transformation
is $A_{\alpha} \to B_{\alpha}$,
$B_{\alpha} \to -A_{\alpha}$.

An important fact is that by a simple rescaling of the independent variables
one is able to write
the group of relations (\ref{p3})
for the function $\F$ in exactly the same form (\ref{elem5}),
so that they differ merely by notation. Therefore we will not distinguish between
these two cases in \cite{MM2} when studying equations, satisfied by
quasiclassical tau-functions of matrix models.

\section{Conclusion}

In the second part of this paper \cite{MM2}
we consider certain examples of applications of the above
methods and discuss a similar
quasiclassical geometric picture, arising already in the context of multidimensional
gauge theories and the AdS/CFT correspondence. We also planning to discuss shortly the
open problems
and speculations how this picture can be treated beyond the quasiclassical limit.

\bigskip\bigskip\noindent
I am grateful to L.Chekhov, V.Kazakov, I.Krichever, A.Losev, A.Mironov, A.Morozov,
N.Nekrasov, D.Vasiliev and A.Zabrodin for collaboration and discussion of various
issues discussed in this paper.
This work was partially supported by RFBR grant 04-01-00642, the grant for
support of Scientific
Schools 1578.2003.2, the NWO project 047.017.015, the ANR-05-BLAN-0029-01
project "Geometry and
Integrability in Mathematical Physics" and the Russian Science Support
Foundation.

\section*{Appendix}
\appendix
\setcounter{equation}0
\section{Rational degenerations
\label{ap:rat}}

Substituting \rf{COMAP} into \rf{complcu} and computing the residues
one finds that the expressions
\be\label{RESEQ}
R_l[F]={\rm res}\left({dw\over w}w^lF(z(w),\tilde z(w))\right) =0
\ee
for $l= -n(n+1),\dots,n(n+1)$ form a triangular system of equations
onto the coefficients $f_{ij}$. It means that each of the equations
\rf{RESEQ} is linear in one of the coefficients, and can be resolved step
by step, starting from the ends of the chain.

For the cubic potential ($n=2$) the solution is
\be
a =  - {   {r^{2}}\over{{  u_2}}}
\\
b = {   {{  u_1}\,r}\over{{  u_2}}}  - 2\,{  {\bar u}_0}
\\
c =  - {  {{  u_1}\,r\,{  {\bar u}_0}}\over{{  u_2}}}
 + {  {\bar u}_0}^{2} - 2\,r\,{  {\bar u}_1} + 3\,{  {r^{2}
\,{  u_0}}\over{{  u_2}}}  - {  {r^{3}\,{  {\bar u}_1}}\over {
{  {\bar u}_2}\,{  u_2}}}
\\
f = r^{2} - 2\,{  u_2}\,{  {\bar u}_2} + 4\,{  u_0}\,{  {\bar u}_0} +
{  {r^{4}}\over{{  {\bar u}_2}\,{  u_2}}}  - {  u_1}\,
{  {\bar u}_1} - 2\,{  {r\,{  {\bar u}_1}\,{  {\bar u}_0}}\over{{
{\bar u}_2}}}  + {  {r^{2}\,{  {\bar u}_1}\,{  u_1}}\over{{  {\bar u}_2
}\,{  u_2}}}  - 2\,{  {r\,{  u_0}\,{  u_1}}\over{
{  u_2}}}
\\
q =  - 3\, {{r^2\,  u_0^2}\over{  u_2}}  + 2u_2{\bar u}_2{\bar u}_0 -
u_2 {\bar u}_1^2 - 2\,{  {r^{2}\,  {\bar u}_1^2}\over{{  {\bar u}_2}}}
 - {  {r^{4}\,{  {\bar u}_0}}\over{{  {\bar u}_2}\,{  u_2}}}
 - 3\,r\,{  {\bar u}_2}\,{  u_1} + {  {\bar u}_0}\,{  u_1}\,{  {\bar u}_1}
\\
+ 4\,
{  u_0}\,r\,{  {\bar u}_1} - 2\,{  {\bar u}_0}^{2}\,{  u_0} - {  {
\,r\,{  {\bar u}_1}}{  u_1}^{2}\over{{  u_2}}}  - r^{2}\,{  {\bar u}_0} + 2\,
{  {r\,\,{  u_1}{  {\bar u}_0}\,{  u_0}}\over{{  u_2}}
}  + 3\,{  {r^{3}\,{  u_1}}\over{{  u_2}}}
\\
 +
{  {r\,{  {\bar u}_1}\,{  {\bar u}_0}^{2}}\over{{  {\bar u}_2}}}  -
{  {r^{2}\,{  {\bar u}_0}\,{  u_1}\,{  {\bar u}_1}}\over{{
{\bar u}_2}\,{  u_2}}} + 2\,{  {{  u_0}\,r^{3}\,{  {\bar u}_1}
}\over{{  {\bar u}_2}\,{  u_2}}}
\\
h =  - {  {r^{6}}\over{{  {\bar u}_2}\,{  u_2}}
}  + {  {r^{2}\,{  {\bar u}_0}^{3}}\over{{  {\bar u}_2}}}  +
{  {\bar u}_2}\,{  {\bar u}_0}\,{  u_1}^{2} - {  {{  u_0}
\,r\,{  {\bar u}_1}\,{  {\bar u}_0}^{2}}\over{{  {\bar u}_2}}}  +
{  u_2}\,{  u_0}\,
{  {\bar u}_1}^{2} - 3\,{  {r^{3}\,{  {\bar u}_0}\,{  {\bar u}_1}
}\over{{  {\bar u}_2}}}  - {  {{  {\bar u}_2}\,{  u_1}^{3}\,r}\over{
{  u_2}}}
\\
 - {  {{  u_2}\,r\,{  {\bar u}_1}^{3}}\over{
{  {\bar u}_2}}}  + {  {r^{2}\,{  u_0}^{3}}\over{{  u_2}
}}  + 2\,{  {r^{4}\,{  {\bar u}_1}\,{  u_1}}\over{{  {\bar u}_2
}\,{  u_2}}}  + {  {{  u_0}\,{  {\bar u}_0}\,{
{\bar u}_1}\,r^{2}\,{  u_1}}\over{{  {\bar u}_2}\,{  u_2}}}  - {
{r^{3}\,{  u_1}\,{  {\bar u}_0}^{2}}\over{{  {\bar u}_2}\,{  u_2}}}  -
{  {r^{3}\,{  {\bar u}_1}\,{  u_0}^{2}}\over{{  {\bar u}_2}\,
{  u_2}}}  - {  {{  u_1}^{2}\,{  {\bar u}_1}^{2}\,r
^{2}}\over{{  {\bar u}_2}\,{  u_2}}}
\\
 + 3\,r^{4} + {  {\bar u}_0}^{2}\,{  u_0}^{2} +
{  {{  u_1}^{2}\,{  {\bar u}_1}\,{  u_0}\,r}\over{{
u_2}}}  - 2\,{  {\bar u}_0}^{2}\,{  u_1}\,r + 2\,{  {
{  u_1}^{2}\,r^{2}\,{  {\bar u}_0}}\over{{  u_2}}}  + {
{{  u_1}\,{  {\bar u}_1}^{2}\,r\,{  {\bar u}_0}}\over{{  {\bar u}_2}}}
+ r^{2}\,{  {\bar u}_0u_0}
\\
 - {  u_1}\,{  {\bar u}_1}\,{  u_0}\,{  {\bar u}_0} +
{  {{  u_0}\,{  {\bar u}_0}\,r^{4}}\over{{  {\bar u}_2}\,{
u_2}}}  + {  u_2}^{2}\,{  {\bar u}_2}^{2} - {  {{
u_1}\,r\,{  {\bar u}_0}\,{  u_0}^{2}}\over{{  u_2}}}
- 2\,r\,{  {\bar u}_1}\,
{  u_0}^{2} + 2\,{  {{  u_0}\,r^{2}\,{  {\bar u}_1}
^{2}}\over{{  {\bar u}_2}}}  - 3\,{  {r^{3}\,{  u_0}\,
{  u_1}}\over{{  u_2}}}
\\
 - {  {\bar u}_1}\,{  u_1}\,{  {\bar u}_2}\,{  u_2} - 3\,r^{2}
\,{  {\bar u}_2}\,{  u_2} + 3\,{  u_2}\,r\,{  {\bar u}_1}\,{  {\bar u}_0} - 2\,
{  {\bar u}_2}\,{  u_2}\,{  {\bar u}_0}\,{  u_0} - {  {\bar u}_1}\,r^{2}\,{  u_1
} + 3\,{  u_0}\,r\,{  {\bar u}_2}\,{  u_1}
\ee
together with the "complex conjugated" expressions for ${\bar a}$,
${\bar b}$, ${\bar c}$ and ${\bar q}$, where one should replace $u_k$
by $\bar u_k$ and vice versa. Resolving \rf{RESEQ} one gets the explicit
description of the rational degeneration of the curve \rf{complcu} in
terms of the coefficients of conformal map \rf{COMAP}. However, in general
situation they are only implicitly defined through the parameters of
the potential $V(z,{\bar z})$.
The simplest
example of such degeneration is given by
\be
\label{racu}
z^2{\bar z}^2 - {1\over 3{\bar t}}z^3 - {1\over 3t}{\bar z}^3 +
\left(t_0+{1\over 9t{\bar t}} - 18t_0^2t{\bar t}\right)
z{\bar z} +
\left(3t_0^2(1-3t{\bar t}) - 27t_0^3t{\bar t} - {t_0\over 9t{\bar t}}\right)
= 0
\ee
and this equation can be resolved via conformal map
\be
z = \sqrt{t_0}w + {3t_0{\bar t}\over w^2},
\ \ \ \
{\bar z} = {\sqrt{t_0}\over w} + 3t_0tw^2
\ee

\setcounter{equation}0
\section{Co-ordinates in multiply-connected case
\label{app:co}}

The existence of a well-defined dual basis of differentials \rf{o},
obeying the orthonormality relation \rf{n3}, is the key feature
of the basis functions ${\sf a}_{k}$ \rf{a}, which makes their moments $\tau_k$
\rf{t1} good local coordinates, in contrast to $M_{n, \alpha}$, since for the
functions $(z-z_{\alpha})^{-n}$ one cannot define the dual basis.

The summation formulas
\be
{dz d\zeta \over \zeta -z}=
\sum_{n=1}^{\infty} d \zeta {\sf a}_{n}(\zeta )
d{\sf b}_{n}(z), \ \ \
|\prod_{\beta=0}^g (z-z_{\beta})| < |\prod_{\beta=0}^g (\zeta-z_{\beta})|
\label{n4}\\
{dz d\zeta\over \zeta -z}=
-\sum_{n=0}^{\infty} d\zeta {\sf a}_{-n}(\zeta )d{\sf b}_{-n}(z),\ \ \
|\prod_{\beta=0}^g (z-z_{\beta})|>|\prod_{\beta=0}^g (\zeta-z_{\beta})|
\ee
which can be checked directly, allow us to repeat arguments
of sect.~\ref{ss:simply}.
Indeed, the Cauchy integral (\ref{C})
\be\label{CC}
C(z)dz=
{dz\over 2\pi i}\oint_{\p {\sf D}} {\bar \zeta d\zeta
\over \zeta -z}
\ee
where the integration now goes along all boundary components,
defines in each of the holes ${\sf D_{\alpha}}$ analytic differentials
$C^{\alpha}(z)dz$ (analogs of $C^{+}(z)dz$ in the simply-connected case).
In the complementary domain ${\sf D^c}$ the
Cauchy integral still defines the differential
$C^{-}(z)dz$ holomorphic everywhere in
${\sf D^c}$ except for infinity where it has a simple pole.
The difference of the boundary values of the Cauchy integral
is equal to $\bar z$:
\be
C^{\alpha}(z)-C^{-} (z)=\bar z\,,
\;\;\;\;\;\; z\in \p {\sf D}_{\alpha}
\ee
From equation (\ref{C1}), which can be written separately for each contour, it follows
that the difference of the boundary values of the derivative of the Cauchy
integral (\ref{CC})
\be
\left(\p_t C^{\alpha} (\zeta)-\p_t C^{-}(\zeta)\right)d\zeta
\ee
is, for all $\alpha$,
a purely imaginary differential on the boundary $\d{\sf D}_{\alpha}$.
The expansion (\ref{n4}) of the Cauchy kernel implies that
if a $t$-deformation preserves all the
moments $\tau_k$, $k\geq 0$, then
$\p_t \bar\zeta d\zeta-\p_t\zeta d\bar\zeta$
extends to a holomorphic
differential in ${\sf D^c}$.

Indeed, for $z$ close enough
to {\em any} of the points $z_{\alpha}$, one can
expand $\p_t C^{\alpha}(z)$ for each $\alpha$ as
\be\label{n5}
\p_t C^{\alpha}(z)dz={1\over 2\pi i}\sum_{k=1}^{\infty} d{\sf b}_{k}
\p_t \left (\oint_{\p {\sf D}}
{\sf a}_{k}(\zeta)\bar \zeta d\zeta \right )=
\sum_{k=1}^{\infty}
\p_t \tau_{k}\ d{\sf b}_{k}(z)
\ee
and conclude that it is vanishes identically provided
$\p_t \tau_k =0$. Hence $\p_t C^{-}(z)dz$ is the desired extension
of $\p_t\bar\zeta d\zeta-\p_t \zeta d\bar\zeta$ and has no
pole at infinity due to $\p_t \tau_0=0$.

Using the Schwarz symmetry principle we extend $\p_t C^{-}(z)dz$ to
a holomorphic differential on the Schottky double. If the variables $S_{\alpha}$
\rf{dual4} are also preserved under the $t$-deformation,
this holomorphic differential has
vanishing periods along all cycles $B_{\alpha}$ and therefore,
it identically vanishes. This completes the proof of the statement, that
any deformation of the domain preserving all
$\tau_k$ and $S_{\alpha}$ is trivial.
In this proof the variables $S_{\alpha}$ were used
only at the last moment, to show that extension of $\p_t C^{-}(z)dz$
as a holomorphic differential
on the Schottky double $\Sigma$ is trivial; instead
the variables $\phi_{\alpha}$ \rf{n0} can be used
in a similar way.

Let us show that if they are preserved under
$t$-deformation then the $A_{\alpha}$-periods of the extension of
$\p_t C^{-}(z)dz$ vanish,
and therefore this extension is identically zero.
Indeed, the variable $\phi_{\alpha}$
(\ref{n0}) can be represented in the form
\be\label{n20}
\phi_{\alpha}=-{2\over\pi}{\rm Re}\int_{0}^{z_{\alpha}}
dz\int_{{\sf D^c}}
\frac{d^2 \zeta}{z-\zeta}
\ee
The differential
$\frac{dz}{\pi}\int_{{\sf D^c}}
\frac{d^2 \zeta}{z-\zeta}$ is equal to
$C^{\alpha}(z)dz$ for $z\in {\sf D}_{\alpha}$
and $(\bar z + C^{-}(z))dz$ for $z\in {\sf D^c}$.
Let $\xi_0$, $\xi_{\alpha}$ be the points where the
integration path from $z=0$ to $z=z_{\alpha}$ intersects
the boundary contours $B_0$, $B_{\alpha}$. Then
\be
\label{n201}
\phi_{\alpha}=-2\, {\rm Re} \left(
\int_{0}^{\xi_0}C^{0} (z)dz +
\int_{\xi_{\alpha}}^{z_{\alpha}}C^{\alpha} (z)dz +
\int_{\xi_0}^{\xi_{\alpha}}(\bar z + C^{-} (z))dz \right)
\ee
It is already shown above that if
a $t$-deformation preserves the variables
$\tau_k$ then all $\p_t C^{\alpha}(z)dz=0$.
Thus vanishing of the $t$-derivative $\p_t \phi_{\alpha} =0$ implies
\be\label{n21}
0=- \p_t \phi_{\alpha} =
2\, {\rm Re}\int_{\xi_0}^{\xi_{\alpha}} \p_tC^{-}(z)dz
\ee
The r.h.s of this equation is just the
$A_{\alpha}$-period of the holomorphic extension
of the differential $\p_tC^{-}(z)dz$.

\end{document}